\documentclass[twocolumn,floatfix,showpacs,prd,aps,tightenlines]{revtex4}
\usepackage{amsmath}
\usepackage{graphicx}
\usepackage{psfrag}
\usepackage{color}
\usepackage{dcolumn}
\usepackage{bm}
\usepackage{longtable}
\usepackage[mathscr]{eucal}
\usepackage{mathrsfs}
\usepackage{xspace}

\def\msun{\rm M_{\odot}}
\def\kms{\rm km \, s^{-1}}

\def\simlt{\mathrel{\rlap{\lower 3pt\hbox{$\sim$}}\raise 2.0pt\hbox{$<$}}}
\def\simgt{\mathrel{\rlap{\lower 3pt\hbox{$\sim$}} \raise 2.0pt\hbox{$>$}}}
\def\lsim{\mathrel{\rlap{\lower 3pt\hbox{$\sim$}}\raise 2.0pt\hbox{$<$}}}
\def\gsim{\mathrel{\rlap{\lower 3pt\hbox{$\sim$}} \raise 2.0pt\hbox{$>$}}}

\def\msunpc3{\msun~{\rm {pc^{-3}}}}
\newcommand{\be}{\begin{equation}}
\newcommand{\ee}{\end{equation}}
\def\kms{{\rm\,km\,s^{-1}}}

\newcommand{\hangup}{{\it hangup kick}\xspace}
\newcommand{\super}{{\it superkick}\xspace}
\newcommand{\cross}{{\it cross kick}\xspace}

\newcommand{\bea}{\begin{eqnarray}}
\newcommand{\eea}{\end{eqnarray}}
\newcommand{\beq}{\begin{equation}}
\newcommand{\eeq}{\end{equation}}
\newcommand{\KMS}{\rm km\ s^{-1}}

\begin{document}

\def\fun#1#2{\lower3.6pt\vbox{\baselineskip0pt\lineskip.9pt
  \ialign{$\mathsurround=0pt#1\hfil##\hfil$\crcr#2\crcr\sim\crcr}}}
\def\lap{\mathrel{\mathpalette\fun <}}
\def\gap{\mathrel{\mathpalette\fun >}}
\def\kms{{\rm km\ s}^{-1}}
\def\vk{V_{\rm recoil}}

\title{Nonlinear Gravitational Recoil from
the Mergers of Precessing Black-Hole Binaries}

\author{
Carlos O. Lousto}
\author{
Yosef Zlochower
}
\affiliation{Center for Computational Relativity and Gravitation,\\
and School of Mathematical Sciences, Rochester Institute of
Technology, 85 Lomb Memorial Drive, Rochester, New York 14623}

\begin{abstract}
We present results from an extensive study of 88 precessing,
equal-mass black-hole binaries with large spins (83 with intrinsic 
spins $|\vec{S}_i/m_i^2|$ of 0.8 and 5 with intrinsic spins of $0.9$),
and use these data to model new
nonlinear contributions to the gravitational recoil imparted to the
merged black hole. We find a new effect, the \cross, that enhances the recoil
for partially aligned binaries beyond the {\em hangup kick}
effect. This has the consequence of increasing the probabilities  
of recoils larger than $2000\ \KMS$ by
nearly a factor two, and,
consequently, of black holes
getting ejected from galaxies, as well as the
observation of large differential redshifts/blueshifts in the cores of
recently merged galaxies. 
\end{abstract}

 \maketitle

\section{Introduction}\label{sec:Introduction}

The studies of black-hole binaries (BHBs) that immediately followed
the 2005 breakthroughs in numerical relativity~\cite{Pretorius:2005gq,
Campanelli:2005dd, Baker:2005vv} soon revealed the importance of spin
to the orbital dynamics~\cite{Campanelli:2006uy}.  One of the most
striking result was the unexpectedly large recoil velocity imparted to
the remnant due to an intense burst of  gravitational radiation
around merger~\cite{Campanelli:2007ew, Gonzalez:2007hi}.  Recoil
velocities as large as $4000\ \KMS$ were predicted for maximally
spinning black holes~\cite{Campanelli:2007cga} (in a configuration
with both spins lying in the orbital plane, known as the {\em
superkick} configuration).
This prediction, which was based on a model for the recoil velocities 
that was linear in the individual spins of 
the
merging holes~\cite{Campanelli:2007ew, Lousto:2009ka}, 
triggered  several astronomical searches for recoiling
supermassive black holes as the byproduct of galaxy collisions,
producing several dozen potential candidates~\cite{Komossa:2008qd,
Shields:2008va,Bogdanovic:2008uz,Comerford:2009ju,Civano:2010es,
Eracleous:2011ua,Tsalmantza:2011ju,Civano:2012bu,Blecha:2012kx}.
See Ref.~\cite{Komossa:2012cy} for a review.

\begin{figure}
  \includegraphics[width=0.9\columnwidth]{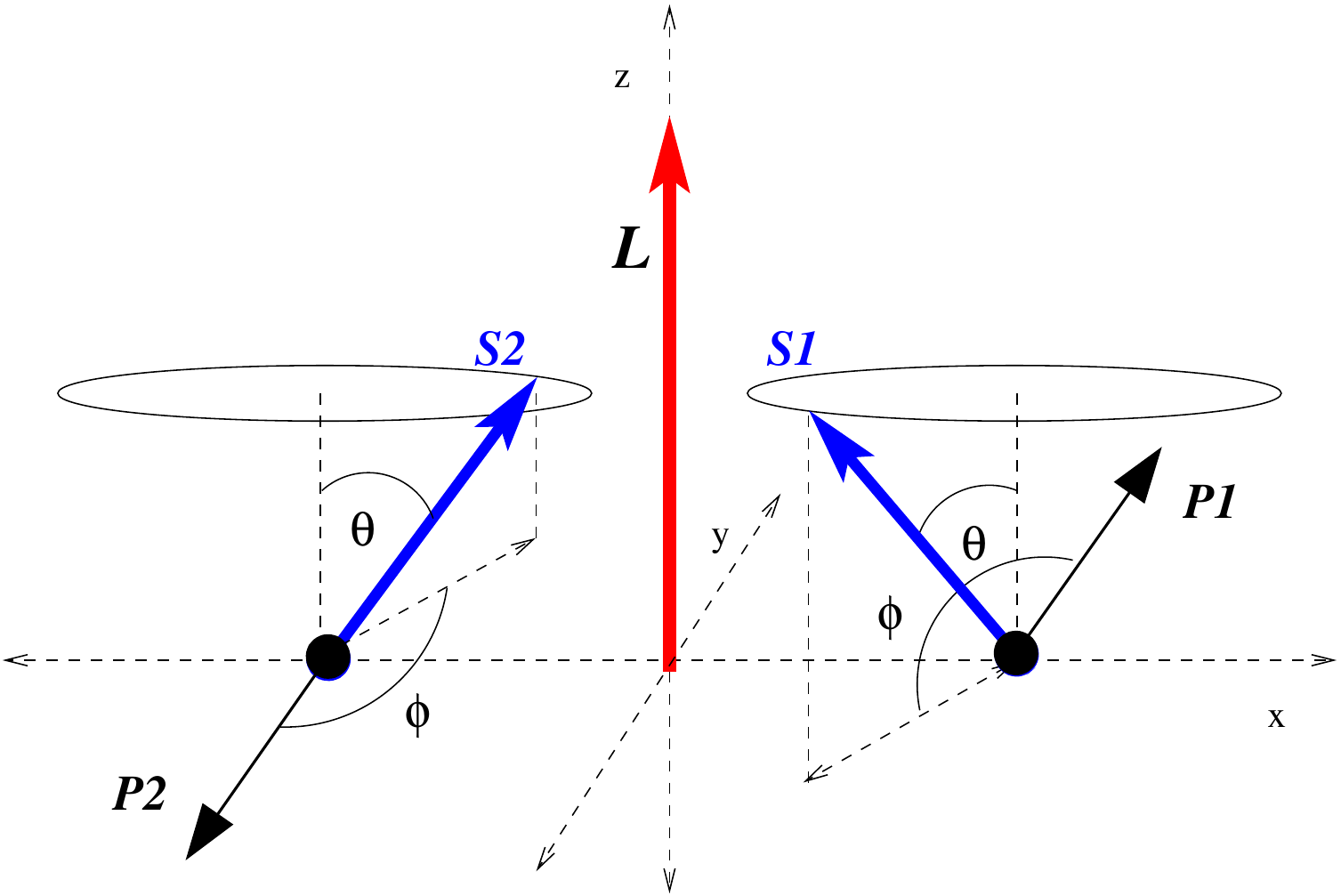}
  \caption{The \hangup configuration. Here $S_{1z} = S_{2 z}$, while
$S_{1x}=-S_{2x}$ and $S_{1y}=-S_{2y}$. The \hangup configurations
are preserved exactly by numerical evolutions.}
 \label{fig:hangup}
\end{figure}

Accretion effects~\cite{Bogdanovic:2007hp,Dotti:2009vz} would tend to
align the spins of the BHs with the orbital angular momentum,
suppressing the {\em superkick} and, apparently, the likelihood of
observing large recoils. We recently found~\cite{Lousto:2011kp, Lousto:2012su}
however, that there are  nonlinear spin couplings that
lead to even larger recoil velocities when the spins are partially
aligned with the orbital angular momentum. These  so-called {\em hangup
kick} recoils can be as large as 
$5000\ \KMS$ (see Fig.~\ref{fig:hangup}).

In this paper we continue our exploration of unexpectedly large
nonlinear contributions to the net recoil~\cite{Lousto:2011kp,
Lousto:2012su}.  Here we concentrate on equal-mass 
BHBs that precess. Our ultimate  goal is to derive an empirical
formula that takes into account all major contributions to the recoil (at
least to the level of a few percent accuracy). This would be a near
hopeless task if we just started with a set of random configurations.
Rather, we propose a program for developing sets of configurations
with exact or approximate symmetries that allow us to model the recoil
term by term.  For example, in the \hangup
configurations~\cite{Lousto:2011kp, Lousto:2012su}, the BHBs can be
described by two parameters, the $z$ component of the total spin
$S_z$, and the in-plane component of $\vec \Delta$ ($\vec \Delta
\propto \vec S_2-\vec
S_1$ in the equal-mass case). If the generic recoil is also a function
of $\Delta_z$, those terms would be suppressed in the \hangup
configurations. Here we continue the exploration by evolving
configurations that activate different possible terms for the recoil.

Not all nonlinear terms in the spins lead to large increases in the
recoil~\cite{Rezzolla:2007xa, Lousto:2010xk}. We therefore need to
perform many diverse simulations to try to elucidate which nonlinear
terms contribute significantly and which can still be ignored.  Here
we explore the effects of precession on recoils.  We perform
simulations of equal-mass, precessing BHBs, but also
discuss the more general unequal-mass case. We also extend the
phenomenological formulas for predicting recoils to include higher
powers of the spins, explicitly including up to fourth order, making
use of discrete symmetry properties of the BHBs.

\section{Higher order recoil velocity expansions}\label{Sec:Symmmetries}

Our approach to the modeling of recoil velocities of merged black
holes is based on the numerical evidence that the vast majority of the
recoil is produced by the anisotropic emission of gravitational
radiation at the very last stage of the merger, i.e., when a common
horizon forms (see, e.g., Fig~\ref{fig:rec_v_time} below). 
In order to model the
dependence of the recoil on the spins of the individual holes (and the
BHB's mass ratio), we were guided by the leading-order post-Newtonian
(PN)
expressions for the instantaneous radiated linear momentum~\cite{Kidder:1995zr}
(higher-order PN couplings can be found in Ref.~\cite{Racine:2008kj}).
As such, we will use the typical PN variables,
individual BH masses $m_1$ and $m_2$, total mass $m = m_1 + m_2$, mass
difference $\delta m = (m_1-m_2)/m$,
mass ratio $q=m_1/m_2$, symmetric mass
ratio $\eta = q/(1+q)^2$, total spin $\vec S = \vec S_1 + \vec S_2$, 
and $\vec \Delta = m(\vec S_2/m_2 - \vec S_1/m_1)$ (where $\vec S_i$ is
the spin of BH $i$). For convenience, we also define dimensionless
spin parameters $\vec \alpha_i = \vec S_i/m_i^2$.
In terms of $\vec \alpha_i$, the values of $\vec S$ and $\vec \Delta$ are given
by $\vec S = m^2 (\vec \alpha_2 + q^2\vec \alpha_1)/(1+q)^2$ and
$\vec \Delta = m^2 (\vec \alpha_2 - q \vec \alpha_1)/(1+q)$.
 Of particular
importance here will be the components of $\vec S$ and $\vec \Delta$
along the direction of angular momentum $\hat L$, which we denote by the
subscript $\|$, and the projection of the vector into the plane
orthogonal to $\hat L$, which we
denote by the subscript $\perp$. Thus, any vector can be written
as $\vec V  = \vec V_\| + \vec V_\perp$, where $\vec V_\| = (\hat
L\cdot \vec V)\hat L$ and $\vec V_\perp = \vec V - \vec V_\|$.

Although the PN approximation is not valid at the moment of
the merger (PN theory does not even account for horizons), our ansatz
is that the parameter dependence in these PN expressions yields a
useful starting point for constructing empirical formulas for the
recoil.  Thus, a PN expression of the form $\vec F(\vec r, \vec P)
\cdot \vec \Delta$, where $\vec F$ is a vector in the orbital plane,
becomes a fitting term on our formula of the form $A \Delta_\perp
\cos(\phi)$, where $\phi$ is the angle between $\vec \Delta_\perp$ and
some weighted {\it averaged} direction of $\vec F$, and  $A$ is a
fitting constant. However, in general, we do not know the  weighted
{\it averaged} direction of $\vec F$ and instead measure the angle
$\phi$ with respect to a fiducial direction $\hat n$ (typically, we
choose $\hat n = \hat r_1 - \hat r_2$, the direction from BH2 to BH1) 
and add an angular fitting
constant $\phi_0$ to the formula, i.e.\ $A \Delta_\perp
\cos(\phi-\phi_0)$. (The value of $\phi_0$ obtained from the fit then
gives the relative orientation between our fiducial $\hat n$ and the
 weighted {\it averaged} direction of $\vec F$).

We then verify that such PN-inspired formulas are accurate
{\it a posteriori} by comparing our predictions to recoil results from
other simulations.  This is the basis of our phenomenological approach
to the modeling of recoil velocities and can be summarized in an
expression of the three components of the linear velocity in the
individual spins of the 
holes~\cite{Campanelli:2007ew, Campanelli:2007cga, Lousto:2009mf}:
\begin{eqnarray}\label{eq:Pempirical}
\vec{V}_{\rm recoil}(q,\vec\alpha)&=&V_m\,\hat{e}_1+
V_\perp(\cos\xi\,\hat{e}_1+\sin\xi\,\hat{e}_2)+V_\|\,\hat{L},\nonumber\\
\end{eqnarray}
where
\begin{eqnarray}\label{eq:Pempiricalcont}
V_m&=&A_m\frac{\eta^2(1-q)}{(1+q)}\left[1+B_m\,\eta\right],\nonumber\\
V_\perp&=&H\frac{\eta^2}{(1+q)}\left[
(1+B_H\,\eta)\,(\alpha_2^\|-q\alpha_1^\|)\right.\nonumber\\
&&\left.+\,H_S\,\frac{(1-q)}{(1+q)^2}\,(\alpha_2^\|+q^2\alpha_1^\|)\right],\nonumber\\
V_\|&=&K\frac{\eta^2}{(1+q)}\Bigg[
(1+B_K\,\eta)
\left|\vec\alpha_2^\perp-q\vec\alpha_1^\perp\right|
\nonumber \\ && \quad \times
\cos(\phi_\Delta-\phi_1)\nonumber\\
&&+\,K_S\,\frac{(1-q)}{(1+q)^2}\,\left|\vec \alpha_2^\perp+q^2\vec \alpha_1^\perp\right|
\nonumber \\ && \quad \times
\cos(\phi_S-\phi_2)\Bigg],
\end{eqnarray}
$\hat{e}_1,\hat{e}_2$ are
orthogonal unit vectors in the orbital plane, and $\xi$ measures the
angle between the unequal mass and spin contribution to the recoil
velocity in the orbital plane. 
 The angles $\phi_{\Delta}$ and $\phi_S$ are defined as the angles
between the in-plane components $\vec \Delta_\perp$  and $\vec
S_\perp$,
 respectively
and a fiducial direction at merger (see Ref.~\cite{Lousto:2008dn} for
a description of the technique). 
Phases $\phi_1$ and $\phi_2$ depend
on the initial separation of the holes for quasicircular orbits.
(Astrophysically realistic evolutions of comparable masses BHs
lead to nearly zero eccentricity mergers.)

Note that the expression for $V_m$ was determined
in Refs.~\cite{Gonzalez:2006md, 1983MNRAS.203.1049F}.
The current estimates for the above parameters
are\cite{Gonzalez:2006md, Lousto:2008dn,
Zlochower:2010sn} : $A_m = 1.2\times 10^{4}\ \KMS$, $B_m = -0.93$, $H
= (6.9\pm0.5)\times 10^{3}\ \KMS$, $K=(5.9\pm0.1)\times 10^4\ \KMS$,
and $\xi \sim 145^\circ$, and $K_S=-4.254$. Here we set $B_H$ and
$B_K$ to zero, which is consistent with the error estimates
in~\cite{Lousto:2009mf}

Additional corrections from the \hangup effect, and a new effect,
which we will dub the \cross effect, are examined here. We note that
these new effects were found using equal-mass BHBs; thus their
dependence on mass ratio is still speculative.

In our recent studies of BHB mergers we found that nonlinear
terms in the spin play an important role in modeling recoil velocities
\cite{Lousto:2011kp}. Here we will
investigate higher-order models for the recoil velocity
based on the symmetry
properties of its components~\cite{Boyle:2007sz}. 
In Boyle {\it et al.},~\cite{Boyle:2007sz} a new method for developing
empirical formulas for the remnant BH properties was proposed. This
new method was based on a Taylor expansion, using symmetry properties 
to limit the total number of terms.  We combine the two
methods by using PN-inspired variables for a Boyle {\it et al.} type 
of expansion. Fundamental to this construction is the 
behavior of the BHB under discrete
operations  such as exchange $(X)$ of the black holes' labels
$(1\longleftrightarrow2)$ and parity $(P)$
$(x\to-x, y\to-y, z\to-z)$. 

\subsection{Comparing Expansion  Variables}
We model higher-order contributions
to the recoil using the PN variables
$\vec{\Delta}$, $\vec{S}$, $\eta$, and $\delta m$~\cite{Racine:2008kj}.
Note that we could use an alternative set of variables, which at first
glance would appear to be simpler,
such as $\vec S^\pm/m^2=(\vec S_1\pm \vec S_2)/m^2$ and drop
the  explicit $\eta$ dependence (which can be reabsorbed in $\delta m$).
For example, the recoil contribution due to unequal mass can
be expressed as
\begin{equation}
V_m=a.\delta m + b.\delta m^3+\cdots,
\end{equation}
and because
\begin{equation}
\eta=(1-\delta m^2)/4,
\end{equation}
this is equivalent to the more usual (See Eq. (\ref{eq:Pempiricalcont}) above)
\begin{equation}
V_m=\eta^2.\delta m.(A+B.\eta+\cdots)
\end{equation}
The same equivalence can be shown for the variable
$\vec\Delta=m.(\vec{S}_2/m_2-\vec{S}_1/m_1)$. That is, since
\begin{equation}
q=(1-\delta m)/(1+\delta m),
\end{equation}
we find
\begin{equation}
\vec\Delta=(\vec{S}_2-\vec{S}_1)
-2.\delta m.(\vec{S}_2+\vec{S}_1)+2.\delta
m^2.(\vec{S}_2-\vec{S}_1)+\cdots,
\end{equation}
and hence $\vec \Delta$ can be reexpressed 
in terms of the alternative spin variables.

\begin{figure}
\includegraphics[width=\columnwidth]{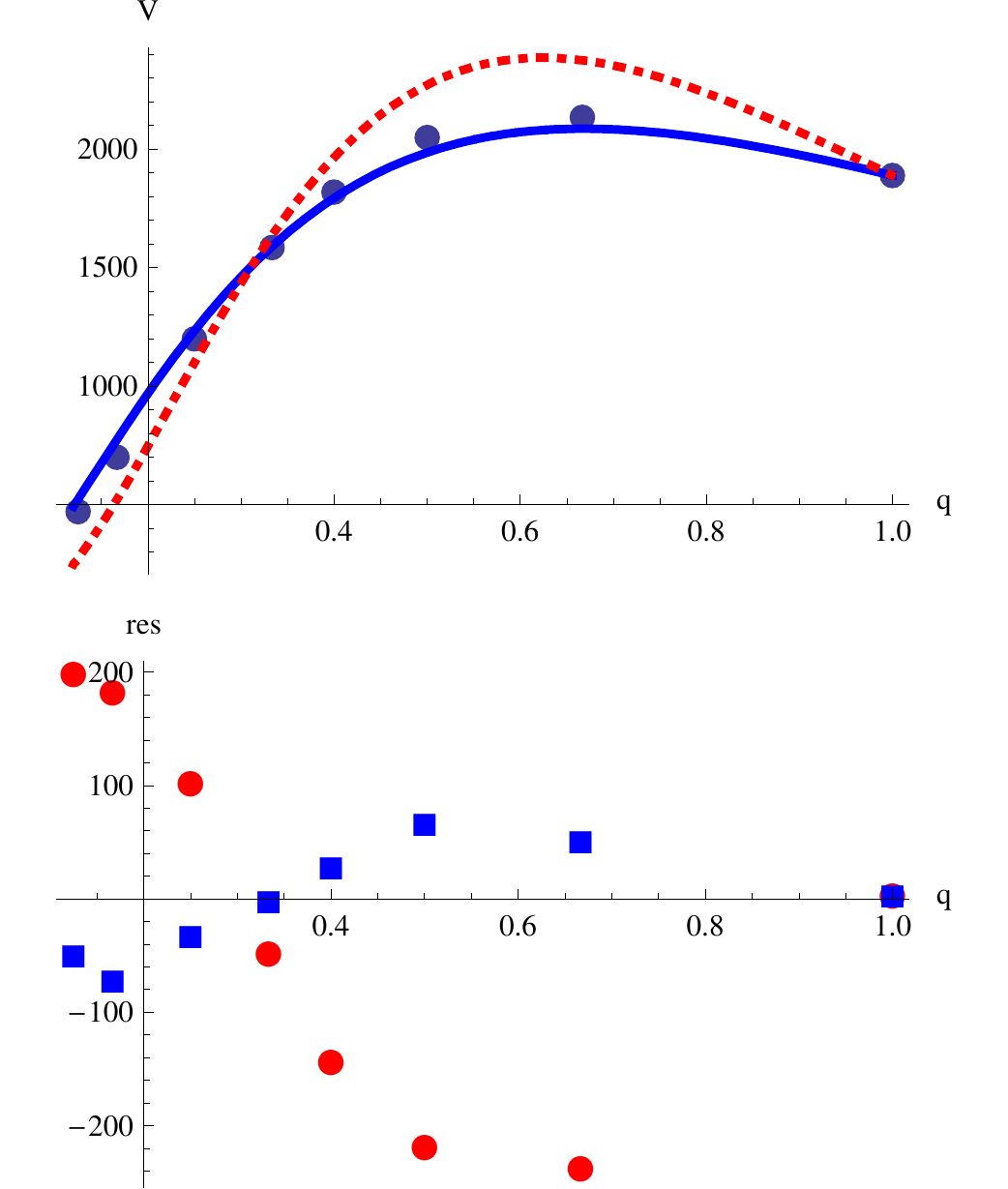}
\caption{ A fit of the data from~\cite{Lousto:2008dn} to determine if the
leading dependence of the recoil is proportional to $\Delta_\perp$
(blue) or $S^-$ (red - dotted). The fits were constructed to reproduce
the same equal-mass limit.}
\label{fig:fit_for_S_D}
\end{figure}

We can investigate which choices of expansion variables give
the best fits with the fewest number of terms.
For example, we can explore if the variables $\vec S$ and $\vec
\Delta$ are really more advantageous then the pair $\vec S_{\pm}$.
To verify that the leading-order contribution to the
out-of-plane recoil is best fit using $\vec \Delta$, we revisit the results
from a previous paper~\cite{Lousto:2008dn}, where we considered the
case of a larger spinning BH,
with spin in the orbital plane, and a smaller nonspinning BH. In
Fig.~\ref{fig:fit_for_S_D},
we show the results of fitting those data to the forms
$$V_\| \approx \frac{K 2^{b-1} \alpha_2 \eta^2}{(1+q)^b} \mbox{ and }
V_\| \approx \frac{K \alpha_2 (4 \eta)^{b'}}{16 (1+q)^2}.$$
The former assumes a
leading $\eta^2$ dependence, but distinguishes between $\Delta_\perp$
[i.e., $\alpha_2/(1+q)$] and $S^-_\perp$ [i.e.,  $\alpha_2/(1+q)^2$],
while the latter is used to find the best leading power of $\eta$
assuming the spin-dependence is proportional to $S^-$ and choosing
functions that reproduce the equal-mass limit.
We find that the best fit parameters are
$b=0.993\pm0.038$ and  $b'=3.3\pm0.2$. 
The results
clearly indicate that the leading-order recoil is best fit by
$\eta^2 \Delta_\perp$.

After finding that the \super effect is best modeled using
$\eta^2 \vec\Delta$ as the spin variable, we can motivate our ansatz for a
leading $\eta^2$ for the remaining spin-dependent terms in
 our empirical formula 
without
appealing to PN theory.
To do this, we examine the particle limit and use perturbation theory.
Perturbative theory applies in the
 small-mass-ratio limit, but is otherwise a relativistic theory, and unlike the 
post-Newtonian approximations, it applies in the strong, highly-dynamical, 
regime. 

 Since the radiative perturbative modes are proportional to $q$,
and the radiation of the linear momentum is proportional
to the surface integrals of squares of these modes, the instantaneous radiated
linear momentum is proportional to $q^2$ (which, by symmetry
considerations,
 generalizes to $\eta^2$).

For the linear-in-spin terms, one can use the decomposition in
Ref.~\cite{Nakano:2010kv}, where the spin of the 
large black hole is considered a
perturbation of a nonrotating, Schwarzschild BH.
Since this is a dipolar (Odd) $\ell=1$ term, it is nonradiative, and, in order
to generate a radiative term, it  must be coupled with an 
$\ell\geq2$ radiative perturbation.
To generate linear momentum, this spin-dependent
radiative mode must couple with a non-spin dependent radiative mode
(otherwise, the spin would enter at quadratic order).
 Again, the leading-order terms in the
recoil are proportional to $q^2$.

\subsection{Symmetry considerations}

A Taylor expansion of a function with  $v$ independent variables 
of a given order of expansion $o$ has 
$n$ terms, where $n$ is given by
\cite{2005mmp..book.....A}
\begin{equation}\label{eq:terms}
n=\frac{(o+v-1)!}{o!\,(v-1)!}.
\end{equation}
However, only certain combinations of variables are allowed.
 In order to take into account the correct combinations of variables
for each component of the recoil velocity at a given order, we consider
the symmetry properties summarized in Table \ref{table:symmetries}.
The possible terms to a given expansion order in spin (i.e., products
of $S$ and $\Delta$) are summarized in
Tables~\ref{table:numberterms}-\ref{table:Vp}.
The terms in Tables~\ref{table:Vz}
and~\ref{table:Vp} are all multiplied by fitting coefficients. 
Note that the coefficients of these terms can depend on higher
powers of $\delta m$ (even powers for terms proportional to $\delta m^0$ and
odd powers for terms proportional to $\delta m$).

\begin{table}
\caption{Symmetry properties of key quantities}
\label{table:symmetries}
\begin{tabular}{|l|c|c|}
\hline\hline
Symmetry & P & X \\
\hline\hline
$S_\perp/m^2=(S_1+S_2)_\perp/m^2$ & -- & -- \\
\hline
$S_\|/m^2=(S_1+S_2)_\|/m^2$ & + & + \\
\hline
$\Delta_\perp/m^2=(S_2/m_2-S_1/m_1)_\perp/m$ & -- & + \\
\hline
$\Delta_\|/m^2=(S_2/m_2-S_1/m_1)_\|/m$ & + & -- \\
\hline
$\hat{n}=\hat{r}_1-\hat{r}_2$ & + & -- \\
\hline
$\delta m=(m_1-m_2)/m$ & + & -- \\
\hline
$V_\perp$ & + & -- \\
\hline
$V_\|$ & -- & + \\
\hline
\end{tabular}
\end{table}

\begin{table}
\caption{Number of possible terms at a given order of expansion (with
respect to $\vec S$ or $\vec \Delta$). Here 1 indicates terms present even in the
equal-mass limit (and/or proportional to even powers of $\delta m$) 
and  $\delta m$ 
indicates terms proportional to $\delta m$ to odd powers.}
\label{table:numberterms}
\begin{tabular}{|c|c|c|c|c|c|c|c|c|c|c|c|}
\hline\hline
Order & 0th & 0th & 1st & 1st & 2nd & 2nd & 3rd & 3rd & 4th & 4th & total\\
\hline
mass & 1 & $\delta m$ & 1 & $\delta m$ & 1 & $\delta m$  & 1 & $\delta m$ & 1 & $\delta m$ & All \\
\hline\hline
$V_\|$ & 0 & 0 & 1 & 1 & 2 & 2 & 5 & 5 & 8 & 8 & 32\\
\hline
$V_\perp$ & 0 & 1 & 1 & 1 & 2 & 4 & 5 & 5 & 8 & 11 & 38\\
\hline
Total & 0 & 1 & 2 & 2 & 4 & 6 & 10 & 10 & 16 & 19 & 70\\
\hline\hline
\end{tabular}
\end{table}

\begin{widetext}

\begin{table}
\caption{Parameter dependence at each order of expansion
 for the off-plane recoil.
 Here 1 indicates terms present even in the
equal-mass limit (and/or proportional to even powers of $\delta m$) 
and  $\delta m$ 
indicates terms proportional to $\delta m$ to odd powers.}
\label{table:Vz}
\begin{tabular}{|c|l|}
\hline\hline
$V_\|$ & 0th order\\
\hline
1 & 0\\
\hline
$\delta m$ & 0\\ 
\hline\hline
$V_\|$ & 1st order\\
\hline
1 & $\Delta_\perp$\\
\hline
$\delta m$ & $S_\perp$\\ 
\hline\hline
$V_\|$ & 2nd order\\
\hline
1 & $\Delta_\perp.S_\|+\Delta_\|.S_\perp$\\
\hline
$\delta m$ & $\Delta_\perp.\Delta_\|+S_\perp.S_\|$\\ 
\hline\hline
$V_\|$ & 3rd order\\
\hline
1 & $\Delta_\|.S_\perp.S_\|+\Delta_\perp.S_\|^2+\Delta_\perp.\Delta_\|^2+\Delta_\perp^3+\Delta_\perp.S_\perp^2$\\
\hline
$\delta m$ & $S_\perp.\Delta_\|^2+S_\perp.S_\|^2+\Delta_\perp.\Delta_\|.S_\|+S_\perp.\Delta_\perp^2+S_\perp^3$\\ 
\hline\hline
$V_\|$ & 4th order\\
\hline
1 & $S_\perp.\Delta_\|^3+\Delta_\perp.S_\|^3+\Delta_\perp.S_\|.\Delta_\|^2+S_\perp.\Delta_\|.S_\|^2+\Delta_\perp^3.S_\|+S_\perp^3.\Delta_\|+\Delta_\perp^2.S_\perp.\Delta_\|+\Delta_\perp.S_\perp^2.S_\|$\\
\hline
$\delta m$ & $\Delta_\perp.\Delta_\|^3+S_\perp.S_\|^3+S_\perp.S_\|.\Delta_\|^2+\Delta_\perp.\Delta_\|.S_\|^2+\Delta_\perp^3.\Delta_\|+S_\perp^3.S_\|+\Delta_\perp^2.S_\perp.S_\|+S_\perp^2.\Delta_\perp.\Delta_\|$\\ 
\hline\hline
\end{tabular}
\end{table}

\begin{table}
\caption{Parameter dependence at each order of expansion for the
in-plane recoil. Here 1 indicates terms present even in the
equal-mass limit (and/or proportional to even powers of $\delta m$) 
and  $\delta m$ 
indicates terms proportional to $\delta m$ to odd powers.}
\label{table:Vp}
\begin{tabular}{|c|l|}
\hline\hline
$V_\perp$ & 0th order\\
\hline
1 & 0\\
\hline
$\delta m$ & 1\\ 
\hline\hline
$V_\perp$ & 1st order\\
\hline
1 & $\Delta_\|$\\
\hline
$\delta m$ & $S_\|$\\ 
\hline\hline
$V_\perp$ & 2nd order\\
\hline
1 & $\Delta_\|.S_\|+\Delta_\perp.S_\perp$\\
\hline
$\delta m$ & $\Delta_\|^2+S_\|^2+\Delta_\perp^2+S_\perp^2$\\
\hline\hline
$V_\perp$ & 3rd order\\
\hline
1 & $\Delta_\perp.S_\perp.S_\|+\Delta_\|.S_\|^2+\Delta_\|.\Delta_\perp^2+\Delta_\|^3+\Delta_\|.S_\perp^2$\\
\hline
$\delta m$ & $S_\|.\Delta_\|^2+S_\|.S_\perp^2+\Delta_\perp.S_\perp.\Delta_\|+S_\|.\Delta_\perp^2+S_\|^3$\\ 
\hline\hline
$V_\perp$ & 4th order\\
\hline
1 & $S_\perp.\Delta_\perp^3+\Delta_\|.S_\|^3+\Delta_\|.S_\|.\Delta_\perp^2+S_\perp.\Delta_\perp.S_\|^2+\Delta_\|^3.S_\|+S_\perp^3.\Delta_\perp+\Delta_\|^2.S_\perp.\Delta_\perp+\Delta_\|.S_\perp^2.S_\|$\\
\hline
$\delta m$ & $\Delta_\perp.\Delta_\|.S_\perp.S_\|+\Delta_\perp^4+\Delta_\|^4+S_\perp^4+S_\|^4+\Delta_\perp^2.\Delta_\|^2+\Delta_\perp^2.S_\perp^2+\Delta_\perp^2.S_\|^2+\Delta_\|^2.S_\perp^2+\Delta_\|^2.S_\|^2+S_\perp^2.S_\|^2$\\
\hline\hline
\end{tabular}
\end{table}

\end{widetext}

Interestingly,  for the $\delta m-$independent terms, we can obtain the
spin-dependence of the in-plane recoil from the spin-dependence of the
out-of-plane recoil via
\begin{equation}
V_\perp=V_\|[\Delta_\perp\longleftrightarrow\Delta_\|],
\end{equation}
while for the $\delta m-$dependent terms that are odd powers in the
spin variables,
we have
\begin{equation}
V_\perp(\delta m)=\delta m.V_\|[S_\perp\longleftrightarrow S_\|].
\end{equation}
On the other hand, for terms proportional to even powers of the
spin
variables,
there are extra terms not present in $V_\|$.

In addition, functionally, the terms proportional to $\delta m$ in 
$V_\|$ can be obtained from the $\delta m$-independent terms in $V_\|$ by 
\begin{equation}
V_\|(\delta m)=\delta m.V_\|[S_\perp\longleftrightarrow \Delta_\perp],
\end{equation}
while for odd powers of the spins {\it only}, the
$\delta m$-dependent terms in $V_\perp$ can be obtained from the
$\delta m$-independent terms via
\begin{equation}
V_\perp(\delta m)=\delta m.V_\perp[S_\|\longleftrightarrow \Delta_\|].
\end{equation}
No such correspondence holds for the $\delta m$-dependent
terms with odd powers in the spin for $V_\perp$.

\subsection{The equal-mass case}
\label{sec:equal}
Using the above properties we find 16 terms up to fourth-order in the spin that
contribute to the off-plane recoil velocity:
\begin{eqnarray}\label{eq:sixteen}
V_\|=&&\Delta_\perp+\Delta_\perp.S_\|+\Delta_\|.S_\perp+\nonumber\\
&&\Delta_\perp.\Delta_\|^2+\Delta_\perp^3+\Delta_\perp.S_\|^2+\nonumber\\
&&\Delta_\perp.S_\perp^2+\Delta_\|.S_\perp.S_\|+\nonumber\\
&&\Delta_\perp^3.S_\|+\Delta_\|.S_\perp^3+\Delta_\|^3.S_\perp+\Delta_\perp.S_\|^3+\nonumber\\
&&\Delta_\|.S_\perp.S_\|^2+\Delta_\|.\Delta_\perp^2.S_\perp+\nonumber\\
&&\Delta_\perp.S_\|.\Delta_\|^2+\Delta_\perp.S_\|.S_\perp^2+\cdots
\end{eqnarray}

We can regroup all these terms (assuming we can collect all $\perp$ terms)
in the following form
\begin{eqnarray}\label{eq:regroup}
V_\|=&&\Delta_\perp.(1+\Delta_\perp^2+\cdots).(1+\Delta_\|^2+\cdots).\nonumber\\
&&.(1+S_\perp^2+\cdots).(1+S_\|+S_\|^2+S_\|^3+\cdots)+\nonumber\\
&&+S_\perp.\Delta_\|.(1+\Delta_\perp^2+\cdots).(1+\Delta_\|^2+\cdots).\nonumber\\
&&.(1+S_\perp^2+\cdots).(1+S_\|+S_\|^2+\cdots).
\end{eqnarray}

The ten terms directly proportional to $\cos(\varphi)$ are those linear
to the subindex $\perp$

\begin{eqnarray}\label{eq:cosphi}
V_\|^{\cos\varphi}=&&\Delta_\perp.(1+\Delta_\|^2+\cdots).(1+S_\|+S_\|^2+S_\|^3+\cdots)+\nonumber\\
	+&&S_\perp.\Delta_\|.(1+\Delta_\|^2+\cdots).(1+S_\|+S_\|^2+\cdots)
\end{eqnarray}
This (symbolic) expression is the one we will use in this paper.

There is a subtlety in the above expansion. Because $S_\perp$ and
$\Delta_\perp$ are
vector quantities, terms like $\Delta_\perp (1+S_z+\cdots)$, etc., should
really be expressed as
$C_1 \vec \Delta\cdot \hat n_1 + C_2 \vec \Delta \cdot \hat n_2 S_z + \cdots$,
where $\hat n_1$ and $\hat n_2$ are unit vectors in the plane,
i.e., not only are there fitting constants $C_1, C_2, \cdots$, but
each coefficient also has its own angular dependence. We will return
to this issue in Sec.~\ref{Sec:Results}.

\section{Numerical Relativity Techniques}\label{Sec:Numerical}

For the black-hole binary (BHB) data presented here, both BHs have the
same mass, but they have different spins.  We use the {\sc TwoPunctures}
thorn~\cite{Ansorg:2004ds} to generate initial puncture
data~\cite{Brandt97b} for the BHB simulations described below. These
data are characterized by mass parameters $m_{p1/2}$, momenta $\vec
p_{1/2}$, spins $\vec S_{1/2}$, and coordinate locations $\vec
x_{1/2}$ of each hole. We obtain parameters for the location,
momentum, and spin of each BH using the 2.5 PN
quasicircular parameters. Here we choose to normalize the PN initial
data such that the total Arnowitt-Deser-Misner (ADM) energy is $1M$.
We obtain parameters $m_{p1/2}$ using an iterative procedure in order
to obtain a system where the two BHs have the same mass and the total
ADM energy is $1M$. This iterative procedure is most efficient when
the horizon masses and ADM energy can be obtained from the initial data
alone. For highly-spinning BHs ($\alpha=S/m^2\gtrsim0.9$),
a relatively large amount of energy
lies outside the BH. This energy is eventually absorbed, changing the
mass of the BH substantially (see, e.g.~\cite{Lousto:2012es}). We
therefore limit the spin of the BHs to $\alpha=\leq 0.8$ for all but a
few simulations.

We evolve these BHB data sets using the {\sc
LazEv}~\cite{Zlochower:2005bj} implementation of the moving puncture
approach~\cite{Campanelli:2005dd,Baker:2005vv} with the conformal
function $W=\sqrt{\chi}=\exp(-2\phi)$ suggested by
Ref.~\cite{Marronetti:2007wz}.  For the runs presented here, we use
centered, eighth-order finite differencing in
space~\cite{Lousto:2007rj} and a fourth-order Runge Kutta time
integrator. (Note that we do not upwind the advection terms.)

Our code uses the {\sc EinsteinToolkit}~\cite{Loffler:2011ay,
einsteintoolkit} / {\sc Cactus}~\cite{cactus_web} /
{\sc Carpet}~\cite{Schnetter-etal-03b}
infrastructure.  The {\sc
Carpet} mesh refinement driver provides a
``moving boxes'' style of mesh refinement. In this approach, refined
grids of fixed size are arranged about the coordinate centers of both
holes.  The {\sc Carpet} code then moves these fine grids about the
computational domain by following the trajectories of the two BHs.

We obtain accurate, convergent waveforms and horizon parameters by
evolving this system in conjunction with a modified 1+log lapse and a
modified Gamma-driver shift
condition~\cite{Alcubierre02a,Campanelli:2005dd,vanMeter:2006vi}, and
an initial lapse $\alpha(t=0) = 2/(1+\psi_{BL}^{4})$, where
$\psi_{BL}$ is the Brill-Lindquist conformal factor and is given by
$$
\psi_{BL} = 1 + \sum_{i=1}^n m_{i}^p / (2 |\vec r- \vec r_i|),
$$
where $\vec r_i$ is the coordinate location of puncture $i$.  The
lapse and shift are evolved with
\begin{subequations}
\label{eq:gauge}
\begin{eqnarray}
(\partial_t - \beta^i \partial_i) \alpha &=& - 2 \alpha K,\\
\partial_t \beta^a &=& (3/4) \tilde \Gamma^a - \eta \beta^a \,,
\label{eq:Bdot}
\end{eqnarray}
\end{subequations}
where we use $\eta=2$ for all simulations presented below.

We use {\sc AHFinderDirect}~\cite{Thornburg2003:AH-finding} to locate
apparent horizons.  We measure the magnitude of the horizon spin using
the {\it isolated horizon} (IH) algorithm detailed in Ref.~\cite{Dreyer02a}.
Note that once we have the horizon spin, we can calculate the horizon
mass via the Christodoulou formula
\begin{equation}
{m_H} = \sqrt{m_{\rm irr}^2 + S_H^2/(4 m_{\rm irr}^2)} \,,
\end{equation}
where $m_{\rm irr} = \sqrt{A/(16 \pi)}$, $A$ is the surface area of
the horizon, and $S_H$ is the spin angular momentum of the BH (in
units of $M^2$).  In the tables below, we use the variation in the
measured horizon irreducible mass and spin during the simulation as a
measure of the error in these quantities.  We measure radiated energy,
linear momentum, and angular momentum, in terms of the radiative Weyl
Scalar $\psi_4$, using the formulas provided in
Refs.~\cite{Campanelli:1998jv,Lousto:2007mh}. However, rather than
using the full $\psi_4$, we decompose it into $\ell$ and $m$ modes and
solve for the radiated linear momentum, dropping terms with $\ell \geq
5$.  The formulas in Refs.~\cite{Campanelli:1998jv,Lousto:2007mh} are
valid at $r=\infty$.  We extract the radiated energy-momentum at
finite radius and extrapolate to $r=\infty$ using both linear and
quadratic extrapolations. We use the difference of these two
extrapolations as a measure of the error.

Both the remnant parameter variation, and the variation in the
extrapolation to infinity of the radiation underestimate the actual
errors in the quantity of interest. However, because quantities like
the total radiated energy can be obtained from either extrapolations
of $\psi_4$ or, quite independently, from the remnant BHs mass, the
difference between these two is a reasonable estimate for the actual
error.

\begin{figure}
\includegraphics[width=\columnwidth]{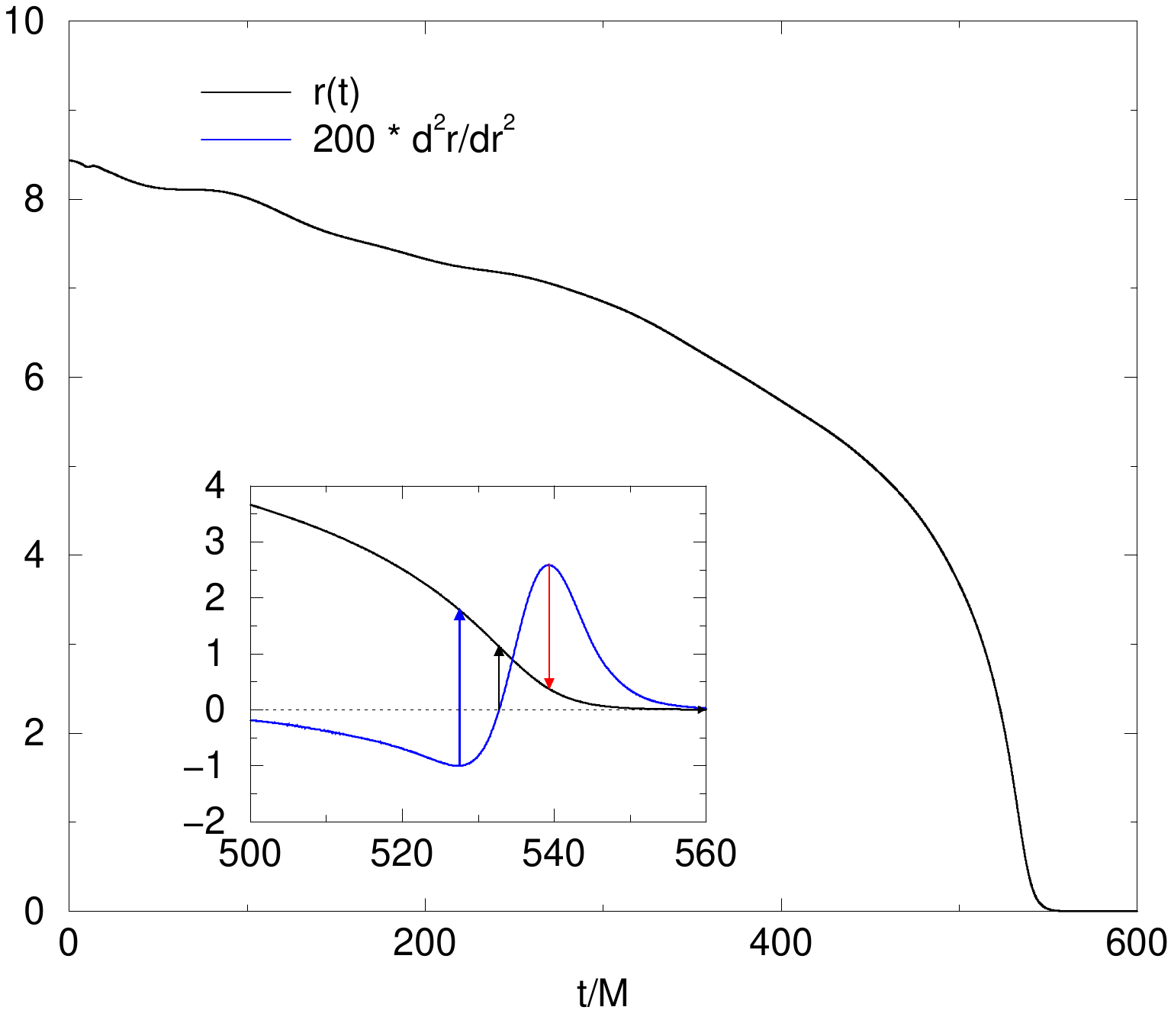}\\
\includegraphics[width=\columnwidth]{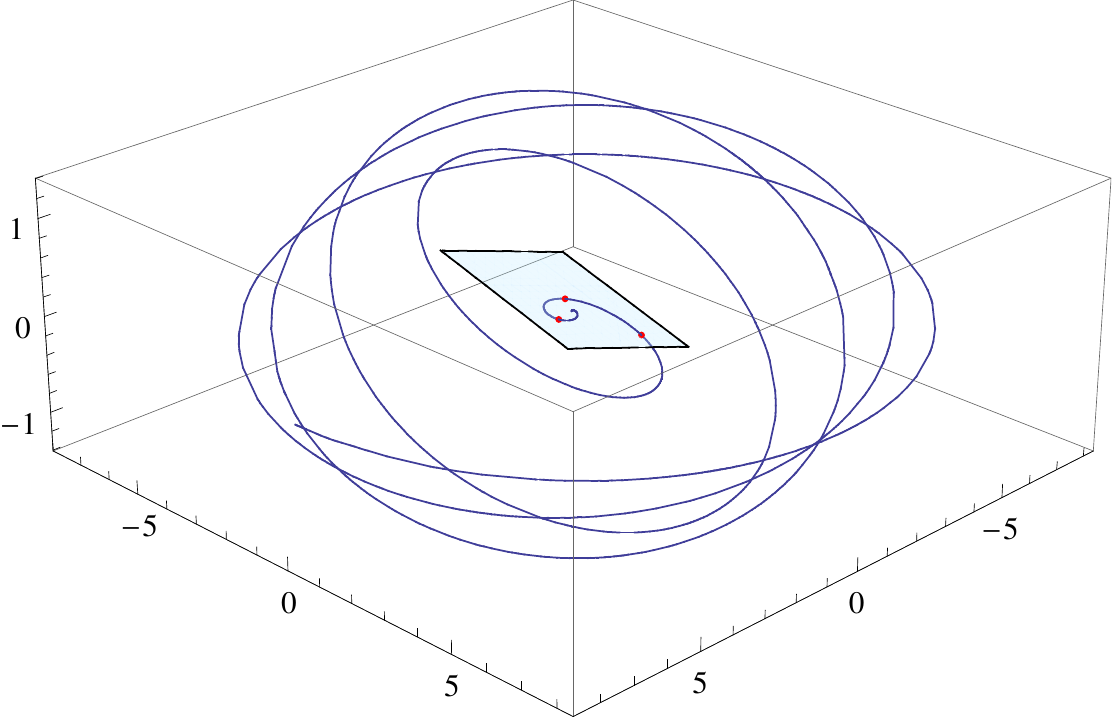}

\caption{Finding the orbital plane near merger. The upper plot
shows the orbital separation $r(t)$ versus time. The inset
shows $r(t)$ near merger and $\ddot r(t)$ (rescaled by 200 for
clarity). The points $\vec r_+$, $\vec r_0$, and $\vec r_-$ correspond
to the times where $\ddot r$ is maximized, zero, and minimized,
respectively
(denoted with arrows here). The plot below
shows the trajectory,  the points $\vec r_+, \vec r_0,
\vec r_-$  (large red dots) and the ``merger'' plane. }
\label{fig:find_plane}
\end{figure}

Our empirical formula will depend on the spins measured with respect
to the orbital plane at merger. In Ref~\cite{Lousto:2008dn} we
described a procedure for determining an approximate plane. This is
based on locating three fiducial points on the BHBs trajectory $\vec
r_+$, $\vec r_0$, and $\vec r_-$, where $\vec r_+$ is the point where
$\ddot r(t)$ ($r(t)$ is the orbital separation) reaches its maximum,
$\vec r_-$ is the point where $\ddot r(t)$ reaches its minimum, and
$\vec r_0$ is the point between the two where $\ddot r(t)=0$.  These
three points can then be used to define an approximate merger plane
(see Fig.~\ref{fig:find_plane}). 
We then need to rotate each trajectory such that the infall directions
all align (as much as possible).
This is accomplished by rotating the system, keeping the merger
plane's orientation fixed, such that the 
vector $\vec r_+ - \vec r_0$  is aligned with the $y$
axis. The azimuthal angle $\varphi$, described below, is measured in
this rotated frame.

\begin{figure}
\includegraphics[width=\columnwidth]{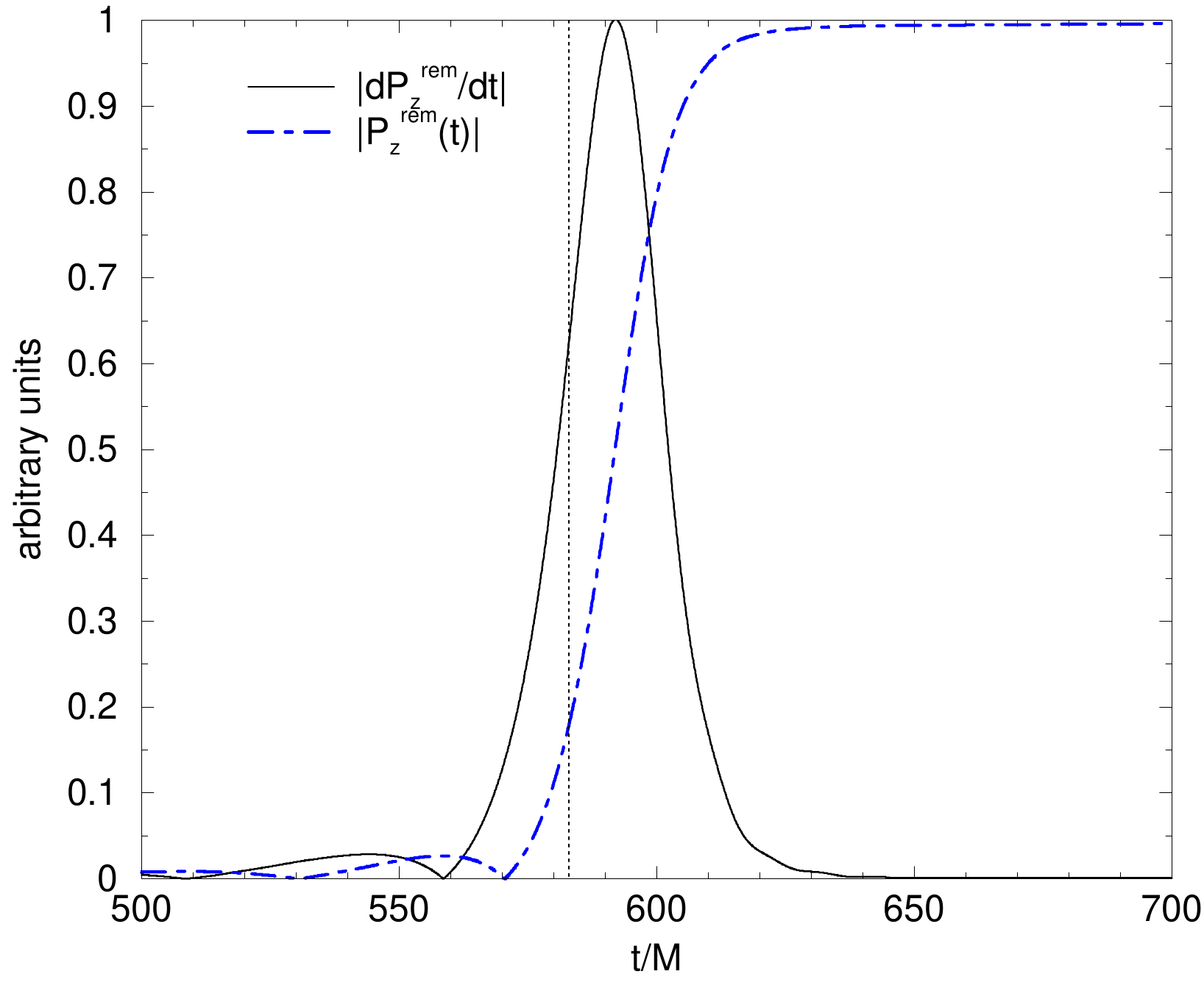}\\

\caption{A plot of the absolute values of $P_z(t)$ (the momentum
imparted to the remnant)  and $dP_z/dt(t)$ for  the NTH45PH30
configuration. The vertical line represents the approximate time of
merger.
}
\label{fig:rec_v_time}
\end{figure}

Our motivation for defining the orbital plane ``at merger'' is the
observation that most of the recoil is generated near (and slightly
after) merger. For example, Fig.~\ref{fig:rec_v_time} shows the recoil
imparted to the remnant BH for the N45PH30 configuration. As seen in
the plot, all but 16\% of the recoil is generated ``post-merger.''

\section{Simulations}\label{sec:Runs}

In this paper we consider four families of equal-mass, precessing,
BHB configurations, which we will denote by S, K, L, and N. Initial
data parameters are given in
Tables~\ref{tab:ID_part1}~and~\ref{tab:ID_part2} (found in
Appendix~\ref{app:ID}).
These configurations are characterized by the spins of the two BHs on
the initial slice. For 83 of the 88 simulations, the intrinsic spin of
each BH in the
binary $\alpha_i=0.8$,
with the exception of the N configurations, where the first BH has
spin $\alpha_1=0.8$ and the second is nonspinning. We also evolved a
set of five N configurations (denoted by N9 below) where the spin of
BH1 is $\alpha_1=0.9$.

For the S configurations, $\vec S_1=-\vec S_2$, i.e., the total
spin $\vec S$ is initially zero,
while for  the K configurations
$S_{1z} = -S_{2z}$ but $S_{1x}=S_{2x}$ and $S_{1y}=S_{2y}$.  The
L configurations have the spin of BH1 entirely in the orbital plane,
while the spin of BH2 is perpendicular to the plane, and finally
the N configurations have BH1 spinning and BH2 nonspinning.
We use the notation zTHxxxPHyyy, where z is N, N9, S, K, or L,
xxx gives the inclination angle $\theta$ of spin of BH1 and yyy gives the orientation
of the spin of BH1 in the initial orbital plane, i.e.\ the azimuthal
angle $\phi$. In order to fit the resulting recoils, we found that we
needed at least six
azimuthal configurations in the interval $[0,180^\circ)$ for each
$\theta$ configuration. This is due to the fact that we need to
separate contributions due to $\cos 3\phi$ from contributions due
to $\cos \phi$ (by symmetry, the recoil out of the plane 
cannot contain terms of the
form $\cos n\phi$ if $n$ is even).

In all cases but N9, the computational domain extended to $\pm400M$, with a
coarsest resolution of $h=4M$ at the outer boundary.
We used 9 levels of refinement, centered
on each puncture, with radii 200, 100, 50, 20, 10, 5, 2, 0.6,
respectively.  For the N9 configurations, the computational domain
extended to $\pm400M$, with a coarsest resolution of
$h=3.33M$, and we used an additional level
of refinement about BH1, with radius 0.35.

\begin{figure}
\includegraphics[width=0.9\columnwidth]{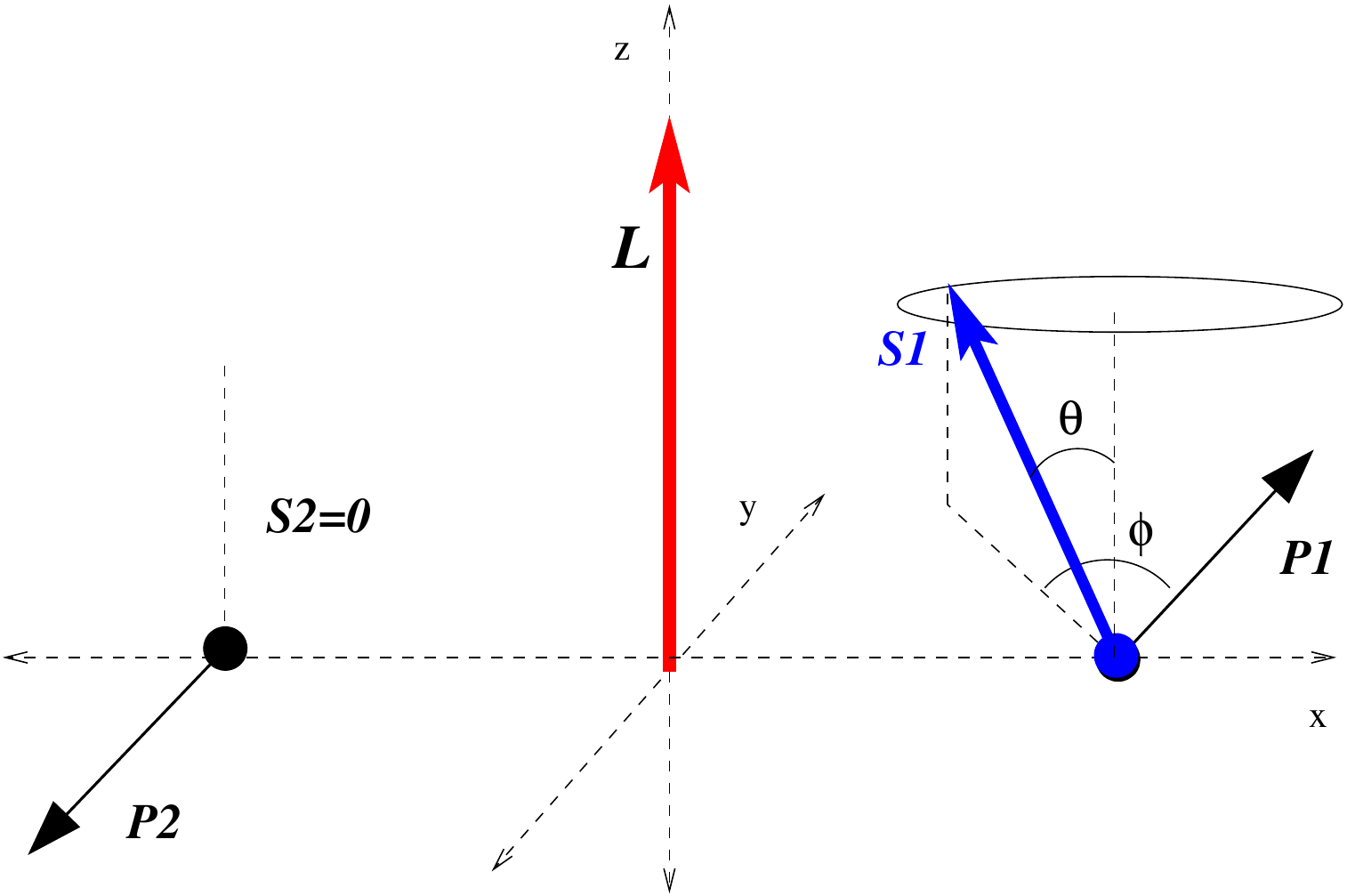}
\caption{The N configuration. These configurations differ from the
\hangup configurations in that BH2 is
nonspinning. Numerical evolutions preserve the N configurations only
approximately.}
\label{fig:N}
\end{figure}

\begin{figure}
\includegraphics[width=0.9\columnwidth]{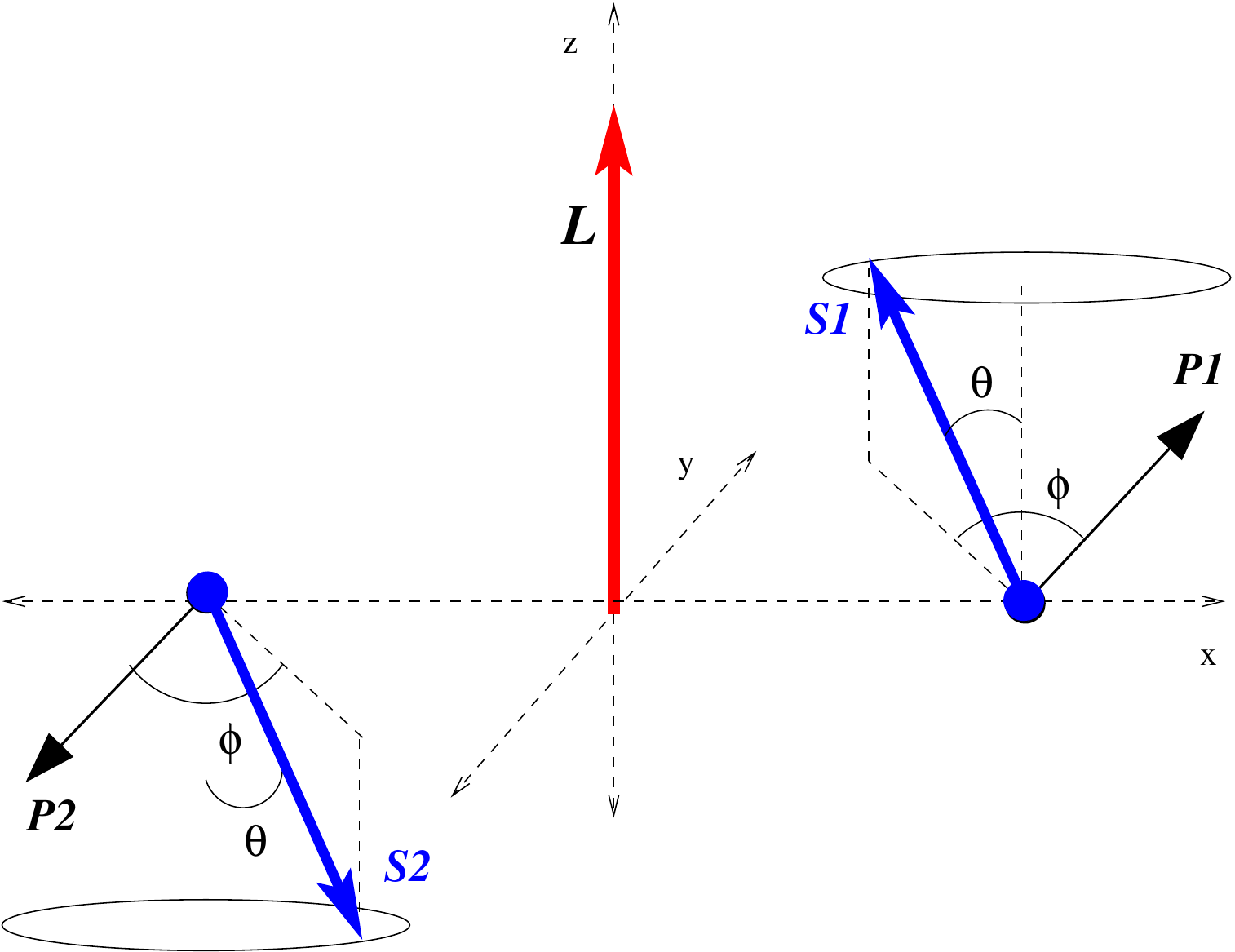}
\caption{The S configuration. These configuration differ from the
\hangup configuration in that $S_{1z}=-S_{2z}$ (and hence $\vec S_1 =
-\vec S_2$), initially. Numerical evolutions preserve the S
configurations only
approximately. }
\label{fig:S}
\end{figure}

\begin{figure}
\includegraphics[width=0.9\columnwidth]{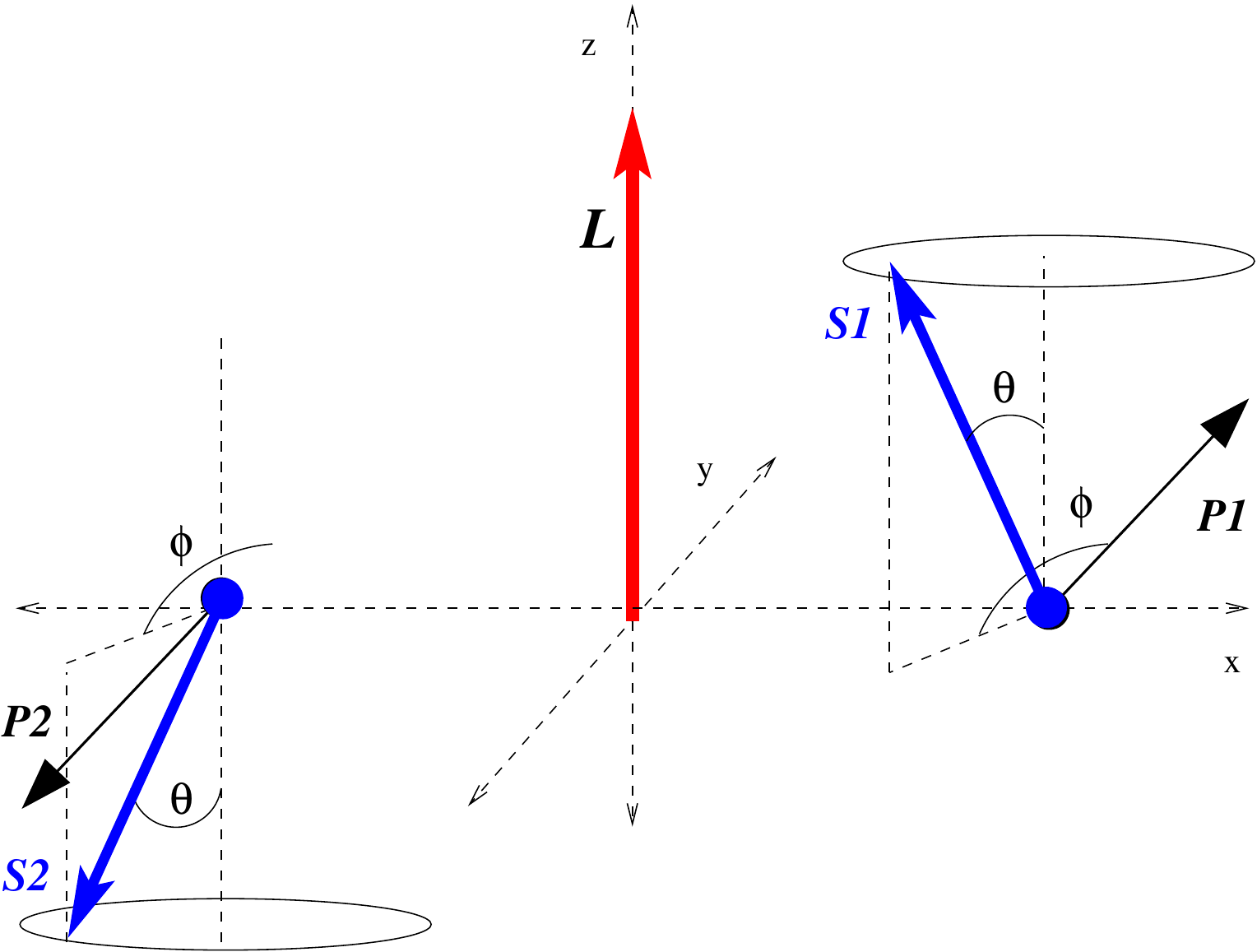}
\caption{The K configuration. These can be thought of as a
modification of the S configurations. There $S_{1z}=-S_{2z}$, while
$S_{1x}=S_{2x}$ and $S_{1y}=S_{2y}$, initially.
}
\label{fig:K}
\end{figure}

\begin{figure}
\includegraphics[width=0.9\columnwidth]{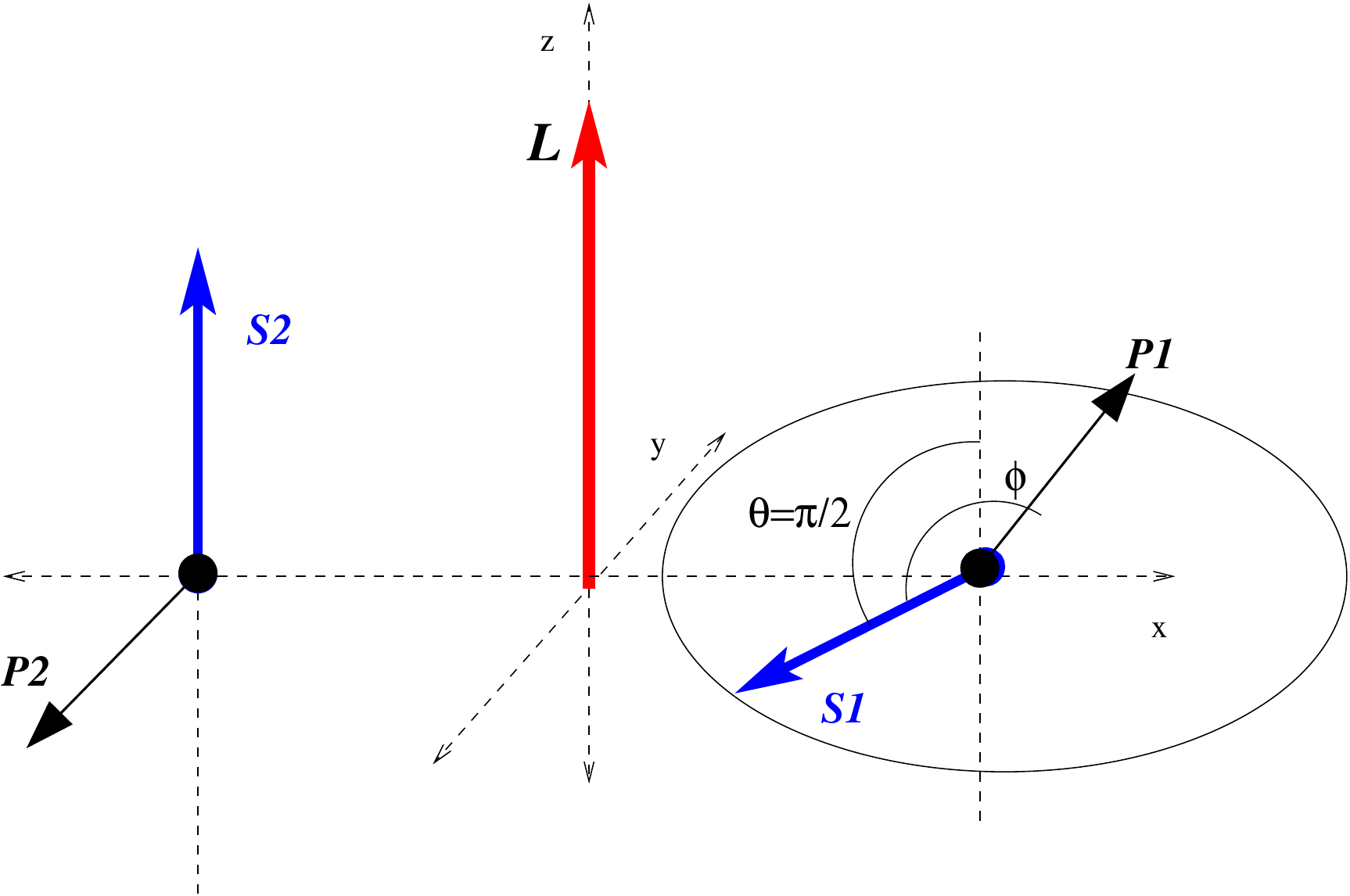}
\caption{The L configuration.  The L configuration is a modification
of the N configuration, where $S_1$ is aligned with the orbital 
angular momentum $\hat z$ and
rather than having $S_2=0$, $\vec S_2$
is varied initially in the orbital plane.}
\label{fig:L}
\end{figure}

We chose these configurations for two main reasons, first each family
of configurations can be described by a single azimuthal
angle parameter $\phi$ and a single polar angle $\theta$, and they activate different terms in our
ansatz for the recoil. The former is necessary in order to reduce the
computational costs. In general, four angular parameters are required
in order to describe the spins at merger (two polar and two
azimuthal). In order to model the polar and azimuthal dependence, we
would need at least $6\times6\times6\times6=1296$ simulations (per
choice of spin magnitude, per choice of mass ratio). By reducing the
dimensionality to two, we only need 36 simulations (per family, per
$\alpha$, per $q$). This reduction only works, however, if the two
parameter family of initial data, maps in a straightforward way to a 2
parameter family of configurations at merger. In particular, we need
the final configuration to be describable by a single azimuthal and
polar angle. We note that this is not the case in general, and that we
used PN simulations to tests the stability of various configurations.
For the N configuration, the mapping to a single azimuthal angle is
automatic because only one BH is spinning, for the other
configurations, we verify that the configuration can be described by a
single angle by comparing four different measurements of the azimuthal
angle. The results are displayed in  Table~\ref{tab:SK_angles}, which
shows that, to within about 3-15 degrees, the configurations are
describable by a single angle.
\begin{table}
\caption{S,  K, and L configuration angles. Here $\phi_1$ is the angle between
the in-plane component of $\vec S_1$ for configuration PH0 and the
corresponding PHXX configuration, while $\phi_2$ is defined using
$\vec S_2$, $\phi_\Delta$ using $\vec \Delta$, and $\phi_S$ using
$\vec S$. These angles agree to within $(3-15)^\circ$, which justifies our
using a single angle in our fitting formula for this configuration. For
the S configurations, $\phi_S$ is ill-defined since $S_\perp$ is very
small.}
\label{tab:SK_angles}
\begin{ruledtabular}
\begin{tabular}{l|llll}
Conf & $\phi_1$ & $\phi_2$ & $\phi_\Delta$ & $\phi_S$ \\
\hline
STH45PH0 & 0 & 0 & 0 & 0\\
STH45PH30 & 33.3518 & 30.9202 & 32.0924 & \\
STH45PH60 & 61.7152 &  58.2499 & 59.9753 &  \\
STH45PH90 & 91.4444 &  88.5921 & 90.0456 &  \\
STH45PH120 & 116.271 &  115.641 & 115.997 &  \\
STH45PH90 & 143.981 &  144.881 & 144.46 &  \\
\hline
KTH45PH0 & 0 & 0 & 0 &0 \\
KTH45PH30 & 28.6011 & 31.0559 & 31.4676 & 28.8129 \\
KTH45PH60 & 39.3732 & 41.6249 & 42.4836 & 39.2248 \\
KTH45PH90 & 58.7044 & 59.2109 & 61.8581 & 56.8317 \\
KTH45PH105 & 75.1457 & 72.9151 & 76.7906 & 71.6609 \\
KTH45PH120 & 95.7513 & 90.7761 & 95.0918 & 91.3502\\
KTH45PH135 & 114.209 & 108.253 & 112.123 & 110.072 \\ 
KTH45PH150 & 135.705 & 130.366 & 133.203 & 132.59\\
KTH45PH165 & 156.52 & 153.385 & 154.838&  154.882 \\
\hline
KTH22.5PH0 & 0 & 0 & 0 &0 \\
KTH22.5PH30 & 29.9 &23.9 &26.8 &25.12\\
KTH22.5PH60 & 52.1 & 42.1 & 45.7 & 46.3 \\
KTH22.5PH90 & 71.4 & 59.8& 62.8 & 71.4\\
KTH22.5PH120 & 101.2 & 92.8 & 92.9 & 102.2 \\
KTH22.5PH150 & 142.5 & 143.4 & 139.9 & 147.5 \\
\hline
LPH0 & 0 & 0 & 0 & 0\\
LPH30 & 17.9 & 20.8 & 17.0 & 24.7 \\
LPH60 & 35.0 & 36.3 & 32.0 & 45.3 \\
LPH90 & 55.2 & 52.3 & 49.8 & 65.3 \\
LPH120 & 87.6 & 80.0 & 81.4 & 93.0 \\
LPH150 & 139.9 & 130.8 & 136.1 & 138.4\\
\end{tabular}
\end{ruledtabular}
\end{table}

\section{Results}

Results from these 88 simulations are given in the tables in
Appendix~\ref{app:Res} below. In
Tables~\ref{tab:remnant_part1}~and~\ref{tab:remnant_part2} we give the
remnant BH mass and spin, as measured using the IH formalism, while in
Tables~\ref{tab:rad_part1}~and~\ref{tab:rad_part2}, we
give the radiated energy, angular momentum, and recoil, as calculated
from the waveform extracted at $60M, 70M, \cdots, 100M$ and then
extrapolated to infinity. We compare these two independent measures in
Tables~\ref{tab:rem_rad_cmp_part1}~and~\ref{tab:rem_rad_cmp_part2}.
Finally, in Tables~\ref{tab:rot_part1}~and~\ref{tab:rot_part2}, we give
the spin of each BH near merger, and the final remnant recoil, in
a rotated frame aligned with averaged orbital angular momentum at
merger. For completeness, we also show in Table~\ref{tab:rot_part3} the value
of $\Delta_\perp$, $S_\perp$, $\Delta_z$, and $S_z$ corresponding
to the BH spin in Table~\ref{tab:rot_part1}.

In order to analyze the results of the present simulations, we use the
techniques developed in~\cite{Lousto:2008dn}. Briefly, we rotate each
configuration such that the trajectories near merger overlap. We then
calculate the spins in this rotated frame. 
This is done separately for each family of constant $\theta$ per
configuration type (S,L,K,N).
The angle $\varphi$ is then
defined to be the angle (at merger) between the spin  of BH1 (the BH
originally located on the positive $x$ axis) for a given
PHyyy configuration and the spin of BH1 in
the corresponding PH0 configuration.  Note that, for a given
family of fixed spin and spin inclination angle $\theta$, the angle
$\varphi$ and $\phi$ differ by a constant, which can be absorbed in
the fitting constants $\phi_1$ and $\phi_3$.  We then fit the recoil
in these sub-families to the form
\begin{equation}
V_{\rm rec} = V_{1}
\cos(\varphi - \phi_1) + V_{3} \cos(3 \varphi - 3 \phi_3).
\label{eq:phifit}
\end{equation}
Our tests indicate that $V_1$ can be obtained accurately with six choices of
the initial $\phi_i$ angles. For example, a fit of all the NTH45PHyyy
configurations gives
$V_1=1349.0\pm9.7$ if we include all twelve angles (see
Table~\ref{tab:rot_part1}), and
$V_1=1346\pm22$ if we include six angles.

\begin{figure}

\includegraphics[width=.49\columnwidth]{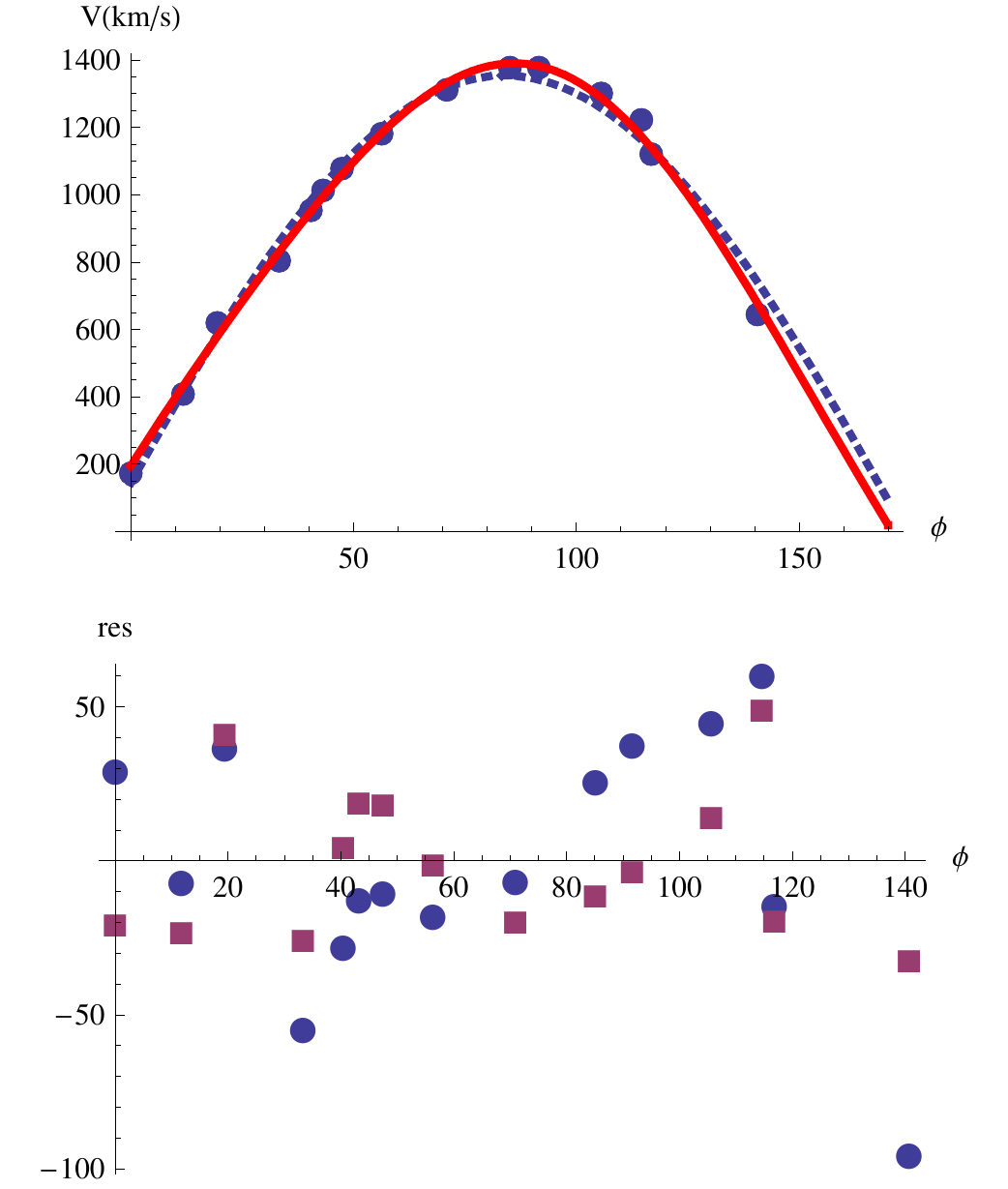}
\includegraphics[width=.49\columnwidth]{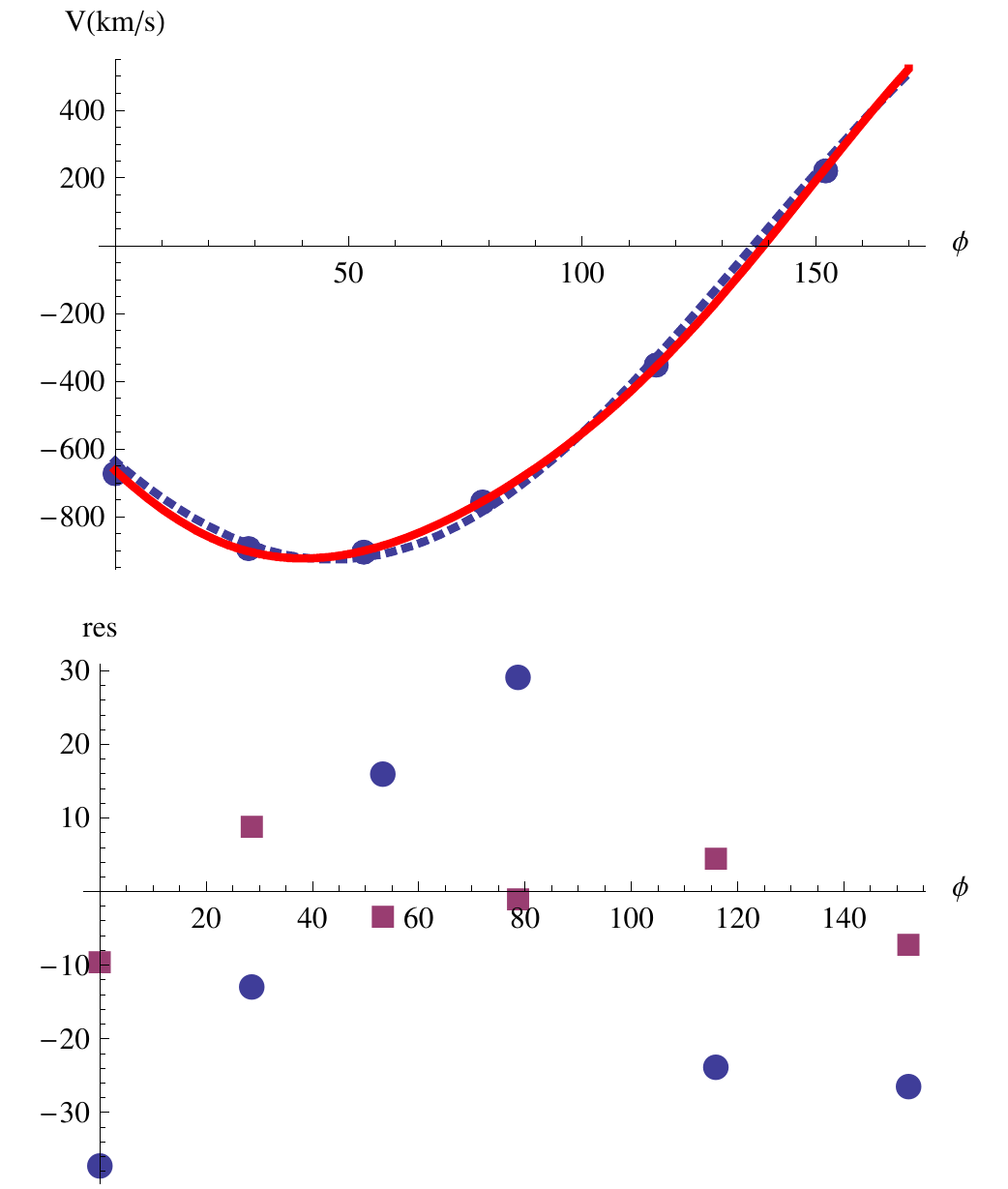}
\caption{A fit of the recoil for the NTH45PHyyy (left) and 
NTH135PHyyy (right) configurations
to $V=V_1 \cos(\varphi-\phi_1)$ (blue-dotted) and  $V=V_1 \cos(\varphi-\phi_1) +
v_3 \cos(3\varphi - 3 \phi_3) $ (red-solid), as well as the residuals (blue circles
and red squares, respectively).
The  remaining scatter is due to the fact that $S_\perp$ and $S_z$
vary from configuration to configuration within the NTH45PHyyy and NTH135PHyyy families.}
\label{fig:fit_to_phi_dep}
\end{figure}

In all cases, $V_{3}$ is much smaller than $V_{1}$.
Results from these fits are given in Table~\ref{tab:fit_to_phi_dep}
and Fig.~\ref{fig:fit_to_phi_dep}.
We note that there are additional approximations inherent in this
procedure. To demonstrate this, consider the formula
\begin{equation}
V_z = \Delta_\perp (A + B S_z +\cdots) + S_\perp \Delta_z (D + E
S_z+\cdots),
\label{eq:ansatz}
\end{equation}  
where $A, B, C, D, E$ are fitting constants.
Even when considering ``symmetric'' configurations like S, K, L, and
N, where
$\vec S$ and $\vec \Delta$ cannot rotate independently, 
each term in Eq.~(\ref{eq:ansatz}) may be maximized at different
azimuthal angles, and  
the formula should really be written as
\begin{eqnarray}
V_z = A \vec \Delta\cdot \hat n_0 + B \vec \Delta \cdot\hat n_1 S_z + \cdots
\nonumber \\
+  D \vec S\cdot \hat n_2 \Delta_z + E
\vec S\cdot\hat n_3  \Delta_z S_z +\cdots,
\label{eq:ansatz_ref}
\end{eqnarray}
where $\hat n_i$ are unit vectors in the orbital plane. If we make the
additional assumption that the coefficient  $A$ dominates this
expression,
then Eq.~(\ref{eq:ansatz_ref}) can be approximated by
\begin{eqnarray}
V_z = \vec \Delta\cdot \hat n_0 (A + B \cos(\vartheta_1)S_z) \nonumber \\
+   \vec S_\perp\cdot \hat n_0 \Delta_z(D \cos(\vartheta_2) +E
\cos(\vartheta_3) S_z),
\label{eq:ansatz_app}
\end{eqnarray}
where $\vartheta_i$ is the angle between $\hat n_i$ and $\hat n_0$.
There will be terms proportional to $\sin \vartheta_i$, but they will be 
${\cal O}(1/A)$, which we will assume to be small enough to ignore.
If, in addition, we assume that the angles $\vartheta_i$ do not vary
significantly between different configurations, then we can replace
the fitting constants in Eq.~(\ref{eq:ansatz_app}) with the product
of the constant and the corresponding $\cos \vartheta_i$. We can then
interpret $V_1$ from Eq.~(\ref{eq:phifit}) as 
$V_1=|\Delta_\perp| (A+ \tilde B S_z) + |S_\perp||\Delta_Z|(\tilde D +
\tilde E S_z)$, where $\tilde B = B \cos \vartheta_1$, etc.
Ultimately, we justify all our approximations by testing the resulting
formula using several different families of configurations.

\begin{table}
\caption{A fit of the recoil for each family of PHYYY configurations
to the form $V_\| = V_1 \cos(\varphi-\phi_1) + V_3 \cos(3\varphi-3\phi_3)$.
Note how the K configurations, which started with $\Delta_\perp =0$,
evolved to configurations with large $\Delta_\perp$.}
\label{tab:fit_to_phi_dep}
\begin{ruledtabular}
\begin{tabular}{l|llll}
CONF & $V_1$ & $V_3$ & $\phi_1$ &$\phi_3$\\
\hline
NTH15 & $539.34\pm2.5$ & $33.2\pm2.3$ & $141.96\pm0.24$ & $297.1\pm1.3$\\
NTH30 & $1002\pm12$ & $43\pm13$ & $126.42\pm0.71$ & $260.3\pm5.3$\\
NTH45 & $1349.0\pm9.7$ & $52\pm12$ & $82.50\pm0.58$ & $337.0\pm3.9$\\
NTH60 & $1542\pm11$ & $34\pm11$ & $20.83\pm0.47$ & $269.2\pm6.8$\\
NTH120 & $1199\pm13$ & $37\pm12$ & $292.79\pm0.54$ & $139.6\pm5.9$\\
NTH135 & $927.5\pm6.4$ & $35.6\pm6.7$ & $226.90\pm0.43$ & $311.3\pm3.6$\\
NTH165 & $312.9\pm6.4$ & $11.6\pm6.2$ & $213.4\pm1.2$ & $189\pm11$\\
\hline
STH45 & $2020\pm 19$ & $50\pm19$ & $291.40\pm0.56$ & $342.3\pm7.3$\\
\hline
KTH45 & $2227\pm12$ & $195\pm12$ & $217.33\pm0.32$ & $155.0\pm1.3$\\
KTH22.5 & $1731\pm25$& $130\pm23$ & $164.49\pm0.75$ & $100.9\pm3.4$\\
\hline
L & $3014\pm21$ & $145\pm18$ & $331.30\pm0.36$ & $263.50\pm2.9$\\
\hline
N9TH55 & $1803.4\pm6.2$ & $27.6\pm6.8$ & $74.89\pm0.14$ &
$102.6\pm2.8$\\
\end{tabular}
\end{ruledtabular}
\end{table}

\begin{figure}
\includegraphics[width=0.49\columnwidth]{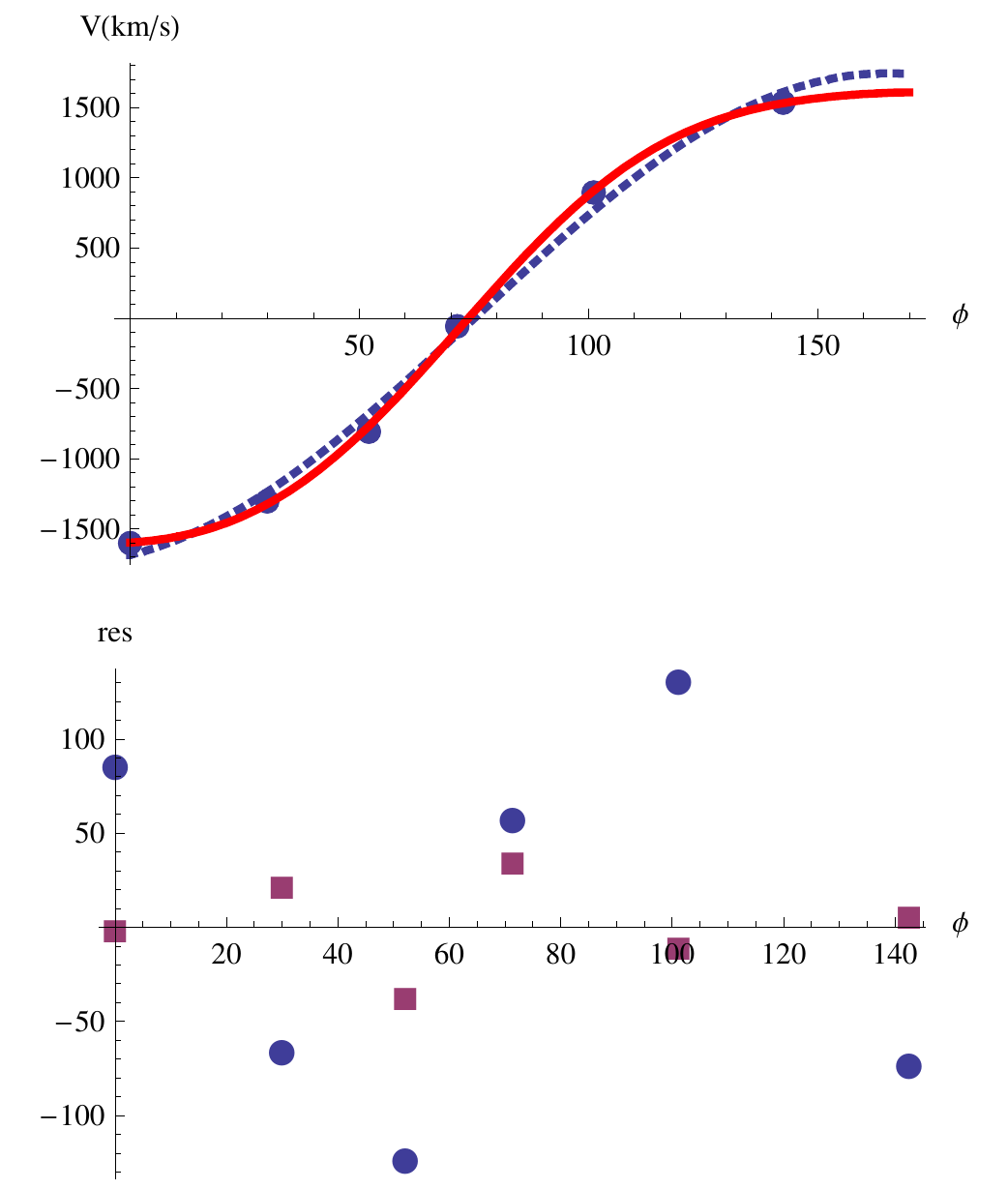}
\includegraphics[width=0.49\columnwidth]{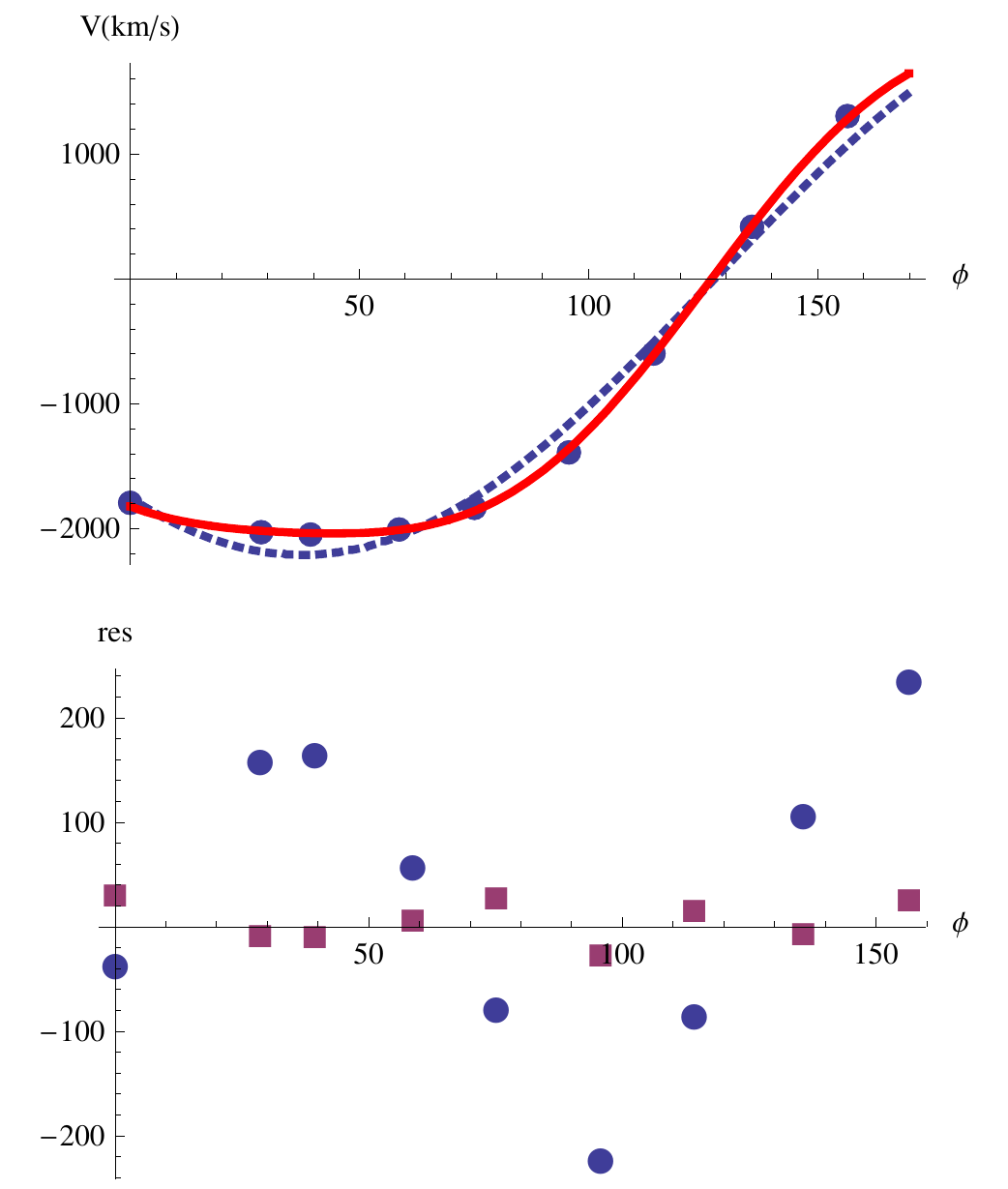}
\caption{Fitting of the KTH22.5PHyyy (left) and KTH45PHyyy (right) configurations
to $V=V_1 \cos(\varphi-\phi_1)$ (blue-dotted) and  $V=V_1 \cos(\varphi-\phi_1) +
v_3 \cos(3\varphi - 3 \phi_3) $ (red-solid), as well as the residuals (blue circles
and red squares, respectively).
Note how strong the
higher-order contributions are compared to the other configurations.}
\end{figure}

The N configurations are the only ones with families of different
$\theta$ (apart from the two angles for the K configuration). 
For these families, 
we fit the data  in Table~\ref{tab:fit_to_phi_dep} to the form
\begin{eqnarray}
V_1=&&C_1 \alpha \sin \theta +
C_2 \alpha^2 \sin\theta \cos\theta+ \nonumber\\
&&C_3\alpha^3 \sin\theta\cos^2\theta + V_{\rm hang},
\label{eq:newkick}
\end{eqnarray} where $V_{\rm
hang}$ is the \hangup~\cite{Lousto:2011kp, Lousto:2012su}, which
has the form
\begin{eqnarray}
V_{\rm hang}=&&3677.76 \alpha \sin\theta + 2481.21
\alpha^2 \sin\theta \cos\theta +\nonumber \\
&& 1792.45 \alpha^3 \sin\theta \cos^2\theta +\nonumber \\ 
&&1506.52 \alpha^4 \sin\theta \cos^3\theta.
\end{eqnarray}
For consistency with our conventions in~\cite{Lousto:2011kp,
Lousto:2012su}, we define $\alpha$ here to be $\alpha_1/2$. 
We note that $S_\|/m^2=(1/2) \alpha$ and $\Delta_\|/m^2
= \alpha$. We also note that, because $\alpha<1/2$ here, the
extrapolation to $\alpha=1$ is more severe than in the original
\hangup configuration. Results from these fits are shown in
Table~\ref{tab:fit_to_th_dep}.  From the table, we can see that the
$C_1$ term is consistent with zero. In subsequent fits, we remove this
term and only include $C_2$ and $C_3$ (we also attempt a fit including
a higher-order $C_4$ term, but this proved to have an unacceptably
large error).

\begin{figure}
\includegraphics[width=0.9\columnwidth]{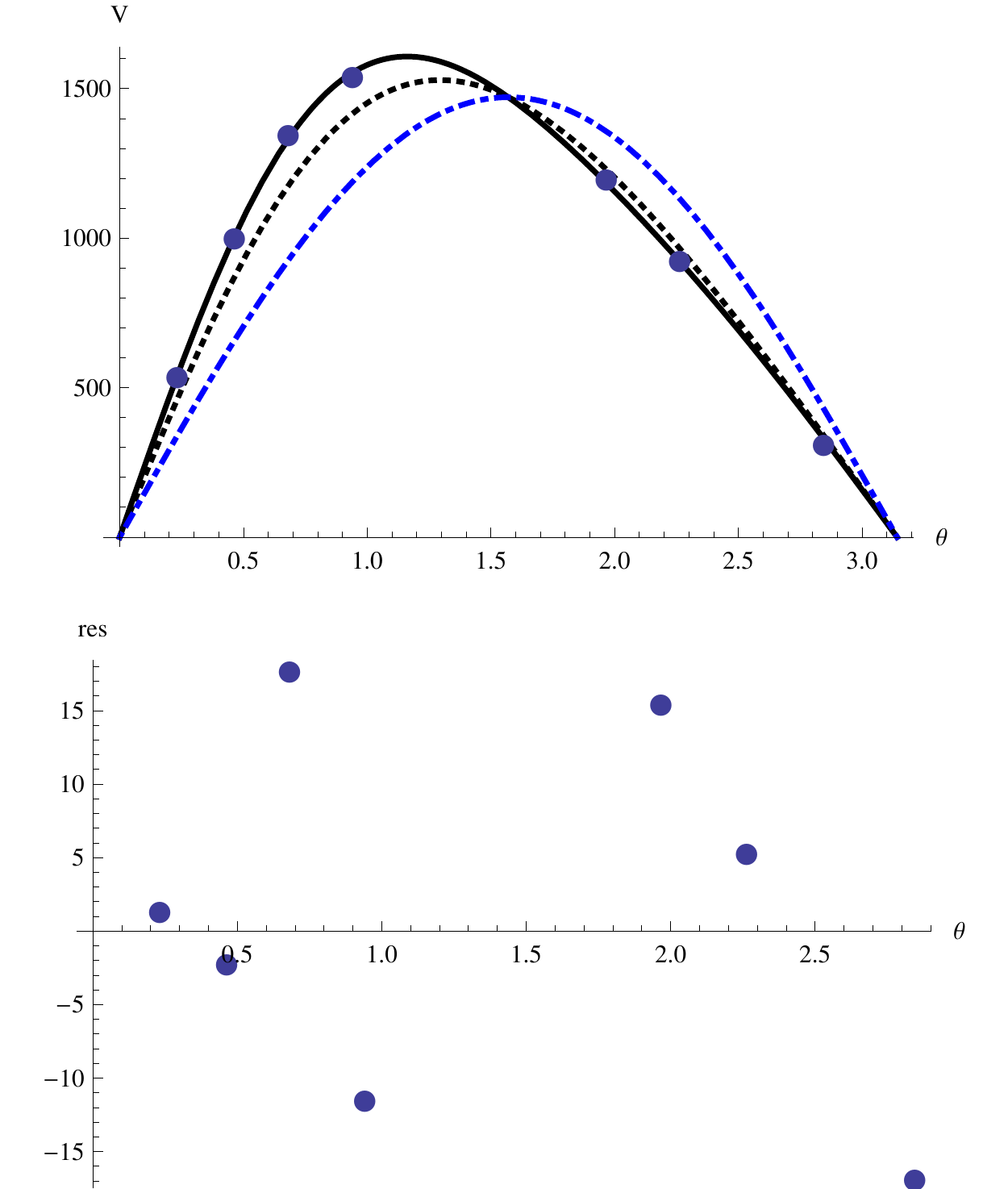}
\caption{A fit of the NTHXXX families, with residuals, and a
comparison with the \hangup formula (black dotted) and
\super (more symmetrical, blue-dotted curve). An excess over both
formulas at small angles is apparent, while a slight deficit with
respect to the \hangup is apparent at large angles.}
\label{fig:a8_fit_and_cmp}
\end{figure}

\begin{table}
\caption{Fit of the N (see Table~\ref{tab:fit_to_phi_dep})
to the form $V_1=C_1 \alpha \sin \theta + C_2 \alpha^2 \sin\theta
\cos\theta+ C_3\alpha^3 \sin\theta\cos^2\theta + C_4 \alpha^4
\sin\theta\cos^3\theta + V_{\rm hang}$. Also included in parentheses
are the fits we obtain when including N9 results.
For the first fit, $C_4$ was set to zero, for the second
$C_1$ (which was found to be consistent with zero) and
$C_4$  were set
to zero. For the third. only $C_1$ was set to zero.
The uncertainty in the $C_4$
coefficients makes using this term in extrapolative formulas
problematic (e.g., the differences in the 
predicted velocity for N9 is under $1\ \KMS$).
 We therefore use the second fit in the analysis below.
The values in parenthesis were obtained from fits that used the N9
results.
}
\label{tab:fit_to_th_dep}
\begin{ruledtabular}
\begin{tabular}{llll}
Coeff. & Correction to \hangup & \hangup term\\
\hline
$C_1$ & $-19\pm21$  $(-21\pm31)$ & 3677.76  \\
$C_2$ & $1124\pm128$ $(1245\pm177)$ & 2481.21\\
$C_3$ & $2961\pm679$ $(3458\pm962)$ & 1792.45\\ 
$C_4$ & 0 & 1506.52\\
\hline
$C_1$ & $0$ & 3677.76  \\
$C_2$ & $1140\pm125$ $(1263\pm168)$ & 2481.21\\
$C_3$ & $2481\pm434$ $(2953\pm573)$ & 1792.45\\ 
$C_4$ & 0 & 1506.52\\
\hline
$C_1$ & $0$ & 3677.76  \\
$C_2$ & $761\pm243$ $(878\pm392)$ & 2481.21\\
$C_3$ & $2281\pm393$ $(2747\pm596)$ & 1792.45\\ 
$C_4$ & $4733\pm2721$ $(4810\pm4432)$ & 1506.52\\
\end{tabular}
\end{ruledtabular}
\end{table}

\section{Modeling the recoil velocity}\label{Sec:Results}

As seen in Sec.~\ref{sec:equal}, when considering the equal-mass case, 
and spin-contributions up through third-order,
the most general formula for the out-of-plane recoil is
\begin{eqnarray}
V_\| = \Delta_\perp (A + B S_\| + C S_\|^2) + \nonumber \\
S_\perp \Delta_\| (D + E S_\|) + \nonumber \\
F \Delta_\perp S_\perp^2 + \nonumber \\
G \Delta_\perp \Delta_\|^2 + \nonumber \\
H \Delta_\perp^3,
\end{eqnarray}
where $A-H$ are fitting constants. The first line is part of the
\hangup recoil, the second and third are new contributions. The
term proportional to $G$ is small (if G were big, then the S
configuration recoils would be significantly different from the \hangup prediction).
The H term is small, as we saw by evolving \super configuration
in~\cite{Lousto:2010xk}. Motivated by the \hangup results,
where the series $A + B S_\| + C S_\|^2$ had similarly larger values
of A, B, and C, we will assume at this point that the term
proportional to E is larger than the term proportional to F.

We therefore interpret the additional terms in
Eq.~(\ref{eq:newkick}) as being proportional to powers of
$\Delta_\|$ and $S_\|$ and
corrections to the \hangup formula have the form
$2 C_2 S_\perp \Delta_\| + 4 C_3 S_\perp \Delta_\| S_\|$. Then, if we {\it
assume} the same $\eta^2$ mass ratio dependence, our ansatz for the
z component of the generic recoil becomes
\begin{eqnarray}
\frac{V_\|}{16 \eta^2}&& = \Delta_\perp \left(3677.76+ 2\times
2481.21 S_\|\right. \nonumber \\
&& \left.+ 4\times 1792.45 S_\|^2+  8\times 1506.52 S_\|^3\right)\nonumber\\
&& +S_\perp  \Delta_\| (2 C_2 + 4 C_3 S_\|). \label{eq:vgen}
\end{eqnarray}
We refer to this  new contribution to the recoil, which has the form
$S_\perp \Delta_\| (2 C_2 + 4 C_3 S_\|)$, as the \cross (since
$S_\perp \Delta_\|$ can be expressed as  $\hat z \cdot \vec S
\times(\hat n \times \vec \Delta)$,
where $\hat n$ is a unit vector in the $xy$ plane). The coefficients
$C_2$ and $C_3$ were determined using only the N and N9 configurations.

We then verify Eq.~(\ref{eq:vgen}), {\em in the equal-mass limit},
by comparing the predictions from
the new formula with the maximum recoil obtained from the S, K, and L
configurations. Our results are given in Table~\ref{tab:fitb}.
The table compares the measured value of $V_1$ for each family with
the predictions of the {\it superkick}, \hangup, and \cross. In all
cases, except S, the \cross provides the most accurate prediction for
$V_1$. The results from the K configurations are particularly
interesting since the measured recoils and the \cross predictions are
both $\sim200\ \KMS$ larger than the \hangup prediction (a 10-16\% effect).
For the S configurations, there is
an ambiguity in the sign of the \cross kick. This is due to the fact that the
\cross correction lies in the same direction as the \hangup. Here, however,
$S_\| = 0.005\pm0.003$ and the small \hangup correction may have the wrong 
sign.
If we assume $S_\|$ is really zero or slightly negative, we find that the
\cross prediction is the most accurate (with a prediction of $2015\
\KMS$). On the other hand, the N9 runs appear to show that there are
still uncertainties in our modeling.

\begin{widetext}

\begin{table}
\caption{Comparison of $V_1$ as fit from the current data and the
predictions of the \super, \hangup, and \cross (new) formulas. Note,
there is an ambiguity in the sign of the \cross correction for the S
configuration (see text). Cross.(B) refer to the \cross
prediction using the  second
set of coefficients from Table~\ref{tab:fit_to_th_dep} (not including
the N9 configurations).  }
\label{tab:fitb}
\begin{ruledtabular}
\begin{tabular}{l|lllllllll}
CONF & $S_\perp/M^2$ & $\Delta_\perp/M^2$ &$S_\|/M^2$&
$\Delta_\|/M^2$ & $V_1$ & Sup. & Hang. & Cross.(B)\\
\hline
NTH15 & $0.046\pm0.004$ & $0.092\pm0.008$ & $0.196\pm0.001$ &
$-0.392\pm0.002$ & $539.4\pm2.3$ & 339.746 & 463.256 & 540\\
NTH30 & $0.090\pm0.007$ & $0.179\pm0.013$ & $0.179\pm0.003$ &
$-0.358\pm0.007$ & $1002\pm12$ &658.497 & 871.282 & 1007\\
NTH45 & $0.126\pm0.008$ & $0.252\pm0.015$ & 
$0.155\pm0.006$ & $-0.311\pm0.013$  & $1349.0\pm9.7$ &926.499 &1176.76 &
1329\\
NTH60 & $0.161\pm0.0073$ & $0.323\pm0.015$ & 
$0.118\pm0.010$ & $-0.235\pm0.020$ & $1542\pm11$ &1186.7 &1413.2 &
1548\\
NTH120 & $0.184\pm0.004$ & $0.368\pm0.008$ & 
$-0.077\pm0.010$ & $0.154\pm0.021$  & $1199\pm13$ &1355.8 &1279 & 1185\\
NTH135 & $0.154\pm0.006$ & $0.308\pm0.011$ & 
$-0.128\pm0.007$ & $0.256\pm0.013$  & $927.5\pm6.4$ &1134.46 &967.015
& 927\\
NTH165 & $0.059\pm0.003$ & $0.118\pm0.007$ & 
$-0.193\pm0.002$ & $0.386\pm0.004$ & $312.9\pm6.4$ & 434.141 & 342.312
& 334\\
\hline
KTH45 & $0.276\pm0.002$ & $0.497\pm.028$ &
$0.054\pm0.021$ & $-0.277\pm0.048$ & $2227\pm12$ &
1826 & 1970 & 2185\\
KTH22.5 & $0.149\pm.003$  & $0.400\pm0.037$ &
$0.021\pm0.008$ & $-0.626\pm0.025$ & $1731\pm25$ &
1470 & 1512 & 1744\\
\hline
L & $0.173\pm0.016$ & $0.551\pm.006$ &
$0.227\pm0.013$ & $0.103\pm0.051$ & $3014\pm21$ &
2026 & 2928 & 3009\\
\hline
STH45 & $0.011\pm0.004$ & $0.552\pm0.004$ & \
$0.005\pm0.003$ & $-0.5760\pm0.0015$  &  $2020\pm 19$ & 2030.15 &
2044.94 &  $2059^*$ \\
\hline
N9TH55 & $0.1642\pm0.0087$ & $0.323\pm0.018$ & $0.151\pm0.010$ &
$-0.297\pm0.019$ & $1803.4\pm6.2$ & 1208.45 & 1522.5 & 1728\\
\end{tabular}
\end{ruledtabular}
\end{table}

\end{widetext}

\section{Discussion}\label{Sec:Discussion}

The discovery that the hangup effect contributes significantly to
the gravitational recoil of merging black hole 
binaries \cite{Lousto:2011kp} implies that
nonlinear spin couplings are crucial in describing those recoils.
Nonlinear couplings come in a variety of combinations, as described in
Sec.~\ref{Sec:Symmmetries}. In order to evaluate which of those terms 
produce the largest contributions to the total recoil, we performed a
large set of new simulations. These 88 simulations
of precessing BHBs allowed us to confirm the relevance of the 
\hangup effect in more generic runs, discover another important
term that we named \cross  that appears in precessing binaries, and
gives more accurate predictions for other families of BHB configurations.

\begin{figure}
\includegraphics[width=\columnwidth]{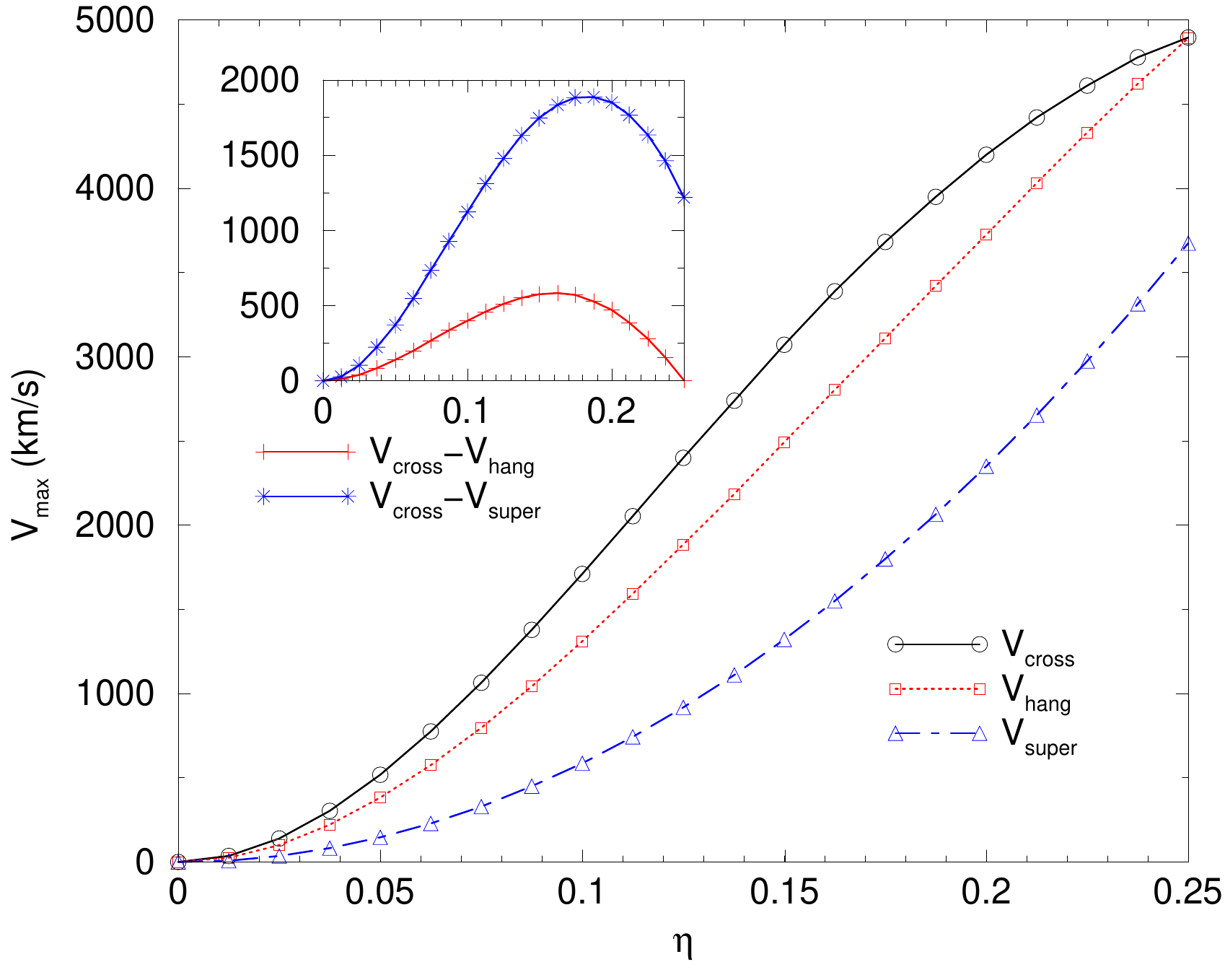}
\caption{The maximum recoil velocity predicted by the \cross,
\hangup, and \super formulas for BHBs with a given mass ratio and maximal spin.
The inset shows the difference between the \cross and \hangup, and
the \cross and \super,
versus symmetric mass ratio.
}
\label{fig:max_recoil}
\end{figure}

While not as dramatic as the \hangup effect, the \cross may prove to
be very important in the non-equal-mass regime.
To help elucidate how this new contribution affects the
recoil (for a given mass ratio), we plot the maximum
recoil for configuration with a given mass ratio and with both BHs
maximally spinning. As shown in Fig.~\ref{fig:max_recoil}, the \cross
enhances the recoil (up to $600\ \KMS$) in the moderate mass-ratio 
range.

\begin{figure}
\includegraphics[width=\columnwidth]{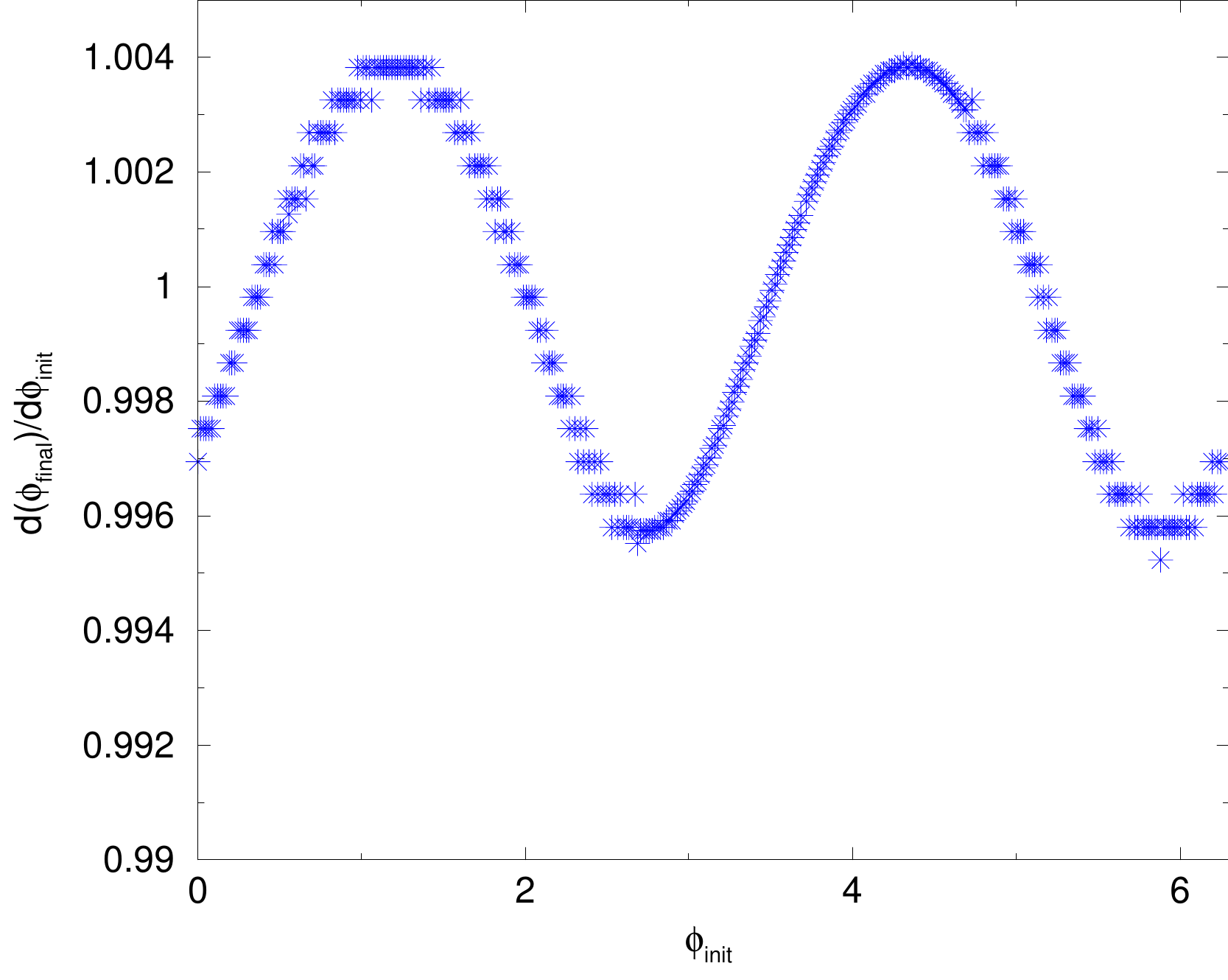}
\caption{A plot of $d\phi_{\rm final}/d\phi_{\rm init}$ versus time
for a set of 360 binaries in a superkick configuration. The initial
separation is $\sim 10M$, while the final separation is $\sim 3M$. The
effect is only 4 parts in 1000.}
\label{fig:dphidphi}
\end{figure}

To see how the \cross contribution to the recoil affects the net probabilities for
large recoils, we revisit the case of the supermassive BH binary with spins
aligned via hot and cold accretion~\cite{Lousto:2012su} and
non-aligned BHBs (i.e., dry mergers).
Briefly, we consider a set of 10 million binaries chosen randomly with
a spin-magnitude
distribution and spin inclination angle distribution taken
from~\cite{Lousto:2012su}, and a mass ratio distribution taken from~\cite{Yu:2011vp, 
Stewart:2008ep,
Hopkins:2009yy}. We assume a uniform distribution of spin directions
in the equatorial plane (see Fig.~\ref{fig:dphidphi}). 
We find an increased probability of 
large recoils ($V > 2000\ \KMS$) by a factor of $\sim 2$ (see
Table~\ref{tab:stats}). However,
to generate these probabilities we used the assumption that all terms in
Eq.~(\ref{eq:vgen}) scale with the mass ratio as  $16 \eta^2$.
 This is a strong assumption that we will revisit in
an upcoming paper. Additionally, we did not take into account new nonlinear
terms proportional to $\delta m $ that may also prove to be important.

As an aside, we note that the  distribution of azimuthal orientations
of the spin quantity $\vec \Delta_\perp$ 
near merger may appear to be nonuniform. This is actually most 
pronounced for the
$\alpha =0.9$ in our original \hangup paper~\cite{Lousto:2012su} (see
Fig. 5 there).
However, this skew appears to be actually due to varying
eccentricity, which leads to different inspiral times for different
starting azimuthal configurations. To help confirm that there is not,
in fact, a strong preference for any particular azimuthal angles, we evolved
a set of 360 superkick configurations (with the spins aligned along
$\phi=0^\circ, 1^\circ, \cdots, 359^\circ$), using 3.5 PN, from a
separation of 10M down to 3M (note, we are not concerned with the accuracy
of PN at 3M, rather, if there is any significant effect predicted by
PN). The distribution of final azimuthal
configurations was flat, with no strong preference or clumping. A plot of
final versus the initial azimuthal angle $\phi$ is shown in
Fig.~\ref{fig:dphidphi}. There is a small sinusoidal effect at the
level of 4 parts in 1000. 

On the other hand, the relative orientation of $\vec S_{1\perp}$ and $\vec
S_{2\perp}$ are correlated due to secular spin-resonant interactions in the
post Newtonian regime.~\cite{Gerosa:2013laa, Berti:2012zp,
Kesden:2010ji, Kesden:2010yp, Schnittman:2004vq} (something
not accounted for in Table~\ref{tab:stats}). One consequence of these
spin interactions is that there is a tendency to drive the in-plane spin
towards alignment or counteralignment~\cite{Schnittman:2004vq,
Kesden:2010yp, Gerosa:2013laa} when the polar angle of the two spins
are different (in a population, the degree of alignment of
counter-alignment scales with
$\langle\theta_1-\theta_2\rangle$).
To model these effects, 
we examine how the recoil probabilities are modified if we
assume the two extreme cases of 
alignment/counteralignment $\vec S_{1\perp} \propto \pm 
\vec S_{2 \perp}$  
of the in-plane component of the spins. Results from these
studies are given in Table~\ref{tab:stat2}. From the table, we can see
that alignment ( $\vec S_{1\perp} \propto
\vec S_{2 \perp}$) suppresses large recoils by a factor of about
4, while counteralignment ( $\vec S_{1\perp} \propto 
-\vec S_{2 \perp}$) increases the probability of large recoils by a
factor of 2-3.

\begin{table}
\caption{Comparison between the predicted probabilities for a
recoil in a given range as from the \hangup and \cross formulas for
hot (top) and cold (middle) accretion and dry mergers (bottom). }
\label{tab:stats}
\begin{ruledtabular}
\begin{tabular}{lll|ll}
Range & P(cross) & P(cross obs) & P(hang)  & P(hang obs)\\
\hline
0-500 & 77.000\%  &91.301\% &  80.871\% & 93.210\%\\
500-1000 & 15.564\% &6.903\% & 13.843\% & 5.623\%\\
1000-2000 & 6.930\% &1.741\%& 5.046\% &1.143\%\\
2000-3000 & 0.498\% & 0.055\% & 0.237\% & 0.025\%\\
3000-4000 & 0.007\% & $3.5\cdot10^{-4}$\% & 0.003\% &$10^{-4}\%$\\
\hline
0-500 & 91.193\%  &97.765\% &  93.657\% & 98.522\%\\
500-1000 & 7.974\% &2.114\% & 5.919\% & 1.423\%\\
1000-2000 & 0.832\% &0.120\%& 0.423\% &0.055\%\\
2000-3000 & 0.002\% & $1.3\cdot10^{-4}$\%& $4.7\cdot10^{-4}$\% & 0 \%\\
3000-4000 & 0\% & 0\% & 0\% &0\%\\
\hline
0-500 & 68.315\%  &86.465\% &  70.229\% & 87.693\%\\
500-1000 & 18.382\% &9.886\% & 18.157\% & 9.251\%\\
1000-2000 & 11.820\% &3.467\%& 10.519\% & 2.924\%\\
2000-3000 & 1.449\% & 0.180\%& 1.074\% & 0.130 \%\\
3000-4000 & 0.034\% & 0.002\% & 0.021\% &0.001\%\\

\end{tabular}
\end{ruledtabular}
\end{table}

\begin{table}
\caption{Comparison between the predicted probabilities 
for a
recoil in a given range as from the \cross formulas for
hot (top) and cold (bottom) accretion when the in-plane components of
the spins are forced to be aligned, antialigned, or are uncorrelated. }
\label{tab:stat2}
\begin{ruledtabular}
\begin{tabular}{llll}
Range & P(aligned) & P(uncorrelated) & P(antialigned)\\
\hline
0-500 & 84.8855\% & 76.9754\% & 71.3786\%\\
500-1000 & 11.2354\% & 15.5831\% & 17.6142\%\\
1000-2000 & 3.7530\% & 6.9365\% & 9.9052\%\\
2000-3000 & 0.1261\% & 0.4978\% & 1.0720\%\\
3000-4000 & 0.0000\% & 0.0072\% & 0.0300\%\\
\hline
0-500 & 95.3182\% & 91.1816\% & 87.0390\%\\
500-1000 & 4.3465\% & 7.9853\% & 11.4147\%\\
1000-2000 & 0.3348\% & 0.8309\% & 1.5400\%\\
2000-3000 & 0.0005\% & 0.0022\% & 0.0064\%\\
3000-4000 & 0.0000\% & 0.0000\% & 0.0000\%\\

\end{tabular}
\end{ruledtabular}
\end{table}

By examining  the N configuration (which have non-zero values for
$S_\|$, $S_\perp$, $\Delta_\|$, and $\Delta_\perp$), we found a new
nonlinear term that amplifies the recoil. We verified this new
effect by examining several other configurations (S, K, L). 
Since the N configurations are generic, in that all relevant spin
parameters are non-trivial, it {\it appears} to be the case that there
is no other large nonlinear contribution to the recoil for equal-mass
BHBs. On the other hand, the unequal-mass regime, which is the
subject of a major research effort by the authors, promises to hold many new
surprises.

\acknowledgments 

The authors thank M. Kesden for discussion on how PN dynamics modify
spin distributions and recoils. M.Campanelli and H.Nakano for 
other discussions and J.Whelan for Eq.~(\ref{eq:terms}).
The authors gratefully acknowledge the NSF for financial support from Grants
PHY-1212426, PHY-1229173,
AST-1028087, PHY-0929114, PHY-0969855, PHY-0903782, OCI-0832606, and
DRL-1136221,  and NASA for financial support from NASA Grant No.
07-ATFP07-0158. Computational resources were provided by the Ranger
system at the Texas Advance Computing Center (XSEDE allocation
TG-PHY060027N), which is supported in part by the NSF, and by
NewHorizons at Rochester Institute of Technology, which was supported
by NSF grant No. PHY-0722703, DMS-0820923 and AST-1028087.

\appendix
\section{Progenitor and Remnant Parameters}
The tables in this appendix provide useful information for modeling
remnant properties and how they relate to the configuration of the
progenitor BHB.
These results can
be used to improve, both quantitatively and qualitatively, the 
empirical formulas that describe the remnant mass and spin 
\cite{Lousto:2009mf, Barausse:2012qz}.
They can also be used in the construction of alternative formulas to model
recoil velocities.
Although we have been able to
accurately model the off-plane recoil and make predictions that fit
new runs, the formulas become increasingly complex to model higher
nonlinear terms in the spins and one can seek a simpler, more compact
formulation of the remnant recoil.

\begin{widetext}
\subsection{Initial Data parameters}\label{app:ID}

${}$
\begin{table}[h!]
\caption{Initial data parameters. In all cases the puncture masses were
chosen such that the total ADM mass of the binary was $1.0\pm10^{-6}M$. Here
the punctures are located at $(x_{1,2},0,0)$ 
with momenta $\pm(0, p,0)$ and spins
$\vec S_1 = (S_x, S_y, S_z)$. For the  N configurations $\vec S_2=0$.
The approximate initial
eccentricities $e_i$,
eccentricities measured over the last orbit $e_f$, and the number of
orbits $N$,
are also given.
}
\label{tab:ID_part1}
\begin{ruledtabular}
\begin{tabular}{l|lllllllllll}
CONF & $m_{p1}/M$ & $m_{p2}/M$ & $x_1/M$ & $x_2/M$  & $p/M$ &
$S_x/M^2$ & $S_y/M^2$ & $S_z/M^2$ & $m_{H1}$ & $m_{H2}$ &
$N_{e_{\rm f}}^{e_{\rm i}}$\\
\hline
NTH15PH0 & $0.307753$ & $0.487740$ & $4.077900$ & $-4.307566$ &
$0.107293$ & $0.000000$ & $0.053084$ & $0.198112$ & $ 0.507321 $  & $
0.504754 $ & $4.5^{0.02}_{0.002}$ \\
NTH15PH30 & $0.307762$ & $0.487741$ & $4.077900$ & $-4.307566$ & $0.107293$ & $-0.026542$ & $0.045972$ & $0.198112$ & $ 0.507322 $  & $ 0.504756 $  \\
NTH15PH60 & $0.307779$ & $0.487745$ & $4.077900$ & $-4.307566$ & $0.107293$ & $-0.045972$ & $0.026542$ & $0.198112$ & $ 0.507321 $  & $ 0.504759 $  \\
NTH15PH90 & $0.307786$ & $0.487747$ & $4.077900$ & $-4.307566$ & $0.107293$ & $-0.053084$ & $0.000000$ & $0.198112$ & $ 0.507321 $  & $ 0.504761 $  \\
NTH15PH120 & $0.307777$ & $0.487745$ & $4.077900$ & $-4.307566$ & $0.107293$ & $-0.045972$ & $-0.026542$ & $0.198112$ & $ 0.507319 $  & $ 0.504759 $  \\
NTH15PH150 & $0.307763$ & $0.487741$ & $4.077900$ & $-4.307566$ & $0.107293$ & $-0.026542$ & $-0.045972$ & $0.198112$ & $ 0.507321 $  & $ 0.504755 $  \\
NTH30PH0 & $0.307652$ & $0.487705$ & $4.098641$ & $-4.304988$ &
$0.107491$ & $0.000000$ & $0.102538$ & $0.177601$ & $ 0.507291 $  & $
0.504702 $ & $4.5^{0.02}_{0.003}$ \\
NTH30PH30 & $0.307686$ & $0.487710$ & $4.098641$ & $-4.304988$ & $0.107491$ & $-0.051269$ & $0.088800$ & $0.177601$ & $ 0.507293 $  & $ 0.504708 $  \\
NTH30PH60 & $0.307744$ & $0.487725$ & $4.098641$ & $-4.304988$ & $0.107491$ & $-0.088800$ & $0.051269$ & $0.177601$ & $ 0.507291 $  & $ 0.504721 $  \\
NTH30PH90 & $0.307775$ & $0.487731$ & $4.098641$ & $-4.304988$ & $0.107491$ & $-0.102538$ & $0.000000$ & $0.177601$ & $ 0.507288 $  & $ 0.504727 $  \\
NTH30PH120 & $0.307745$ & $0.487725$ & $4.098641$ & $-4.304988$ & $0.107491$ & $-0.088800$ & $-0.051269$ & $0.177601$ & $ 0.507287 $  & $ 0.504720 $  \\
NTH30PH150 & $0.307685$ & $0.487711$ & $4.098641$ & $-4.304988$ &
$0.107491$ & $-0.051269$ & $-0.088800$ & $0.177601$ & $ 0.507289 $  &
$ 0.504707 $ & $4.5^{0.02}_{0.002}$ \\
NTH45PH0 & $0.307512$ & $0.487654$ & $4.131521$ & $-4.300556$ & $0.107804$ & $0.000000$ & $0.144983$ & $0.144983$ & $ 0.507241 $  & $ 0.504625 $  \\
NTH45PH10 & $0.307522$ & $0.487655$ & $4.131521$ & $-4.300556$ & $0.107804$ & $-0.025176$ & $0.142780$ & $0.144983$ & $ 0.507244 $  & $ 0.504626 $  \\
NTH45PH20 & $0.307544$ & $0.487659$ & $4.131521$ & $-4.300556$ & $0.107804$ & $-0.049587$ & $0.136239$ & $0.144983$ & $ 0.507244 $  & $ 0.504631 $  \\
NTH45PH30 & $0.307584$ & $0.487664$ & $4.131521$ & $-4.300556$ & $0.107804$ & $-0.072491$ & $0.125559$ & $0.144983$ & $ 0.507246 $  & $ 0.504635 $  \\
NTH45PH40 & $0.307615$ & $0.487676$ & $4.131521$ & $-4.300556$ & $0.107804$ & $-0.093193$ & $0.111063$ & $0.144983$ & $ 0.507242 $  & $ 0.504647 $  \\
NTH45PH45 & $0.307636$ & $0.487681$ & $4.131521$ & $-4.300556$ & $0.107804$ & $-0.102518$ & $0.102518$ & $0.144983$ & $ 0.507242 $  & $ 0.504651 $  \\
NTH45PH50 & $0.307658$ & $0.487685$ & $4.131521$ & $-4.300556$ & $0.107804$ & $-0.111063$ & $0.093193$ & $0.144983$ & $ 0.507241 $  & $ 0.504655 $  \\
NTH45PH60 & $0.307700$ & $0.487693$ & $4.131521$ & $-4.300556$ & $0.107804$ & $-0.125559$ & $0.072491$ & $0.144983$ & $ 0.507240 $  & $ 0.504663 $  \\
NTH45PH75 & $0.307746$ & $0.487703$ & $4.131521$ & $-4.300556$ & $0.107804$ & $-0.140042$ & $0.037524$ & $0.144983$ & $ 0.507238 $  & $ 0.504671 $  \\
NTH45PH90 & $0.307757$ & $0.487708$ & $4.131521$ & $-4.300556$ & $0.107804$ & $-0.144983$ & $0.000000$ & $0.144983$ & $ 0.507233 $  & $ 0.504675 $  \\
NTH45PH975 & $0.307755$ & $0.487707$ & $4.131521$ & $-4.300556$ & $0.107804$ & $-0.143742$ & $-0.018924$ & $0.144983$ & $ 0.507233 $  & $ 0.504673 $  \\
NTH45PH120 & $0.307699$ & $0.487694$ & $4.131521$ & $-4.300556$ & $0.107804$ & $-0.125559$ & $-0.072491$ & $0.144983$ & $ 0.507230 $  & $ 0.504660 $  \\
NTH45PH1125 & $0.307723$ & $0.487700$ & $4.131521$ & $-4.300556$ & $0.107804$ & $-0.133947$ & $-0.055482$ & $0.144983$ & $ 0.507229 $  & $ 0.504666 $  \\
NTH45PH1275 & $0.307668$ & $0.487688$ & $4.131521$ & $-4.300556$ & $0.107804$ & $-0.115022$ & $-0.088260$ & $0.144983$ & $ 0.507229 $  & $ 0.504655 $  \\
NTH45PH150 & $0.307581$ & $0.487665$ & $4.131521$ & $-4.300556$ & $0.107804$ & $-0.072491$ & $-0.125559$ & $0.144983$ & $ 0.507236 $  & $ 0.504634 $  \\
NTH60PH0 & $0.307374$ & $0.487595$ & $4.174170$ & $-4.294196$ &
$0.108211$ & $0.000000$ & $0.177524$ & $0.102493$ & $ 0.507180 $  & $
0.504534 $ & $4^{0.02}_{0.003}$ \\
NTH60PH30 & $0.307465$ & $0.487616$ & $4.174170$ & $-4.294196$ & $0.108211$ & $-0.088762$ & $0.153740$ & $0.102493$ & $ 0.507184 $  & $ 0.504556 $  \\
NTH60PH60 & $0.307656$ & $0.487656$ & $4.174170$ & $-4.294196$ & $0.108211$ & $-0.153740$ & $0.088762$ & $0.102493$ & $ 0.507180 $  & $ 0.504592 $  \\
NTH60PH90 & $0.307743$ & $0.487679$ & $4.174170$ & $-4.294196$ & $0.108211$ & $-0.177524$ & $0.000000$ & $0.102493$ & $ 0.507170 $  & $ 0.504611 $  \\
NTH60PH120 & $0.307652$ & $0.487657$ & $4.174170$ & $-4.294196$ & $0.108211$ & $-0.153740$ & $-0.088762$ & $0.102493$ & $ 0.507164 $  & $ 0.504589 $  \\
NTH60PH150 & $0.307465$ & $0.487617$ & $4.174170$ & $-4.294196$ & $0.108211$ & $-0.088762$ & $-0.153740$ & $0.102493$ & $ 0.507169 $  & $ 0.504552 $  \\
NTH120PH0 & $0.307974$ & $0.488280$ & $4.661137$ & $-4.542010$ &
$0.105032$ & $0.000000$ & $0.177129$ & $-0.102265$ & $ 0.506644 $  & $
0.504014 $  & $3.5^{0.02}_{0.004}$\\
NTH120PH30 & $0.308059$ & $0.488303$ & $4.661137$ & $-4.542010$ & $0.105032$ & $-0.088564$ & $0.153398$ & $-0.102265$ & $ 0.506646 $  & $ 0.504037 $  \\
NTH120PH60 & $0.308241$ & $0.488342$ & $4.661137$ & $-4.542010$ & $0.105032$ & $-0.153398$ & $0.088564$ & $-0.102265$ & $ 0.506642 $  & $ 0.504073 $  \\
NTH120PH90 & $0.308327$ & $0.488364$ & $4.661137$ & $-4.542010$ & $0.105032$ & $-0.177129$ & $0.000000$ & $-0.102265$ & $ 0.506634 $  & $ 0.504090 $  \\
NTH120PH120 & $0.308241$ & $0.488342$ & $4.661137$ & $-4.542010$ & $0.105032$ & $-0.153398$ & $-0.088564$ & $-0.102265$ & $ 0.506630 $  & $ 0.504068 $  \\
NTH120PH150 & $0.308058$ & $0.488303$ & $4.661137$ & $-4.542010$ & $0.105032$ & $-0.088564$ & $-0.153398$ & $-0.102265$ & $ 0.506635 $  & $ 0.504032 $  \\
NTH135PH0 & $0.308101$ & $0.488288$ & $4.701321$ & $-4.532307$ &
$0.105392$ & $0.000000$ & $0.144599$ & $-0.144599$ & $ 0.506603 $  & $
0.504004 $  & $3.5^{0.02}_{0.004}$\\
NTH135PH30 & $0.308159$ & $0.488302$ & $4.701321$ & $-4.532307$ & $0.105392$ & $-0.072299$ & $0.125226$ & $-0.144599$ & $ 0.506602 $  & $ 0.504019 $  \\
NTH135PH60 & $0.308281$ & $0.488329$ & $4.701321$ & $-4.532307$ & $0.105392$ & $-0.125226$ & $0.072299$ & $-0.144599$ & $ 0.506600 $  & $ 0.504043 $  \\
NTH135PH90 & $0.308342$ & $0.488343$ & $4.701321$ & $-4.532307$ & $0.105392$ & $-0.144599$ & $0.000000$ & $-0.144599$ & $ 0.506597 $  & $ 0.504053 $  \\
NTH135PH120 & $0.308282$ & $0.488329$ & $4.701321$ & $-4.532307$ & $0.105392$ & $-0.125226$ & $-0.072299$ & $-0.144599$ & $ 0.506591 $  & $ 0.504039 $  \\
NTH135PH150 & $0.308161$ & $0.488301$ & $4.701321$ & $-4.532307$ & $0.105392$ & $-0.072299$ & $-0.125226$ & $-0.144599$ & $ 0.506595 $  & $ 0.504014 $  \\
NTH165PH0 & $0.308524$ & $0.488582$ & $4.854936$ & $-4.625367$ &
$0.104087$ & $0.000000$ & $0.052895$ & $-0.197408$ & $ 0.506462 $  & $
0.503917 $  & $3.5^{0.02}_{0.005}$\\
NTH165PH30 & $0.308529$ & $0.488584$ & $4.854936$ & $-4.625367$ & $0.104087$ & $-0.026448$ & $0.045809$ & $-0.197408$ & $ 0.506461 $  & $ 0.503920 $  \\
NTH165PH60 & $0.308549$ & $0.488587$ & $4.854936$ & $-4.625367$ & $0.104087$ & $-0.045809$ & $0.026448$ & $-0.197408$ & $ 0.506463 $  & $ 0.503922 $  \\
NTH165PH90 & $0.308553$ & $0.488590$ & $4.854936$ & $-4.625367$ & $0.104087$ & $-0.052895$ & $0.000000$ & $-0.197408$ & $ 0.506461 $  & $ 0.503925 $  \\
NTH165PH120 & $0.308545$ & $0.488588$ & $4.854936$ & $-4.625367$ & $0.104087$ & $-0.045809$ & $-0.026448$ & $-0.197408$ & $ 0.506461 $  & $ 0.503923 $  \\
NTH165PH150 & $0.308529$ & $0.488584$ & $4.854936$ & $-4.625367$ & $0.104087$ & $-0.026448$ & $-0.045809$ & $-0.197408$ & $ 0.506460 $  & $ 0.503919 $  \\
\end{tabular}
\end{ruledtabular}
\end{table}

\begin{table}
\caption{Initial data parameters. In all cases the puncture masses were
chosen such that the total ADM mass of the binary was $1.0\pm10^{-6}M$. Here
the punctures are located at $(x_{1,2},0,0)$ 
with momenta $\pm(0, p,0)$ and spins
$\vec S_1 = (S_x, S_y, S_z)$. For the S configurations the second
BH spin is given by $\vec S_2 =
-\vec S_1$, while for the K configurations it is
given by $\vec S_2 = (S_x, S_y,
-S_z)$. Finally, for the L configurations, $\vec S_2 = (0,0,|\vec
S_1|)$.
The approximate initial
eccentricities $e_i$,
eccentricities measured over the last orbit $e_f$, and the number of
orbits $N$,
are also given.
} \label{tab:ID_part2}
\begin{ruledtabular}
\begin{tabular}{l|lllllllllll}
CONF & $m_{p1}/M$ & $m_{p2}/M$ & $x_1/M$ & $x_2/M$  & $p/M$ &
$S_x/M^2$ & $S_y/M^2$ & $S_z/M^2$ & $m_{H1}$ & $m_{H2}$ &
$N_{e_{\rm f}}^{e_{\rm i}}$\\
\hline
STH45PH0 & $0.302923$ & $0.303004$ & $4.194252$ & $-4.534000$ &
$0.107508$ & $0.000000$ & $0.144813$ & $0.144813$ & $ 0.505422 $  & $
0.505411 $ &  $3.5^{0.02}_{0.004}$ \\
STH45PH30 & $0.303051$ & $0.303137$ & $4.194252$ & $-4.534000$ & $0.107508$ & $-0.072406$ & $0.125412$ & $0.144813$ & $ 0.505446 $  & $ 0.505436 $  \\
STH45PH60 & $0.303313$ & $0.303398$ & $4.194252$ & $-4.534000$ & $0.107508$ & $-0.125412$ & $0.072406$ & $0.144813$ & $ 0.505482 $  & $ 0.505474 $  \\
STH45PH90 & $0.303442$ & $0.303531$ & $4.194252$ & $-4.534000$ & $0.107508$ & $-0.144813$ & $0.000000$ & $0.144813$ & $ 0.505496 $  & $ 0.505488 $  \\
STH45PH120 & $0.303312$ & $0.303400$ & $4.194252$ & $-4.534000$ & $0.107508$ & $-0.125412$ & $-0.072406$ & $0.144813$ & $ 0.505468 $  & $ 0.505461 $  \\
STH45PH150 & $0.303051$ & $0.303137$ & $4.194252$ & $-4.534000$ & $0.107508$ & $-0.072406$ & $-0.125412$ & $0.144813$ & $ 0.505433 $  & $ 0.505424 $  \\
\hline
KTH45PH0 & $0.303010$ & $0.303092$ & $4.187859$ & $-4.527100$ &
$0.107491$ & $0.000000$ & $0.144822$ & $0.144822$ & $ 0.505460 $  & $
0.505454 $ & $3.5^{0.02}_{0.004}$\\
\end{tabular}
\end{ruledtabular}
\end{table}

\clearpage
\subsection{Remnant Properties and Radiated Energy-Momentum}
\label{app:Res}
\begin{table}[h!]
\caption{Remnant horizon properties using the IH formalism. Quoted errors are calculated from the
variation of IH quantities with time. See
Tables~\ref{tab:rem_rad_cmp_part1} and ~\ref{tab:rem_rad_cmp_part2} for
more realistic measures of the true error.}
\label{tab:remnant_part1}
\begin{ruledtabular}
\begin{tabular}{l|llllllllll}
CONF & $m_{H}/M$ & $\alpha$ & $S_x/M$ & $S_y/M$ & $S_z/M$\\
\hline
NTH15PH0 & $ 0.949002\pm2.5\cdot10^{-5} $  & $ 0.800400\pm2.9\cdot10^{-5} $  & $ -0.0059188\pm5\cdot10^{-3} $  & $ 0.0229940\pm5\cdot10^{-3} $  & $ 0.72045\pm3\cdot10^{-4} $  \\
NTH15PH30 & $ 0.949096\pm2.5\cdot10^{-5} $  & $ 0.800619\pm3.1\cdot10^{-5} $  & $ -0.0087750\pm5\cdot10^{-7} $  & $ 0.0327727\pm2\cdot10^{-6} $  & $ 0.72039\pm4\cdot10^{-5} $  \\
NTH15PH60 & $ 0.949177\pm2.5\cdot10^{-5} $  & $ 0.800721\pm2.8\cdot10^{-5} $  & $ -0.0193710\pm1\cdot10^{-6} $  & $ 0.0200562\pm1\cdot10^{-6} $  & $ 0.72086\pm4\cdot10^{-5} $  \\
NTH15PH90 & $ 0.949169\pm2.6\cdot10^{-5} $  & $ 0.800619\pm3.4\cdot10^{-5} $  & $ -0.0247011\pm5\cdot10^{-4} $  & $ 0.0037424\pm2\cdot10^{-3} $  & $ 0.72086\pm4\cdot10^{-5} $  \\
NTH15PH120 & $ 0.949076\pm2.6\cdot10^{-5} $  & $ 0.800397\pm3.3\cdot10^{-5} $  & $ -0.0231286\pm1\cdot10^{-2} $  & $ -0.0123739\pm7\cdot10^{-3} $  & $ 0.72048\pm4\cdot10^{-4} $  \\
NTH15PH150 & $ 0.948988\pm2.5\cdot10^{-5} $  & $ 0.800289\pm2.9\cdot10^{-5} $  & $ -0.0271884\pm1\cdot10^{-2} $  & $ -0.0324823\pm7\cdot10^{-3} $  & $ 0.71948\pm1\cdot10^{-3} $  \\
NTH30PH0 & $ 0.950352\pm1.1\cdot10^{-5} $  & $ 0.793089\pm8.7\cdot10^{-6} $  & $ 0.0102968\pm3\cdot10^{-7} $  & $ 0.0606032\pm2\cdot10^{-6} $  & $ 0.71365\pm2\cdot10^{-5} $  \\
\end{tabular}
\end{ruledtabular}
\end{table}

\begin{table}
\caption{Remnant horizon properties using the IH formalism
(continued). Quoted errors are calculated from the
variation of IH quantities with time. See
Tables~\ref{tab:rem_rad_cmp_part1} and \ref{tab:rem_rad_cmp_part2} for
more realistic measures of the true error.}
\label{tab:remnant_part2}
\begin{ruledtabular}
\begin{tabular}{l|llllllllll}
CONF & $m_{H}/M$ & $\alpha$ & $S_x/M$ & $S_y/M$ & $S_z/M$\\
\hline
STH45PH0 & $ 0.960386\pm1.2\cdot10^{-5} $  & $ 0.681328\pm5.5\cdot10^{-5} $  & $ -0.0029258\pm2\cdot10^{-7} $  & $ -0.0112825\pm7\cdot10^{-7} $  & $ 0.62831\pm4\cdot10^{-5} $  \\
STH45PH30 & $ 0.961370\pm1.1\cdot10^{-5} $  & $ 0.683685\pm3.8\cdot10^{-5} $  & $ -0.0049041\pm2\cdot10^{-7} $  & $ -0.0106043\pm4\cdot10^{-7} $  & $ 0.63178\pm3\cdot10^{-5} $  \\
STH45PH60 & $ 0.961447\pm1.2\cdot10^{-5} $  & $ 0.682989\pm4.3\cdot10^{-5} $  & $ -0.0059768\pm3\cdot10^{-7} $  & $ -0.0059344\pm3\cdot10^{-7} $  & $ 0.63128\pm3\cdot10^{-5} $  \\
STH45PH90 & $ 0.960884\pm1.3\cdot10^{-5} $  & $ 0.680984\pm5.8\cdot10^{-5} $  & $ 0.0043894\pm3\cdot10^{-7} $  & $ 0.0118551\pm7\cdot10^{-7} $  & $ 0.62862\pm4\cdot10^{-5} $  \\
STH45PH120 & $ 0.960038\pm1.5\cdot10^{-5} $  & $ 0.678902\pm5.9\cdot10^{-5} $  & $ -0.0020680\pm2\cdot10^{-7} $  & $ -0.0083257\pm6\cdot10^{-7} $  & $ 0.62567\pm4\cdot10^{-5} $  \\
STH45PH150 & $ 0.959700\pm1.6\cdot10^{-5} $  & $ 0.678717\pm6.1\cdot10^{-5} $  & $ -0.0006975\pm9\cdot10^{-8} $  & $ 0.0130117\pm8\cdot10^{-7} $  & $ 0.62498\pm4\cdot10^{-5} $  \\
\hline
KTH45PH0 & $ 0.959306\pm4.6\cdot 10^{-6} $  & $ 0.733285\pm1.5\cdot 10^{-5} $  & $
0.0096842\pm3\cdot 10^{-7} $  & $ 0.1670485\pm4\cdot 10^{-6} $  & $ 0.65374\pm2\cdot 10^{-5} $
\\
\end{tabular}
\end{ruledtabular}
\end{table}

\begin{table}
\caption{Radiated mass, angular momentum, and the
remnant recoil (in original frame) as calculated from $\psi_4$. Errors
quoted are from differences between to extrapolation to $r=\infty$. See
Tables~\ref{tab:rem_rad_cmp_part1} and \ref{tab:rem_rad_cmp_part2} for more accurate measurement of the error. }
\label{tab:rad_part1}
\begin{ruledtabular}
\begin{tabular}{l|llllllllll}
CONF & $\delta M_{\rm rad}$ & $\delta J_{x}$ & $\delta J_{y}$ & $\delta J_{z}$ & $V_{x}$ & $V_{y}$ & $V_{z}$\\
\hline
NTH15PH0 & $ 0.0494\pm0.0001 $  & $ -0.0015\pm0.0008 $  & $ 0.0199\pm0.0023 $  & $ 0.3690\pm0.0023 $  & $ 75\pm22 $  & $ 113\pm5 $  & $ -456\pm2 $  \\
NTH15PH30 & $ 0.0493\pm0.0001 $  & $ -0.0090\pm0.0004 $  & $ 0.0154\pm0.0005 $  & $ 0.3687\pm0.0024 $  & $ 75\pm21 $  & $ 121\pm5 $  & $ -248\pm2 $  \\
NTH15PH60 & $ 0.0492\pm0.0001 $  & $ -0.0140\pm0.0016 $  & $ 0.0068\pm0.0015 $  & $ 0.3685\pm0.0025 $  & $ 67\pm21 $  & $ 126\pm5 $  & $ 21\pm1 $  \\
NTH15PH90 & $ 0.0492\pm0.0001 $  & $ -0.0153\pm0.0022 $  & $ -0.0035\pm0.0029 $  & $ 0.3690\pm0.0021 $  & $ 60\pm21 $  & $ 125\pm4 $  & $ 286\pm1 $  \\
NTH15PH120 & $ 0.0493\pm0.0001 $  & $ -0.0126\pm0.0022 $  & $ -0.0129\pm0.0037 $  & $ 0.3689\pm0.0025 $  & $ 59\pm21 $  & $ 117\pm5 $  & $ 482\pm2 $  \\
NTH15PH150 & $ 0.0494\pm0.0001 $  & $ -0.0065\pm0.0017 $  & $ -0.0189\pm0.0035 $  & $ 0.3692\pm0.0023 $  & $ 68\pm22 $  & $ 111\pm5 $  & $ 546\pm2 $  \\
NTH30PH0 & $ 0.0481\pm0.0001 $  & $ -0.0040\pm0.0021 $  & $ 0.0377\pm0.0036 $  & $ 0.3588\pm0.0028 $  & $ 129\pm21 $  & $ 41\pm6 $  & $ -548\pm3 $  \\
\end{tabular}
\end{ruledtabular}
\end{table}

\begin{table}
\caption{Radiated mass, angular momentum, and the
remnant recoil (in original frame) as calculated from $\psi_4$. Errors
quoted are from differences between to extrapolation to $r=\infty$. See
Tables~\ref{tab:rem_rad_cmp_part1} and \ref{tab:rem_rad_cmp_part2} for more accurate measurement of the error. }
\label{tab:rad_part2}
\begin{ruledtabular}
\begin{tabular}{l|llllllllll}
CONF & $\delta M_{\rm rad}$ & $\delta J_{x}$ & $\delta J_{y}$ & $\delta J_{z}$ & $V_{x}$ & $V_{y}$ & $V_{z}$\\
\hline
STH45PH0 & $ 0.0380\pm0.0001 $  & $ -0.0039\pm0.0022 $  & $ 0.0033\pm0.0029 $  & $ 0.3026\pm0.0035 $  & $ -61\pm1 $  & $ 255\pm3 $  & $ 783\pm3 $  \\
STH45PH30 & $ 0.0371\pm0.0001 $  & $ -0.0046\pm0.0034 $  & $ 0.0018\pm0.0009 $  & $ 0.2988\pm0.0036 $  & $ -14\pm1 $  & $ 261\pm2 $  & $ -471\pm3 $  \\
STH45PH60 & $ 0.0371\pm0.0001 $  & $ -0.0047\pm0.0043 $  & $ 0.0013\pm0.0008 $  & $ 0.3000\pm0.0032 $  & $ -16\pm2 $  & $ 265\pm2 $  & $ -1335\pm6 $  \\
STH45PH90 & $ 0.0376\pm0.0001 $  & $ -0.0035\pm0.0040 $  & $ 0.0006\pm0.0010 $  & $ 0.3036\pm0.0025 $  & $ -47\pm3 $  & $ 265\pm1 $  & $ -1835\pm8 $  \\
STH45PH120 & $ 0.0384\pm0.0001 $  & $ -0.0003\pm0.0017 $  & $ -0.0016\pm0.0011 $  & $ 0.3065\pm0.0025 $  & $ -70\pm3 $  & $ 262\pm1 $  & $ -2004\pm8 $  \\
STH45PH150 & $ 0.0386\pm0.0001 $  & $ 0.0028\pm0.0009 $  & $ -0.0032\pm0.0030 $  & $ 0.3068\pm0.0030 $  & $ -89\pm1 $  & $ 254\pm2 $  & $ -1707\pm5 $  \\
\hline
KTH45PH0 & $ 0.0389\pm0.0001 $  & $ -0.0026\pm0.0029 $  & $
0.1083\pm0.0020 $  & $ 0.2797\pm0.0025 $  & $ 147\pm2 $  & $ -486\pm1
$  & $ -1777\pm4 $  \\
\end{tabular}
\end{ruledtabular}
\end{table}

\begin{table}
\caption{Comparison between remnant horizon properties and radiated
quantities. Differences between the two is a much better measurement
of the true error.}
\label{tab:rem_rad_cmp_part1}
\begin{ruledtabular}
\begin{tabular}{l|ll|ll|ll|ll|ll}
CONF & $\delta M_{\rm rad}$ & $\delta M_{\rm rem}$ &  $\delta J_{x {\rm rad}}$ & $\delta J_{x{\rm rem}}$ & $\delta J_{y {\rm rad}}$ & $\delta J_{y{\rm rem}}$ &
$\delta J_{z {\rm rad}}$ & $\delta J_{z{\rm rem}}$ &\\
\hline
NTH15PH0 & $ 0.0494$  & $ 0.0510$  & $ -0.0015$  & $ 0.0059$  & $ 0.0199$  & $ 0.0301$  & $ 0.3690 $  & $ 0.3774$  \\
NTH15PH30 & $ 0.0493$  & $ 0.0509$  & $ -0.0090$  & $ -0.0178$  & $ 0.0154$  & $ 0.0132$  & $ 0.3687 $  & $ 0.3774$  \\
NTH15PH60 & $ 0.0492$  & $ 0.0508$  & $ -0.0140$  & $ -0.0266$  & $ 0.0068$  & $ 0.0065$  & $ 0.3685 $  & $ 0.3770$  \\
NTH15PH90 & $ 0.0492$  & $ 0.0508$  & $ -0.0153$  & $ -0.0284$  & $ -0.0035$  & $ -0.0037$  & $ 0.3690 $  & $ 0.3770$  \\
NTH15PH120 & $ 0.0493$  & $ 0.0509$  & $ -0.0126$  & $ -0.0228$  & $ -0.0129$  & $ -0.0142$  & $ 0.3689 $  & $ 0.3773$  \\
NTH15PH150 & $ 0.0494$  & $ 0.0510$  & $ -0.0065$  & $ 0.0006$  & $ -0.0189$  & $ -0.0135$  & $ 0.3692 $  & $ 0.3783$  \\
NTH30PH0 & $ 0.0481$  & $ 0.0496$  & $ -0.0040$  & $ -0.0103$  & $ 0.0377$  & $ 0.0419$  & $ 0.3588 $  & $ 0.3673$  \\
\end{tabular}
\end{ruledtabular}
\end{table}

\begin{table}
\caption{Comparison between remnant horizon properties and radiated
quantities. Differences between the two is a much better measurement
of the true error.}
\label{tab:rem_rad_cmp_part2}
\begin{ruledtabular}
\begin{tabular}{l|ll|ll|ll|ll|ll}
CONF & $\delta M_{\rm rad}$ & $\delta M_{\rm rem}$ &  $\delta J_{x {\rm rad}}$ & $\delta J_{x{\rm rem}}$ & $\delta J_{y {\rm rad}}$ & $\delta J_{y{\rm rem}}$ &
$\delta J_{z {\rm rad}}$ & $\delta J_{z{\rm rem}}$ &\\
\hline
STH45PH0 & $ 0.0380$  & $ 0.0396$  & $ -0.0039$  & $ 0.0029$  & $ 0.0033$  & $ 0.0113$  & $ 0.3026 $  & $ 0.3100$  \\
STH45PH30 & $ 0.0371$  & $ 0.0386$  & $ -0.0046$  & $ 0.0049$  & $ 0.0018$  & $ 0.0106$  & $ 0.2988 $  & $ 0.3066$  \\
STH45PH60 & $ 0.0371$  & $ 0.0386$  & $ -0.0047$  & $ 0.0060$  & $ 0.0013$  & $ 0.0059$  & $ 0.3000 $  & $ 0.3071$  \\
STH45PH90 & $ 0.0376$  & $ 0.0391$  & $ -0.0035$  & $ -0.0044$  & $ 0.0006$  & $ -0.0119$  & $ 0.3036 $  & $ 0.3097$  \\
STH45PH120 & $ 0.0384$  & $ 0.0400$  & $ -0.0003$  & $ 0.0021$  & $ -0.0016$  & $ 0.0083$  & $ 0.3065 $  & $ 0.3127$  \\
STH45PH150 & $ 0.0386$  & $ 0.0403$  & $ 0.0028$  & $ 0.0007$  & $ -0.0032$  & $ -0.0130$  & $ 0.3068 $  & $ 0.3134$  \\
\hline
KTH45PH0 & $ 0.0389$  & $ 0.0407$  & $ -0.0026$  & $ -0.0097$  & $
0.1083$  & $ 0.1226$  & $ 0.2797 $  & $ 0.2830$  \\
\end{tabular}
\end{ruledtabular}
\end{table}

\begin{table}
\caption{BH spins during final plunge, recoil velocity,
and the angle between $\vec\Delta_\perp$ for PHYYY and $\vec\Delta_\perp$ of the corresponding PH0 configuration; all calculated
in 
a rotated frame where the infall occurs in the $xy$ plane.}
\label{tab:rot_part1}
\begin{ruledtabular}
\begin{tabular}{l|llllllllll}
CONF & $S_{x1}$ & $S_{y1}$ & $S_{z1}$ & $S_{x2}$ & $S_{y2}$ & $S_{z2}$ & $V_{x}$ & $V_{y}$ & $V_{z}$ & $\varphi$ \\
\hline
NTH15PH0 & $ 0.0319 $  & $ -0.0412 $  & $ 0.2000 $  & $ 0.0000 $  & $ 0.0000 $  & $ 0.0000 $  & $ 77 $  & $ -110 $  & $ -457 $  & $ 0$  \\
NTH15PH30 & $ 0.0454 $  & $ -0.0238 $  & $ 0.1998 $  & $ 0.0000 $  & $ 0.0000 $  & $ 0.0000 $  & $ 79 $  & $ -110 $  & $ -252 $  & $ 24.5$  \\
NTH15PH60 & $ 0.0470 $  & $ -0.0014 $  & $ 0.2012 $  & $ 0.0000 $  & $ 0.0000 $  & $ 0.0000 $  & $ 82 $  & $ -117 $  & $ 15 $  & $ 50.5$  \\
NTH15PH90 & $ 0.0372 $  & $ 0.0210 $  & $ 0.2022 $  & $ 0.0000 $  & $ 0.0000 $  & $ 0.0000 $  & $ 85 $  & $ -126 $  & $ 279 $  & $ 81.7$  \\
NTH15PH120 & $ -0.0249 $  & $ -0.0216 $  & $ 0.2030 $  & $ 0.0000 $  & $ 0.0000 $  & $ 0.0000 $  & $ -116 $  & $ -98 $  & $ 476 $  & $ 122.8$  \\
NTH15PH150 & $ -0.0083 $  & $ 0.0475 $  & $ 0.2009 $  & $ 0.0000 $  & $ 0.0000 $  & $ 0.0000 $  & $ 79 $  & $ -117 $  & $ 543 $  & $ 152.1$  \\
NTH30PH0 & $ 0.0818 $  & $ -0.0569 $  & $ 0.1806 $  & $ 0.0000 $  & $ 0.0000 $  & $ 0.0000 $  & $ 62 $  & $ -82 $  & $ -555 $  & $ 0$  \\
\end{tabular}
\end{ruledtabular}
\end{table}

\begin{table}
\caption{BH spins during final plunge, recoil velocity,
and the angle between $\vec\Delta_\perp$ for PHYYY and $\vec\Delta_\perp$ of the corresponding PH0 configuration; all calculated
in 
a rotated frame where the infall occurs in the $xy$ plane.}
\label{tab:rot_part2}
\begin{ruledtabular}
\begin{tabular}{l|llllllllll}
CONF & $S_{x1}$ & $S_{y1}$ & $S_{z1}$ & $S_{x2}$ & $S_{y2}$ & $S_{z2}$ & $V_{x}$ & $V_{y}$ & $V_{z}$ & $\varphi$ \\
\hline
STHPH0 & $ -0.1274 $  & $ 0.0542 $  & $ 0.1520 $  & $ 0.1323 $  & $ -0.0633 $  & $ -0.1432 $  & $ 78 $  & $ -252 $  & $ 783 $  & $ 0 $ \\
STHPH30 & $ -0.1346 $  & $ -0.0245 $  & $ 0.1531 $  & $ 0.1453 $  & $ 0.0137 $  & $ -0.1435 $  & $ 43 $  & $ -255 $  & $ -472 $  & $ 33.4 $ \\
STHPH60 & $ -0.1086 $  & $ -0.0869 $  & $ 0.1521 $  & $ 0.1205 $  & $ 0.0774 $  & $ -0.1449 $  & $ 21 $  & $ -258 $  & $ -1336 $  & $ 61.7 $ \\
STHPH90 & $ -0.0518 $  & $ -0.1308 $  & $ 0.1499 $  & $ 0.0642 $  & $ 0.1262 $  & $ -0.1464 $  & $ 38 $  & $ -258 $  & $ -1836 $  & $ 91.4 $ \\
STHPH120 & $ 0.0080 $  & $ -0.1428 $  & $ 0.1486 $  & $ -0.0002 $  & $ 0.1422 $  & $ -0.1480 $  & $ 39 $  & $ -262 $  & $ -2005 $  & $ 116.3 $ \\
STHPH150 & $ 0.0729 $  & $ -0.1216 $  & $ 0.1493 $  & $ -0.0707 $  & $ 0.1259 $  & $ -0.1458 $  & $ 60 $  & $ -260 $  & $ -1707 $  & $ 144. $ \\
\hline
KTH45PH0 & $ -0.1815 $  & $ -0.0207 $  & $ 0.0933 $  & $ -0.0487 $  & $
0.1872 $  & $ -0.0674 $  & $ 233 $  & $ -379 $  & $ -1794 $  & $ 0 $
\\
\end{tabular}
\end{ruledtabular}
\end{table}

\begin{table}
\caption{Comparing the measured value of the z component of the
recoil to values of $\Delta_\perp$, $S_\perp$, $\Delta_z$, and $S_z$
in 
a rotated frame where the infall occurs in the $xy$ plane. These data
are equivalent to the data given in Table~\ref{tab:rot_part2}.
}
   \label{tab:rot_part3}
\begin{ruledtabular}
\begin{tabular}{l|llllllllll}
  CONF & $\Delta_\perp$ & $S_\perp$ & $ \Delta_z$ & $S_z$ &  $V_{z}$ & $\varphi$ \\
\hline
STHPH0 & 0.554431 & 0.0100482 & -0.574235 & 0.00849741 & $ 783 $  & $ 0 $ \\
STHPH30 & 0.549469 & 0.014804 & -0.577022 & 0.00931534 &  $ -472 $  & $ 33.4 $ \\
STHPH60 & 0.548492 & 0.0148527 & -0.577713 & 0.00691211 &  $ -1336 $  & $ 61.7 $ \\
STHPH90 & 0.548418 & 0.0128664 & -0.576338 & 0.0034721 & $ -1836 $  & $ 91.4 $ \\
STHPH120 &  0.554563 & 0.00763799 & -0.576841 & 0.000557287 & $ -2005 $  & $ 116.3 $ \\
STHPH150 &  0.556667 & 0.00463136 & -0.574073 & 0.00338651 & $ -1707 $  & $ 144. $ \\
\hline
KTH45PH0 & 0.482864 & 0.278006 & -0.314672 & 0.025327 &  $ -1794 $  & $ 0 $
\\
\end{tabular}
\end{ruledtabular}
\end{table}
\clearpage
\end{widetext}

\bibliographystyle{apsrev}
\bibliography{../../../../Bibtex/references,local}

\begin{thebibliography}{59}
\expandafter\ifx\csname natexlab\endcsname\relax\def\natexlab#1{#1}\fi
\expandafter\ifx\csname bibnamefont\endcsname\relax
  \def\bibnamefont#1{#1}\fi
\expandafter\ifx\csname bibfnamefont\endcsname\relax
  \def\bibfnamefont#1{#1}\fi
\expandafter\ifx\csname citenamefont\endcsname\relax
  \def\citenamefont#1{#1}\fi
\expandafter\ifx\csname url\endcsname\relax
  \def\url#1{\texttt{#1}}\fi
\expandafter\ifx\csname urlprefix\endcsname\relax\def\urlprefix{URL }\fi
\providecommand{\bibinfo}[2]{#2}
\providecommand{\eprint}[2][]{\url{#2}}

\bibitem[{\citenamefont{Pretorius}(2005)}]{Pretorius:2005gq}
\bibinfo{author}{\bibfnamefont{F.}~\bibnamefont{Pretorius}},
  \bibinfo{journal}{Phys. Rev. Lett.} \textbf{\bibinfo{volume}{95}},
  \bibinfo{pages}{121101} (\bibinfo{year}{2005}), \eprint{gr-qc/0507014}.

\bibitem[{\citenamefont{Campanelli
  et~al.}(2006{\natexlab{a}})\citenamefont{Campanelli, Lousto, Marronetti, and
  Zlochower}}]{Campanelli:2005dd}
\bibinfo{author}{\bibfnamefont{M.}~\bibnamefont{Campanelli}},
  \bibinfo{author}{\bibfnamefont{C.~O.} \bibnamefont{Lousto}},
  \bibinfo{author}{\bibfnamefont{P.}~\bibnamefont{Marronetti}},
  \bibnamefont{and}
  \bibinfo{author}{\bibfnamefont{Y.}~\bibnamefont{Zlochower}},
  \bibinfo{journal}{Phys. Rev. Lett.} \textbf{\bibinfo{volume}{96}},
  \bibinfo{pages}{111101} (\bibinfo{year}{2006}{\natexlab{a}}),
  \eprint{gr-qc/0511048}.

\bibitem[{\citenamefont{Baker et~al.}(2006)\citenamefont{Baker, Centrella,
  Choi, Koppitz, and van Meter}}]{Baker:2005vv}
\bibinfo{author}{\bibfnamefont{J.~G.} \bibnamefont{Baker}},
  \bibinfo{author}{\bibfnamefont{J.}~\bibnamefont{Centrella}},
  \bibinfo{author}{\bibfnamefont{D.-I.} \bibnamefont{Choi}},
  \bibinfo{author}{\bibfnamefont{M.}~\bibnamefont{Koppitz}}, \bibnamefont{and}
  \bibinfo{author}{\bibfnamefont{J.}~\bibnamefont{van Meter}},
  \bibinfo{journal}{Phys. Rev. Lett.} \textbf{\bibinfo{volume}{96}},
  \bibinfo{pages}{111102} (\bibinfo{year}{2006}), \eprint{gr-qc/0511103}.

\bibitem[{\citenamefont{Campanelli
  et~al.}(2006{\natexlab{b}})\citenamefont{Campanelli, Lousto, and
  Zlochower}}]{Campanelli:2006uy}
\bibinfo{author}{\bibfnamefont{M.}~\bibnamefont{Campanelli}},
  \bibinfo{author}{\bibfnamefont{C.~O.} \bibnamefont{Lousto}},
  \bibnamefont{and}
  \bibinfo{author}{\bibfnamefont{Y.}~\bibnamefont{Zlochower}},
  \bibinfo{journal}{Phys. Rev.} \textbf{\bibinfo{volume}{D74}},
  \bibinfo{pages}{041501(R)} (\bibinfo{year}{2006}{\natexlab{b}}),
  \eprint{gr-qc/0604012}.

\bibitem[{\citenamefont{Campanelli
  et~al.}(2007{\natexlab{a}})\citenamefont{Campanelli, Lousto, Zlochower, and
  Merritt}}]{Campanelli:2007ew}
\bibinfo{author}{\bibfnamefont{M.}~\bibnamefont{Campanelli}},
  \bibinfo{author}{\bibfnamefont{C.~O.} \bibnamefont{Lousto}},
  \bibinfo{author}{\bibfnamefont{Y.}~\bibnamefont{Zlochower}},
  \bibnamefont{and} \bibinfo{author}{\bibfnamefont{D.}~\bibnamefont{Merritt}},
  \bibinfo{journal}{Astrophys. J.} \textbf{\bibinfo{volume}{659}},
  \bibinfo{pages}{L5} (\bibinfo{year}{2007}{\natexlab{a}}),
  \eprint{gr-qc/0701164}.

\bibitem[{\citenamefont{Gonz\'alez
  et~al.}(2007{\natexlab{a}})\citenamefont{Gonz\'alez, Hannam, Sperhake,
  Brugmann, and Husa}}]{Gonzalez:2007hi}
\bibinfo{author}{\bibfnamefont{J.~A.} \bibnamefont{Gonz\'alez}},
  \bibinfo{author}{\bibfnamefont{M.~D.} \bibnamefont{Hannam}},
  \bibinfo{author}{\bibfnamefont{U.}~\bibnamefont{Sperhake}},
  \bibinfo{author}{\bibfnamefont{B.}~\bibnamefont{Brugmann}}, \bibnamefont{and}
  \bibinfo{author}{\bibfnamefont{S.}~\bibnamefont{Husa}},
  \bibinfo{journal}{Phys. Rev. Lett.} \textbf{\bibinfo{volume}{98}},
  \bibinfo{pages}{231101} (\bibinfo{year}{2007}{\natexlab{a}}),
  \eprint{gr-qc/0702052}.

\bibitem[{\citenamefont{Campanelli
  et~al.}(2007{\natexlab{b}})\citenamefont{Campanelli, Lousto, Zlochower, and
  Merritt}}]{Campanelli:2007cga}
\bibinfo{author}{\bibfnamefont{M.}~\bibnamefont{Campanelli}},
  \bibinfo{author}{\bibfnamefont{C.~O.} \bibnamefont{Lousto}},
  \bibinfo{author}{\bibfnamefont{Y.}~\bibnamefont{Zlochower}},
  \bibnamefont{and} \bibinfo{author}{\bibfnamefont{D.}~\bibnamefont{Merritt}},
  \bibinfo{journal}{Phys. Rev. Lett.} \textbf{\bibinfo{volume}{98}},
  \bibinfo{pages}{231102} (\bibinfo{year}{2007}{\natexlab{b}}),
  \eprint{gr-qc/0702133}.

\bibitem[{\citenamefont{Lousto et~al.}(2010{\natexlab{a}})\citenamefont{Lousto,
  Nakano, Zlochower, and Campanelli}}]{Lousto:2009ka}
\bibinfo{author}{\bibfnamefont{C.~O.} \bibnamefont{Lousto}},
  \bibinfo{author}{\bibfnamefont{H.}~\bibnamefont{Nakano}},
  \bibinfo{author}{\bibfnamefont{Y.}~\bibnamefont{Zlochower}},
  \bibnamefont{and}
  \bibinfo{author}{\bibfnamefont{M.}~\bibnamefont{Campanelli}},
  \bibinfo{journal}{Phys. Rev.} \textbf{\bibinfo{volume}{D81}},
  \bibinfo{pages}{084023} (\bibinfo{year}{2010}{\natexlab{a}}),
  \eprint{0910.3197}.

\bibitem[{\citenamefont{Komossa et~al.}(2008)\citenamefont{Komossa, Zhou, and
  Lu}}]{Komossa:2008qd}
\bibinfo{author}{\bibfnamefont{S.}~\bibnamefont{Komossa}},
  \bibinfo{author}{\bibfnamefont{H.}~\bibnamefont{Zhou}}, \bibnamefont{and}
  \bibinfo{author}{\bibfnamefont{H.}~\bibnamefont{Lu}},
  \bibinfo{journal}{Astrop. J. Letters} \textbf{\bibinfo{volume}{678}},
  \bibinfo{pages}{L81} (\bibinfo{year}{2008}), \eprint{0804.4585}.

\bibitem[{\citenamefont{{Shields} and {Bonning}}(2008)}]{Shields:2008va}
\bibinfo{author}{\bibfnamefont{G.~A.} \bibnamefont{{Shields}}}
  \bibnamefont{and} \bibinfo{author}{\bibfnamefont{E.~W.}
  \bibnamefont{{Bonning}}}, \bibinfo{journal}{Astrophys. J.}
  \textbf{\bibinfo{volume}{682}}, \bibinfo{pages}{758} (\bibinfo{year}{2008}),
  \eprint{0802.3873}.

\bibitem[{\citenamefont{Bogdanovic et~al.}(2009)\citenamefont{Bogdanovic,
  Eracleous, and Sigurdsson}}]{Bogdanovic:2008uz}
\bibinfo{author}{\bibfnamefont{T.}~\bibnamefont{Bogdanovic}},
  \bibinfo{author}{\bibfnamefont{M.}~\bibnamefont{Eracleous}},
  \bibnamefont{and}
  \bibinfo{author}{\bibfnamefont{S.}~\bibnamefont{Sigurdsson}},
  \bibinfo{journal}{Astrophys. J.} \textbf{\bibinfo{volume}{697}},
  \bibinfo{pages}{288} (\bibinfo{year}{2009}), \eprint{0809.3262}.

\bibitem[{\citenamefont{Civano et~al.}(2010)}]{Civano:2010es}
\bibinfo{author}{\bibfnamefont{F.}~\bibnamefont{Civano}} \bibnamefont{et~al.},
  \bibinfo{journal}{Astrophys. J.} \textbf{\bibinfo{volume}{717}},
  \bibinfo{pages}{209} (\bibinfo{year}{2010}), \eprint{1003.0020}.

\bibitem[{\citenamefont{Eracleous et~al.}(2012)\citenamefont{Eracleous,
  Boroson, Halpern, and Liu}}]{Eracleous:2011ua}
\bibinfo{author}{\bibfnamefont{M.}~\bibnamefont{Eracleous}},
  \bibinfo{author}{\bibfnamefont{T.~A.} \bibnamefont{Boroson}},
  \bibinfo{author}{\bibfnamefont{J.~P.} \bibnamefont{Halpern}},
  \bibnamefont{and} \bibinfo{author}{\bibfnamefont{J.}~\bibnamefont{Liu}},
  \bibinfo{journal}{The Astrophysical Journal Supplement}
  \textbf{\bibinfo{volume}{201}}, \bibinfo{eid}{23} (\bibinfo{year}{2012}),
  \eprint{1106.2952}.

\bibitem[{\citenamefont{Tsalmantza et~al.}(2011)\citenamefont{Tsalmantza,
  Decarli, Dotti, and Hogg}}]{Tsalmantza:2011ju}
\bibinfo{author}{\bibfnamefont{P.}~\bibnamefont{Tsalmantza}},
  \bibinfo{author}{\bibfnamefont{R.}~\bibnamefont{Decarli}},
  \bibinfo{author}{\bibfnamefont{M.}~\bibnamefont{Dotti}}, \bibnamefont{and}
  \bibinfo{author}{\bibfnamefont{D.~W.} \bibnamefont{Hogg}},
  \bibinfo{journal}{Astrophys. J.} \textbf{\bibinfo{volume}{738}},
  \bibinfo{pages}{20} (\bibinfo{year}{2011}), \eprint{1106.1180}.

\bibitem[{\citenamefont{Comerford et~al.}(2009)\citenamefont{Comerford,
  Griffith, Gerke, Cooper, Newman et~al.}}]{Comerford:2009ju}
\bibinfo{author}{\bibfnamefont{J.~M.} \bibnamefont{Comerford}},
  \bibinfo{author}{\bibfnamefont{R.~L.} \bibnamefont{Griffith}},
  \bibinfo{author}{\bibfnamefont{B.~F.} \bibnamefont{Gerke}},
  \bibinfo{author}{\bibfnamefont{M.~C.} \bibnamefont{Cooper}},
  \bibinfo{author}{\bibfnamefont{J.~A.} \bibnamefont{Newman}},
  \bibnamefont{et~al.}, \bibinfo{journal}{Astrophys. J.}
  \textbf{\bibinfo{volume}{702}}, \bibinfo{pages}{L82} (\bibinfo{year}{2009}),
  \eprint{0906.3517}.

\bibitem[{\citenamefont{Civano et~al.}(2012)\citenamefont{Civano, Elvis,
  Lanzuisi, Aldcroft, Trichas et~al.}}]{Civano:2012bu}
\bibinfo{author}{\bibfnamefont{F.}~\bibnamefont{Civano}},
  \bibinfo{author}{\bibfnamefont{M.}~\bibnamefont{Elvis}},
  \bibinfo{author}{\bibfnamefont{G.}~\bibnamefont{Lanzuisi}},
  \bibinfo{author}{\bibfnamefont{T.}~\bibnamefont{Aldcroft}},
  \bibinfo{author}{\bibfnamefont{M.}~\bibnamefont{Trichas}},
  \bibnamefont{et~al.}, \bibinfo{journal}{Astrophys. J.}
  \textbf{\bibinfo{volume}{752}}, \bibinfo{pages}{49} (\bibinfo{year}{2012}),
  \eprint{1205.0815}.

\bibitem[{\citenamefont{Blecha et~al.}(2012)\citenamefont{Blecha, Civano,
  Elvis, and Loeb}}]{Blecha:2012kx}
\bibinfo{author}{\bibfnamefont{L.}~\bibnamefont{Blecha}},
  \bibinfo{author}{\bibfnamefont{F.}~\bibnamefont{Civano}},
  \bibinfo{author}{\bibfnamefont{M.}~\bibnamefont{Elvis}}, \bibnamefont{and}
  \bibinfo{author}{\bibfnamefont{A.}~\bibnamefont{Loeb}}
  (\bibinfo{year}{2012}), \eprint{1205.6202}.

\bibitem[{\citenamefont{Komossa}(2012)}]{Komossa:2012cy}
\bibinfo{author}{\bibfnamefont{S.}~\bibnamefont{Komossa}},
  \bibinfo{journal}{Adv. Astron.} \textbf{\bibinfo{volume}{2012}},
  \bibinfo{pages}{364973} (\bibinfo{year}{2012}), \eprint{1202.1977}.

\bibitem[{\citenamefont{Bogdanovic et~al.}(2007)\citenamefont{Bogdanovic,
  Reynolds, and Miller}}]{Bogdanovic:2007hp}
\bibinfo{author}{\bibfnamefont{T.}~\bibnamefont{Bogdanovic}},
  \bibinfo{author}{\bibfnamefont{C.~S.} \bibnamefont{Reynolds}},
  \bibnamefont{and} \bibinfo{author}{\bibfnamefont{M.~C.}
  \bibnamefont{Miller}}, \bibinfo{journal}{Astrophys. J.}
  \textbf{\bibinfo{volume}{661}}, \bibinfo{pages}{L147} (\bibinfo{year}{2007}),
  \eprint{astro-ph/0703054}.

\bibitem[{\citenamefont{{Dotti} et~al.}(2010)\citenamefont{{Dotti},
  {Volonteri}, {Perego}, {Colpi}, {Ruszkowski}, and {Haardt}}}]{Dotti:2009vz}
\bibinfo{author}{\bibfnamefont{M.}~\bibnamefont{{Dotti}}},
  \bibinfo{author}{\bibfnamefont{M.}~\bibnamefont{{Volonteri}}},
  \bibinfo{author}{\bibfnamefont{A.}~\bibnamefont{{Perego}}},
  \bibinfo{author}{\bibfnamefont{M.}~\bibnamefont{{Colpi}}},
  \bibinfo{author}{\bibfnamefont{M.}~\bibnamefont{{Ruszkowski}}},
  \bibnamefont{and} \bibinfo{author}{\bibfnamefont{F.}~\bibnamefont{{Haardt}}},
  \bibinfo{journal}{mnras} \textbf{\bibinfo{volume}{402}}, \bibinfo{pages}{682}
  (\bibinfo{year}{2010}), \eprint{0910.5729}.

\bibitem[{\citenamefont{Lousto and
  Zlochower}(2011{\natexlab{a}})}]{Lousto:2011kp}
\bibinfo{author}{\bibfnamefont{C.~O.} \bibnamefont{Lousto}} \bibnamefont{and}
  \bibinfo{author}{\bibfnamefont{Y.}~\bibnamefont{Zlochower}},
  \bibinfo{journal}{Phys. Rev. Lett.} \textbf{\bibinfo{volume}{107}},
  \bibinfo{pages}{231102} (\bibinfo{year}{2011}{\natexlab{a}}),
  \eprint{1108.2009}.

\bibitem[{\citenamefont{Lousto et~al.}(2012{\natexlab{a}})\citenamefont{Lousto,
  Zlochower, Dotti, and Volonteri}}]{Lousto:2012su}
\bibinfo{author}{\bibfnamefont{C.~O.} \bibnamefont{Lousto}},
  \bibinfo{author}{\bibfnamefont{Y.}~\bibnamefont{Zlochower}},
  \bibinfo{author}{\bibfnamefont{M.}~\bibnamefont{Dotti}}, \bibnamefont{and}
  \bibinfo{author}{\bibfnamefont{M.}~\bibnamefont{Volonteri}},
  \bibinfo{journal}{Phys. Rev.} \textbf{\bibinfo{volume}{D85}},
  \bibinfo{pages}{084015} (\bibinfo{year}{2012}{\natexlab{a}}),
  \eprint{1201.1923}.

\bibitem[{\citenamefont{Rezzolla et~al.}(2008)}]{Rezzolla:2007xa}
\bibinfo{author}{\bibfnamefont{L.}~\bibnamefont{Rezzolla}}
  \bibnamefont{et~al.}, \bibinfo{journal}{Astrophys. J.}
  \textbf{\bibinfo{volume}{679}}, \bibinfo{pages}{1422} (\bibinfo{year}{2008}),
  \eprint{arXiv:0708.3999 [gr-qc]}.

\bibitem[{\citenamefont{Lousto and
  Zlochower}(2011{\natexlab{b}})}]{Lousto:2010xk}
\bibinfo{author}{\bibfnamefont{C.~O.} \bibnamefont{Lousto}} \bibnamefont{and}
  \bibinfo{author}{\bibfnamefont{Y.}~\bibnamefont{Zlochower}},
  \bibinfo{journal}{Phys. Rev.} \textbf{\bibinfo{volume}{D83}},
  \bibinfo{pages}{024003} (\bibinfo{year}{2011}{\natexlab{b}}),
  \eprint{1011.0593}.

\bibitem[{\citenamefont{Kidder}(1995)}]{Kidder:1995zr}
\bibinfo{author}{\bibfnamefont{L.~E.} \bibnamefont{Kidder}},
  \bibinfo{journal}{Phys. Rev.} \textbf{\bibinfo{volume}{D52}},
  \bibinfo{pages}{821} (\bibinfo{year}{1995}), \eprint{gr-qc/9506022}.

\bibitem[{\citenamefont{Racine et~al.}(2009)\citenamefont{Racine, Buonanno, and
  Kidder}}]{Racine:2008kj}
\bibinfo{author}{\bibfnamefont{E.}~\bibnamefont{Racine}},
  \bibinfo{author}{\bibfnamefont{A.}~\bibnamefont{Buonanno}}, \bibnamefont{and}
  \bibinfo{author}{\bibfnamefont{L.~E.} \bibnamefont{Kidder}},
  \bibinfo{journal}{Phys. Rev.} \textbf{\bibinfo{volume}{D80}},
  \bibinfo{pages}{044010} (\bibinfo{year}{2009}), \eprint{0812.4413}.

\bibitem[{\citenamefont{Lousto et~al.}(2010{\natexlab{b}})\citenamefont{Lousto,
  Campanelli, Zlochower, and Nakano}}]{Lousto:2009mf}
\bibinfo{author}{\bibfnamefont{C.~O.} \bibnamefont{Lousto}},
  \bibinfo{author}{\bibfnamefont{M.}~\bibnamefont{Campanelli}},
  \bibinfo{author}{\bibfnamefont{Y.}~\bibnamefont{Zlochower}},
  \bibnamefont{and} \bibinfo{author}{\bibfnamefont{H.}~\bibnamefont{Nakano}},
  \bibinfo{journal}{Class. Quant. Grav.} \textbf{\bibinfo{volume}{27}},
  \bibinfo{pages}{114006} (\bibinfo{year}{2010}{\natexlab{b}}),
  \eprint{0904.3541}.

\bibitem[{\citenamefont{Lousto and Zlochower}(2009)}]{Lousto:2008dn}
\bibinfo{author}{\bibfnamefont{C.~O.} \bibnamefont{Lousto}} \bibnamefont{and}
  \bibinfo{author}{\bibfnamefont{Y.}~\bibnamefont{Zlochower}},
  \bibinfo{journal}{Phys. Rev.} \textbf{\bibinfo{volume}{D79}},
  \bibinfo{pages}{064018} (\bibinfo{year}{2009}), \eprint{0805.0159}.

\bibitem[{\citenamefont{Gonz\'alez
  et~al.}(2007{\natexlab{b}})\citenamefont{Gonz\'alez, Sperhake, Brugmann,
  Hannam, and Husa}}]{Gonzalez:2006md}
\bibinfo{author}{\bibfnamefont{J.~A.} \bibnamefont{Gonz\'alez}},
  \bibinfo{author}{\bibfnamefont{U.}~\bibnamefont{Sperhake}},
  \bibinfo{author}{\bibfnamefont{B.}~\bibnamefont{Brugmann}},
  \bibinfo{author}{\bibfnamefont{M.}~\bibnamefont{Hannam}}, \bibnamefont{and}
  \bibinfo{author}{\bibfnamefont{S.}~\bibnamefont{Husa}},
  \bibinfo{journal}{Phys. Rev. Lett.} \textbf{\bibinfo{volume}{98}},
  \bibinfo{pages}{091101} (\bibinfo{year}{2007}{\natexlab{b}}),
  \eprint{gr-qc/0610154}.

\bibitem[{\citenamefont{{Fitchett}}(1983)}]{1983MNRAS.203.1049F}
\bibinfo{author}{\bibfnamefont{M.~J.} \bibnamefont{{Fitchett}}},
  \bibinfo{journal}{MNRAS} \textbf{\bibinfo{volume}{203}},
  \bibinfo{pages}{1049} (\bibinfo{year}{1983}).

\bibitem[{\citenamefont{Zlochower et~al.}(2011)\citenamefont{Zlochower,
  Campanelli, and Lousto}}]{Zlochower:2010sn}
\bibinfo{author}{\bibfnamefont{Y.}~\bibnamefont{Zlochower}},
  \bibinfo{author}{\bibfnamefont{M.}~\bibnamefont{Campanelli}},
  \bibnamefont{and} \bibinfo{author}{\bibfnamefont{C.~O.}
  \bibnamefont{Lousto}}, \bibinfo{journal}{Class. Quant. Grav.}
  \textbf{\bibinfo{volume}{28}}, \bibinfo{pages}{114015}
  (\bibinfo{year}{2011}), \eprint{1011.2210}.

\bibitem[{\citenamefont{Boyle et~al.}(2008)\citenamefont{Boyle, Kesden, and
  Nissanke}}]{Boyle:2007sz}
\bibinfo{author}{\bibfnamefont{L.}~\bibnamefont{Boyle}},
  \bibinfo{author}{\bibfnamefont{M.}~\bibnamefont{Kesden}}, \bibnamefont{and}
  \bibinfo{author}{\bibfnamefont{S.}~\bibnamefont{Nissanke}},
  \bibinfo{journal}{Phys. Rev. Lett.} \textbf{\bibinfo{volume}{100}},
  \bibinfo{pages}{151101} (\bibinfo{year}{2008}), \eprint{0709.0299}.

\bibitem[{\citenamefont{Nakano et~al.}(2011)\citenamefont{Nakano, Campanelli,
  Lousto, and Zlochower}}]{Nakano:2010kv}
\bibinfo{author}{\bibfnamefont{H.}~\bibnamefont{Nakano}},
  \bibinfo{author}{\bibfnamefont{M.}~\bibnamefont{Campanelli}},
  \bibinfo{author}{\bibfnamefont{C.~O.} \bibnamefont{Lousto}},
  \bibnamefont{and}
  \bibinfo{author}{\bibfnamefont{Y.}~\bibnamefont{Zlochower}},
  \bibinfo{journal}{Class. Quant. Grav.} \textbf{\bibinfo{volume}{28}},
  \bibinfo{pages}{134005} (\bibinfo{year}{2011}), \eprint{1011.2767}.

\bibitem[{\citenamefont{{Arfken} and {Weber}}(2005)}]{2005mmp..book.....A}
\bibinfo{author}{\bibfnamefont{G.~B.} \bibnamefont{{Arfken}}} \bibnamefont{and}
  \bibinfo{author}{\bibfnamefont{H.~J.} \bibnamefont{{Weber}}},
  \emph{\bibinfo{title}{{Mathematical methods for physicists 6th ed.}}}
  (\bibinfo{year}{2005}).

\bibitem[{\citenamefont{Ansorg et~al.}(2004)\citenamefont{Ansorg, Br\"ugmann,
  and Tichy}}]{Ansorg:2004ds}
\bibinfo{author}{\bibfnamefont{M.}~\bibnamefont{Ansorg}},
  \bibinfo{author}{\bibfnamefont{B.}~\bibnamefont{Br\"ugmann}},
  \bibnamefont{and} \bibinfo{author}{\bibfnamefont{W.}~\bibnamefont{Tichy}},
  \bibinfo{journal}{Phys. Rev.} \textbf{\bibinfo{volume}{D70}},
  \bibinfo{pages}{064011} (\bibinfo{year}{2004}), \eprint{gr-qc/0404056}.

\bibitem[{\citenamefont{Brandt and Br{\"u}gmann}(1997)}]{Brandt97b}
\bibinfo{author}{\bibfnamefont{S.}~\bibnamefont{Brandt}} \bibnamefont{and}
  \bibinfo{author}{\bibfnamefont{B.}~\bibnamefont{Br{\"u}gmann}},
  \bibinfo{journal}{Phys. Rev. Lett.} \textbf{\bibinfo{volume}{78}},
  \bibinfo{pages}{3606} (\bibinfo{year}{1997}), \eprint{gr-qc/9703066}.

\bibitem[{\citenamefont{Lousto et~al.}(2012{\natexlab{b}})\citenamefont{Lousto,
  Nakano, Zlochower, Mundim, and Campanelli}}]{Lousto:2012es}
\bibinfo{author}{\bibfnamefont{C.~O.} \bibnamefont{Lousto}},
  \bibinfo{author}{\bibfnamefont{H.}~\bibnamefont{Nakano}},
  \bibinfo{author}{\bibfnamefont{Y.}~\bibnamefont{Zlochower}},
  \bibinfo{author}{\bibfnamefont{B.~C.} \bibnamefont{Mundim}},
  \bibnamefont{and}
  \bibinfo{author}{\bibfnamefont{M.}~\bibnamefont{Campanelli}},
  \bibinfo{journal}{Phys. Rev.} \textbf{\bibinfo{volume}{D85}},
  \bibinfo{pages}{124013} (\bibinfo{year}{2012}{\natexlab{b}}),
  \eprint{1203.3223}.

\bibitem[{\citenamefont{Zlochower et~al.}(2005)\citenamefont{Zlochower, Baker,
  Campanelli, and Lousto}}]{Zlochower:2005bj}
\bibinfo{author}{\bibfnamefont{Y.}~\bibnamefont{Zlochower}},
  \bibinfo{author}{\bibfnamefont{J.~G.} \bibnamefont{Baker}},
  \bibinfo{author}{\bibfnamefont{M.}~\bibnamefont{Campanelli}},
  \bibnamefont{and} \bibinfo{author}{\bibfnamefont{C.~O.}
  \bibnamefont{Lousto}}, \bibinfo{journal}{Phys. Rev.}
  \textbf{\bibinfo{volume}{D72}}, \bibinfo{pages}{024021}
  (\bibinfo{year}{2005}), \eprint{gr-qc/0505055}.

\bibitem[{\citenamefont{Marronetti et~al.}(2008)\citenamefont{Marronetti,
  Tichy, Br{\"u}gmann, Gonzalez, and Sperhake}}]{Marronetti:2007wz}
\bibinfo{author}{\bibfnamefont{P.}~\bibnamefont{Marronetti}},
  \bibinfo{author}{\bibfnamefont{W.}~\bibnamefont{Tichy}},
  \bibinfo{author}{\bibfnamefont{B.}~\bibnamefont{Br{\"u}gmann}},
  \bibinfo{author}{\bibfnamefont{J.}~\bibnamefont{Gonzalez}}, \bibnamefont{and}
  \bibinfo{author}{\bibfnamefont{U.}~\bibnamefont{Sperhake}},
  \bibinfo{journal}{Phys. Rev.} \textbf{\bibinfo{volume}{D77}},
  \bibinfo{pages}{064010} (\bibinfo{year}{2008}), \eprint{0709.2160}.

\bibitem[{\citenamefont{Lousto and Zlochower}(2008)}]{Lousto:2007rj}
\bibinfo{author}{\bibfnamefont{C.~O.} \bibnamefont{Lousto}} \bibnamefont{and}
  \bibinfo{author}{\bibfnamefont{Y.}~\bibnamefont{Zlochower}},
  \bibinfo{journal}{Phys. Rev.} \textbf{\bibinfo{volume}{D77}},
  \bibinfo{pages}{024034} (\bibinfo{year}{2008}), \eprint{0711.1165}.

\bibitem[{\citenamefont{Loffler et~al.}(2012)\citenamefont{Loffler, Faber,
  Bentivegna, Bode, Diener et~al.}}]{Loffler:2011ay}
\bibinfo{author}{\bibfnamefont{F.}~\bibnamefont{Loffler}},
  \bibinfo{author}{\bibfnamefont{J.}~\bibnamefont{Faber}},
  \bibinfo{author}{\bibfnamefont{E.}~\bibnamefont{Bentivegna}},
  \bibinfo{author}{\bibfnamefont{T.}~\bibnamefont{Bode}},
  \bibinfo{author}{\bibfnamefont{P.}~\bibnamefont{Diener}},
  \bibnamefont{et~al.}, \bibinfo{journal}{Class. Quant. Grav.}
  \textbf{\bibinfo{volume}{29}}, \bibinfo{pages}{115001}
  (\bibinfo{year}{2012}), \eprint{1111.3344}.

\bibitem[{ein()}]{einsteintoolkit}
\bibinfo{note}{Einstein Toolkit home page: {\tt http://einsteintoolkit.org}}.

\bibitem[{cac()}]{cactus_web}
\bibinfo{note}{Cactus Computational Toolkit home page: {\tt
  http://cactuscode.org}}.

\bibitem[{\citenamefont{Schnetter et~al.}(2004)\citenamefont{Schnetter, Hawley,
  and Hawke}}]{Schnetter-etal-03b}
\bibinfo{author}{\bibfnamefont{E.}~\bibnamefont{Schnetter}},
  \bibinfo{author}{\bibfnamefont{S.~H.} \bibnamefont{Hawley}},
  \bibnamefont{and} \bibinfo{author}{\bibfnamefont{I.}~\bibnamefont{Hawke}},
  \bibinfo{journal}{Class. Quant. Grav.} \textbf{\bibinfo{volume}{21}},
  \bibinfo{pages}{1465} (\bibinfo{year}{2004}), \eprint{gr-qc/0310042}.

\bibitem[{\citenamefont{Alcubierre et~al.}(2003)\citenamefont{Alcubierre,
  Br\"ugmann, Diener, Koppitz, Pollney, Seidel, and Takahashi}}]{Alcubierre02a}
\bibinfo{author}{\bibfnamefont{M.}~\bibnamefont{Alcubierre}},
  \bibinfo{author}{\bibfnamefont{B.}~\bibnamefont{Br\"ugmann}},
  \bibinfo{author}{\bibfnamefont{P.}~\bibnamefont{Diener}},
  \bibinfo{author}{\bibfnamefont{M.}~\bibnamefont{Koppitz}},
  \bibinfo{author}{\bibfnamefont{D.}~\bibnamefont{Pollney}},
  \bibinfo{author}{\bibfnamefont{E.}~\bibnamefont{Seidel}}, \bibnamefont{and}
  \bibinfo{author}{\bibfnamefont{R.}~\bibnamefont{Takahashi}},
  \bibinfo{journal}{Phys. Rev.} \textbf{\bibinfo{volume}{D67}},
  \bibinfo{pages}{084023} (\bibinfo{year}{2003}), \eprint{gr-qc/0206072}.

\bibitem[{\citenamefont{van Meter et~al.}(2006)\citenamefont{van Meter, Baker,
  Koppitz, and Choi}}]{vanMeter:2006vi}
\bibinfo{author}{\bibfnamefont{J.~R.} \bibnamefont{van Meter}},
  \bibinfo{author}{\bibfnamefont{J.~G.} \bibnamefont{Baker}},
  \bibinfo{author}{\bibfnamefont{M.}~\bibnamefont{Koppitz}}, \bibnamefont{and}
  \bibinfo{author}{\bibfnamefont{D.-I.} \bibnamefont{Choi}},
  \bibinfo{journal}{Phys. Rev.} \textbf{\bibinfo{volume}{D73}},
  \bibinfo{pages}{124011} (\bibinfo{year}{2006}), \eprint{gr-qc/0605030}.

\bibitem[{\citenamefont{Thornburg}(2004)}]{Thornburg2003:AH-finding}
\bibinfo{author}{\bibfnamefont{J.}~\bibnamefont{Thornburg}},
  \bibinfo{journal}{Class. Quant. Grav.} \textbf{\bibinfo{volume}{21}},
  \bibinfo{pages}{743} (\bibinfo{year}{2004}), \eprint{gr-qc/0306056}.

\bibitem[{\citenamefont{Dreyer et~al.}(2003)\citenamefont{Dreyer, Krishnan,
  Shoemaker, and Schnetter}}]{Dreyer02a}
\bibinfo{author}{\bibfnamefont{O.}~\bibnamefont{Dreyer}},
  \bibinfo{author}{\bibfnamefont{B.}~\bibnamefont{Krishnan}},
  \bibinfo{author}{\bibfnamefont{D.}~\bibnamefont{Shoemaker}},
  \bibnamefont{and}
  \bibinfo{author}{\bibfnamefont{E.}~\bibnamefont{Schnetter}},
  \bibinfo{journal}{Phys. Rev.} \textbf{\bibinfo{volume}{D67}},
  \bibinfo{pages}{024018} (\bibinfo{year}{2003}), \eprint{gr-qc/0206008}.

\bibitem[{\citenamefont{Campanelli and Lousto}(1999)}]{Campanelli:1998jv}
\bibinfo{author}{\bibfnamefont{M.}~\bibnamefont{Campanelli}} \bibnamefont{and}
  \bibinfo{author}{\bibfnamefont{C.~O.} \bibnamefont{Lousto}},
  \bibinfo{journal}{Phys. Rev.} \textbf{\bibinfo{volume}{D59}},
  \bibinfo{pages}{124022} (\bibinfo{year}{1999}), \eprint{gr-qc/9811019}.

\bibitem[{\citenamefont{Lousto and Zlochower}(2007)}]{Lousto:2007mh}
\bibinfo{author}{\bibfnamefont{C.~O.} \bibnamefont{Lousto}} \bibnamefont{and}
  \bibinfo{author}{\bibfnamefont{Y.}~\bibnamefont{Zlochower}},
  \bibinfo{journal}{Phys. Rev.} \textbf{\bibinfo{volume}{D76}},
  \bibinfo{pages}{041502(R)} (\bibinfo{year}{2007}), \eprint{gr-qc/0703061}.

\bibitem[{\citenamefont{Yu et~al.}(2011)\citenamefont{Yu, Lu, Mohayaee, and
  Colin}}]{Yu:2011vp}
\bibinfo{author}{\bibfnamefont{Q.}~\bibnamefont{Yu}},
  \bibinfo{author}{\bibfnamefont{Y.}~\bibnamefont{Lu}},
  \bibinfo{author}{\bibfnamefont{R.}~\bibnamefont{Mohayaee}}, \bibnamefont{and}
  \bibinfo{author}{\bibfnamefont{J.}~\bibnamefont{Colin}},
  \bibinfo{journal}{Astrophys. J.} \textbf{\bibinfo{volume}{738}},
  \bibinfo{pages}{92} (\bibinfo{year}{2011}), \eprint{1105.1963}.

\bibitem[{\citenamefont{Stewart et~al.}(2009)\citenamefont{Stewart, Bullock,
  Barton, and Wechsler}}]{Stewart:2008ep}
\bibinfo{author}{\bibfnamefont{K.~R.} \bibnamefont{Stewart}},
  \bibinfo{author}{\bibfnamefont{J.~S.} \bibnamefont{Bullock}},
  \bibinfo{author}{\bibfnamefont{E.~J.} \bibnamefont{Barton}},
  \bibnamefont{and} \bibinfo{author}{\bibfnamefont{R.~H.}
  \bibnamefont{Wechsler}}, \bibinfo{journal}{Astrophys. J.}
  \textbf{\bibinfo{volume}{702}}, \bibinfo{pages}{1005} (\bibinfo{year}{2009}),
  \eprint{0811.1218}.

\bibitem[{\citenamefont{Hopkins et~al.}(2010)\citenamefont{Hopkins, Bundy,
  Croton, Hernquist, Keres et~al.}}]{Hopkins:2009yy}
\bibinfo{author}{\bibfnamefont{P.~F.} \bibnamefont{Hopkins}},
  \bibinfo{author}{\bibfnamefont{K.}~\bibnamefont{Bundy}},
  \bibinfo{author}{\bibfnamefont{D.}~\bibnamefont{Croton}},
  \bibinfo{author}{\bibfnamefont{L.}~\bibnamefont{Hernquist}},
  \bibinfo{author}{\bibfnamefont{D.}~\bibnamefont{Keres}},
  \bibnamefont{et~al.}, \bibinfo{journal}{Astrophys. J.}
  \textbf{\bibinfo{volume}{715}}, \bibinfo{pages}{202} (\bibinfo{year}{2010}),
  \eprint{0906.5357}.

\bibitem[{\citenamefont{Gerosa et~al.}(2013)\citenamefont{Gerosa, Kesden,
  Berti, O'Shaughnessy, and Sperhake}}]{Gerosa:2013laa}
\bibinfo{author}{\bibfnamefont{D.}~\bibnamefont{Gerosa}},
  \bibinfo{author}{\bibfnamefont{M.}~\bibnamefont{Kesden}},
  \bibinfo{author}{\bibfnamefont{E.}~\bibnamefont{Berti}},
  \bibinfo{author}{\bibfnamefont{R.}~\bibnamefont{O'Shaughnessy}},
  \bibnamefont{and} \bibinfo{author}{\bibfnamefont{U.}~\bibnamefont{Sperhake}}
  (\bibinfo{year}{2013}), \eprint{1302.4442}.

\bibitem[{\citenamefont{Berti et~al.}(2012)\citenamefont{Berti, Kesden, and
  Sperhake}}]{Berti:2012zp}
\bibinfo{author}{\bibfnamefont{E.}~\bibnamefont{Berti}},
  \bibinfo{author}{\bibfnamefont{M.}~\bibnamefont{Kesden}}, \bibnamefont{and}
  \bibinfo{author}{\bibfnamefont{U.}~\bibnamefont{Sperhake}},
  \bibinfo{journal}{Phys.Rev.} \textbf{\bibinfo{volume}{D85}},
  \bibinfo{pages}{124049} (\bibinfo{year}{2012}), \eprint{1203.2920}.

\bibitem[{\citenamefont{Kesden et~al.}(2010{\natexlab{a}})\citenamefont{Kesden,
  Sperhake, and Berti}}]{Kesden:2010ji}
\bibinfo{author}{\bibfnamefont{M.}~\bibnamefont{Kesden}},
  \bibinfo{author}{\bibfnamefont{U.}~\bibnamefont{Sperhake}}, \bibnamefont{and}
  \bibinfo{author}{\bibfnamefont{E.}~\bibnamefont{Berti}},
  \bibinfo{journal}{Astrophys. J.} \textbf{\bibinfo{volume}{715}},
  \bibinfo{pages}{1006} (\bibinfo{year}{2010}{\natexlab{a}}),
  \eprint{1003.4993}.

\bibitem[{\citenamefont{Kesden et~al.}(2010{\natexlab{b}})\citenamefont{Kesden,
  Sperhake, and Berti}}]{Kesden:2010yp}
\bibinfo{author}{\bibfnamefont{M.}~\bibnamefont{Kesden}},
  \bibinfo{author}{\bibfnamefont{U.}~\bibnamefont{Sperhake}}, \bibnamefont{and}
  \bibinfo{author}{\bibfnamefont{E.}~\bibnamefont{Berti}},
  \bibinfo{journal}{Phys. Rev.} \textbf{\bibinfo{volume}{D81}},
  \bibinfo{pages}{084054} (\bibinfo{year}{2010}{\natexlab{b}}),
  \eprint{1002.2643}.

\bibitem[{\citenamefont{Schnittman}(2004)}]{Schnittman:2004vq}
\bibinfo{author}{\bibfnamefont{J.~D.} \bibnamefont{Schnittman}},
  \bibinfo{journal}{Phys. Rev.} \textbf{\bibinfo{volume}{D70}},
  \bibinfo{pages}{124020} (\bibinfo{year}{2004}), \eprint{astro-ph/0409174}.

\bibitem[{\citenamefont{Barausse et~al.}(2012)\citenamefont{Barausse, Morozova,
  and Rezzolla}}]{Barausse:2012qz}
\bibinfo{author}{\bibfnamefont{E.}~\bibnamefont{Barausse}},
  \bibinfo{author}{\bibfnamefont{V.}~\bibnamefont{Morozova}}, \bibnamefont{and}
  \bibinfo{author}{\bibfnamefont{L.}~\bibnamefont{Rezzolla}},
  \bibinfo{journal}{Astrophys. J.} \textbf{\bibinfo{volume}{758}},
  \bibinfo{pages}{63} (\bibinfo{year}{2012}), \eprint{1206.3803}.

\end{thebibliography}

\end{document}